\documentclass{aa}

\usepackage[utf8]{inputenc}
\usepackage{color}
\usepackage{amsmath}
\usepackage{multirow}
\usepackage{xspace}
\usepackage{comment}
\usepackage{ulem}
\usepackage{hyperref}% Clickable links 
\hypersetup{% Links colors parametrisation
    colorlinks=true,
    linkcolor=black,
    filecolor=blue,      
    urlcolor=blue,
    citecolor=blue,
    breaklinks=true,
}

\newcommand\who{WhoSGlAd\xspace}

\newcommand\egg{EGGMiMoSA\xspace}
\newcommand\eggac{\textbf{E}xtracting \textbf{G}uesses about \textbf{G}iants via \textbf{Mi}xed-\textbf{Mo}des \textbf{S}pectrum \textbf{A}djustment\xspace}

\DeclareMathOperator*{\dprime}{\prime \prime}

\newcommand{\algn} [1] {
\begin{align} #1
\end{align}}

\let\originaleqref\eqref
\renewcommand{\eqref}{Eq.~\originaleqref}

\makeatletter

\newcommand{\checknextarg}{\@ifnextchar\bgroup{\gobblenextarg}{}}
\newcommand{\gobblenextarg}[1]{\@ifnextchar\bgroup{, (\ref{#1})\gobblenextarg}{ and (\ref{#1})}}
\makeatother

\title{Asteroseismology of evolved stars with \egg}
\titlerunning{Asteroseismology of evolved stars with \egg}
\subtitle{I. Theoretical mixed-mode patterns from the subgiant to the RGB phase}
\author{M.~Farnir
\and C. Pin\c{c}on
\and M-A.~Dupret
\and A. Noels
\and R. Scuflaire}
\institute{Institut d’Astrophysique et G\'eophysique de l’Universit\'e de Li\`ege, All\'ee du 6 ao\^ut 17, 4000 Li\`ege, Belgium \\ \email{martin.farnir@uliege.be}\label{ULg}}
\date{Received 17/05/2021 /
Accepted 30/06/2021}

\abstract {In the context of an ever increasing amount of highly precise data, thanks to the numerous space-borne missions, came a revolution in stellar physics. This data allowed asteroseismology to thrive and improve our general knowledge of stars. Important results were obtained about giant stars owing to the presence of `mixed modes' in their oscillation spectra. These modes carry information about the whole stellar interior, enabling the comprehensive characterisation of their structure.}
{The current study is part of a series of papers that provide a technique to coherently and robustly analyse the mixed-modes frequency spectra and characterise the stellar structure of stars on both the subgiant branch and red-giant branch (RGB). In this paper we aim at defining seismic indicators, relevant of the stellar structure,  as well as studying their evolution along a grid of models.} {The proposed method, \protect{\egg}, relies on the asymptotic description of mixed modes. It defines appropriate initial guesses for the parameters of the asymptotic formulation and uses a Levenberg-Marquardt minimisation scheme in order to adjust the complex mixed-modes pattern in a fast and robust way.}{We are able to follow the evolution of the mixed-modes parameters along a grid of models from the subgiant phase to the RGB bump, therefore extending previous works. We show the impact of the stellar mass and composition on the evolution of these parameters. We observe that the evolution of the period spacing $\Delta\pi_1$, pressure offset $\epsilon_p$, gravity offset $\epsilon_g$, and coupling factor $q$ as a function of the large frequency separation $\Delta\nu$ is little affected by the chemical composition and that it follows two different regimes depending on the evolutionary stage. On the subgiant branch, the stellar models display a moderate core-envelope density contrast. Therefore, the evolution of $\Delta \pi_1$, $\epsilon_p$, $\epsilon_g$, and $q$ significantly changes with the stellar mass. Furthermore, we demonstrate that, for a given metallicity and with proper measurements of the period spacing $\Delta \pi_1$ and large frequency separation $\Delta \nu$, we may unambiguously constrain the stellar mass, radius and age of a subgiant star. Conversely, as the star reaches the red-giant branch, the core-envelope density contrast becomes very large. Consequently, the evolution of $\epsilon_p$, $\epsilon_g$ and $q$ as a function of $\Delta\nu$ becomes independent of the stellar mass. This is also true for $\Delta \pi_1$ in stars with masses $\lesssim 1.8M_\odot$ because of core electron degeneracy. This degeneracy in $\Delta \pi_1$ is lifted for higher masses, again allowing for a precise measurement of the stellar age. Overall, our computations qualitatively agree with previous observed and theoretical studies.}{The method provides automated measurements of the adjusted parameters along a grid of models and opens the way to the precise seismic characterisation of both subgiants and red giants. In the following papers of the series, we will explore further refinements to the technique as well as its application to observed stars.}

\begin{document}
\maketitle

\section{Introduction}
Red giant and subgiant stars constitute essential ingredients to our understanding of the Universe. Indeed, such stars are very bright and may therefore be observed at large distances and in great numbers. Firstly, the determination of their properties is crucial to galactic archaeology, which is aimed at tracing the structural and dynamical evolution of the Milky Way \citep[e.g.][]{2017AN....338..644M}.
Secondly, these stars are key targets with regard to the precise characterisation of stellar structure and evolution. In the recent decades, the data of unprecedented quality collected by the CoRoT \citep{2009IAUS..253...71B} and {\it Kepler} \citep{2010AAS...21510101B} spacecrafts have enabled us to make a sizeable leap towards the characterisation of red giants and subgiants, thanks to the detection of mixed modes \citep{2011Natur.471..608B}. Even though their detection is recent, their theoretical existence and detectability was predicted early on \citep{1974A&A....36..107S,2009A&A...506...57D}. These modes exhibit a twofold nature: they behave as pressure modes in the outermost regions of the star, with the pressure gradient as the restoring force, and as gravity modes in the innermost regions, with the buoyancy being the restoring force. Both cavities are coupled through an evanescent region, the properties of which determine the coupling strength \citep[e.g.][]{2017A&ARv..25....1H}. These modes constitute a unique opportunity to probe the entire stellar structure as they propagate from the surface to the core. It is not the case in solar-type stars that exhibit pure pressure modes, that propagate in an outer pressure cavity. Consequently, it is only information about the outermost layers of these stars that may be retrieved.

The coupling between gravity modes (g-modes) and pressure modes (p-modes) leads to complex behaviours, that evolve in tandem with the star. On the main sequence, a solar-like oscillator presents a p-modes spectrum that displays significant regularity in frequency. At first order, oscillation modes of a given spherical degree $l$ are separated by a constant quantity, the large separation $\Delta\nu$ \citep{1980ApJS...43..469T,1986HiA.....7..283G}. The observed frequency range is almost constant and lies around the frequency of maximum power, $\nu_{\textrm{max}}$. As the star evolves along the subgiant branch, $\nu_{\textrm{max}}$ decreases. At some point, the observed frequencies of the p-modes become so small that they can couple with the g-modes and create so-called mixed modes. This leads to the phenomenon called avoided crossings \citep{1971PASJ...23..485O,1977A&A....58...41A}. This creates a bumping of the frequency spacing of the modes, perturbing the apparent regularity of the spectrum.
% The frequency of gravity modes increases along with the changes in the shape and size of the evanescent region, allowing for the coupling of an increasing number of central g-modes with superficial p-modes. This coupling leads to avoided crossings along the evolution: when g- and p-modes approach one another in frequency, they exchange nature instead of crossing \citep{1975PASJ...27..237O,1977A&A....58...41A}. This perturbs the regular pattern of a pure pressure spectrum. When a g-mode is encountered, we observe a periodic and local decrease of the separation between two successive modes with the same spherical degree. This decrease is referred to as the mode bumping. Fig. \ref{Fig:Dnuth} provides an illustration of this phenomenon for a theoretical spectrum.
Later on, during the red giant phase, as $\nu_{\textrm{max}}$ continues to decrease the frequency pattern is composed of a large number of modes that behave, at leading order, as gravity modes with a constant separation between successive mode periods, the period spacing $\Delta\pi_1$ \citep{1980ApJS...43..469T}. Again, because of the coupling between p and g-modes, this regularity is disturbed and mode bumping appears, the local period spacing between consecutive modes decreases when encountering p-modes. % This is illustrated in Fig. \ref{Fig:DPth}.
Despite the apparent complexity exhibited by mixed modes, several studies have demonstrated that their frequency pattern can be described via a limited number of parameters.

On the one hand, \cite{2011A&A...535A..91D} described avoided crossings via a series of coupled harmonic oscillators, mimicking the coupling between p- and g-modes. This approach was later used by \cite{2012ApJ...745L..33B} who demonstrated on a grid of subgiants that the coupling strength was predominantly function of the mass. Furthermore, they noted that it should increase right before the transition to the red giant phase. However, linking this approach to the stellar structure is not straightforward. 

On the other hand, to exploit the physical knowledge we have about the stellar structure, many authors rely on the asymptotic description of mixed modes \citep{1979PASJ...31...87S,2016PASJ...68...91T}, which assumes that the oscillating modes are of a short wavelength compared to the variations in the stellar structure (i.e., the modes radial order is large). In this formalism, the resonance condition takes the following form
\begin{equation}
\tan \theta_p = q \tan \theta_g,
\label{Eq:Shi}
\end{equation}
where $\theta_p$ and $\theta_g$ are phase terms describing the propagation of the modes in the pressure and gravity cavities, respectively, and $q$ is the coupling factor describing the level of interaction between both cavities. In this general form, the analytical expressions of these parameters directly depend on the stellar structure properties and the frequency. Based on observations, \cite{2012A&A...540A.143M,2015A&A...584A..50M} proposed explicit formulations for both phases of dipolar modes, which are the most observed:
\begin{equation}
\theta_p = \pi \left(\frac{\nu}{\Delta\nu} - \epsilon_p\right),
\label{Eq:tp}
\end{equation}
\begin{equation}
\theta_g = \pi \left(\frac{1}{\nu\Delta\pi_1} - \epsilon_g + \frac{1}{2}\right).
\label{Eq:tg}
\end{equation}
We present here the gravity phase with an opposite sign for the $1/2$ term. Assuming in addition that $q$ is independent of the frequency, the asymptotic expression is then a function of $5$ frequency-independent parameters (henceforth referred to as the `mixed-modes parameters'): the large separation $\Delta\nu$, the period spacing $\Delta\pi_1$, the pressure offset $\epsilon_p$, the gravity offset $\epsilon_g$, and the coupling factor $q$. Solving Eq. \eqref{Eq:Shi} for $\nu$ provides the theoretical asymptotic frequencies of the dipolar modes. Under the form given by Eqs. \eqref{Eq:Shi}-\eqref{Eq:tg}, the asymptotic formulation has already been shown to be a very powerful tool that allowed us to interpret both observed and model data as functions of the stellar structure.

Indeed, the asymptotic formulation has successfully been applied to adjust observed data in several studies. For example, \citet{2015A&A...584A..50M} use the asymptotic formulation along with a carefully defined variable such that it restores the regularity in the oscillation spectrum and eases its adjustment, the so-called period stretching. This technique was then used by \citet{2016A&A...588A..87V} and \cite{2017A&A...600A...1M,2018A&A...618A.109M} to generate an automated adjustment of a large sample of giant stars. These studies provided an accurate measurement of $\Delta \pi_1$ and $q$ in more than 5000 stars. They were also able to measure $\epsilon_g$ in several hundreds of red giant stars. In addition, the asymptotic formalism was shown to be valid on the subgiant branch. For example, Eqs. \eqref{Eq:Shi}-\eqref{Eq:tg} were also fitted for about 40 stars observed by {\it Kepler} for which we could measure the mixed-mode parameters \citep{2014A&A...572L...5M,2020A&A...642A.226A}.

In order to interpret the observed variations in these parameters, numerous authors took interest in the mixed-modes oscillation spectra from a theoretical point of view, most of them using a grid-based approach. These studies provide invaluable insight on the evolution of the mixed-modes parameters with the stellar parameters. Namely, \cite{2014MNRAS.444.3622J}, \cite{2018A&A...610A..80H}, and \cite{2020MNRAS.495..621J} provided adjustments for $q$ on theoretical frequency spectra computed from red giant stellar models. These studies showed that the decrease observed in the value of $q$ during the evolution along the red giant branch is correlated with the increase in the size of the evanescent region. \cite{2020A&A...634A..68P} demonstrated by means of analytical models that the thickening of this region on the red giant branch actually results from its migration to the radiative core towards the base of the convective envelope. This fact also explains the variations observed in the measurement of the gravity offset \citep{2019A&A...626A.125P}. Other studies demonstrated the interest of the period spacing and large frequency separation as constraints to the stellar structure. Indeed, measuring both $\Delta\pi_1$ and $\Delta\nu$ allows us to distinguish between core helium burning and hydrogen shell burning stages, which are otherwise indistinguishable \citep{2011Natur.471..608B,2014A&A...572L...5M}. This is due to the fact that the core density greatly differs in these stages, therefore impacting the value of $\Delta\pi_1$ \citep{2010ApJ...721L.182M}. 
Also, by measuring the mass of the core in core helium burning stars models, \citet{2013ApJ...766..118M} demonstrated the possibility to constrain the convective overshooting in intermediate mass stars, the amount and nature of which greatly impacts the central stellar composition as well as the duration of the main sequence, directly influencing the inferred stellar age. 

%Other studies further demonstrated that the information carried by the asymptotic description is of prime importance. 

%However, some aspects are not covered by the asymptotic formulation as given by Eq. \eqref{Eq:Shi}. For example, in the case of evolved red giants, it may be necessary to introduce a dependency of the coupling factor on the frequencies in order to properly account for the pattern of the spectrum \citep{2019MNRAS.490..909C}. It may even come as a necessity to consider departures from the asymptotic. In the case of sharp variations in the stellar structure, the asymptotic expansion is not valid locally and the sharp feature leaves an oscillating signal in the observed frequencies, the glitch. For example, \cite{2015ApJ...805..127C} account in their study for buoyancy glitches caused by the composition gradient left by the onset of central helium burning. Other additional information, not included in the asymptotic formulation, may be retrieved about the structure of giants from their oscillation spectra. For example, it was also shown that the small amplitude signature left by the stellar rotation in the spectrum allowed to infer that the stellar core rotates much slower than is expected because of its contraction \citep{2012Natur.481...55B,2014A&A...564A..27D,2018A&A...616A..24G}. \cite{2012A&A...544L...4E} demonstrate that shear-induced mixing and meridional circulation alone are not sufficient to account for the rotational splittings of dipolar mixed-modes. They argue the necessity to include an additional mixing process during the post-main sequence.

All the aforementioned works have demonstrated the high potential of mixed modes to probe and characterise the properties of evolved stars. However, all the information carried by seismic data still remains to be exploited in full. In particular, previous theoretical works mainly focused on the red giant branch and on only one parameter at a time. It is thus necessary to extend these works to the subgiant branch and to account for all the mixed-mode parameters together in a robust and convenient way. Consequently, the present paper is part of a series aiming at providing a method to precisely adjust the mixed-modes pattern of evolved solar-like stars, either extracted from observed seismic data or predicted by a pulsation code, and to tightly constrain the stellar structure. We present in this paper the seismic method we developed, namely: \eggac (\egg), which relies on the asymptotic formulation (Eq. \eqref{Eq:Shi}). In this method, the adjustment is performed thanks to the use of appropriate initial estimations of the five parameters of the asymptotic formulation and a Levenberg-Marquardt minimisation scheme. In the current paper, the aim is to depict the evolution of the five mixed-mode parameters across a grid of models of different masses and chemical compositions, extending from the subgiant phase to the red-giant phase. We insist that our goal is not to provide a detection and identification of mixed modes but rather to asses the relevance of the five mixed-modes parameters as probes of the stellar structure. Therefore, we do not pretend to replace identification methods of the likes of \cite{2015A&A...584A..50M}, as our method should come as a secondary step to such techniques in order to put constraints on stellar models.

This paper is structured as follows. We first present the method and its fitting procedure in Sect. \ref{Sec:Met}. In Sect. \ref{Sec:Ind}, we demonstrate the ability of the technique to properly account for mixed-mode spectra and display the evolution of the adjusted parameters with stellar evolution, mass, and composition. This is followed by a discussion in Sect. \ref{Sec:Dis} and we present out the conclusions in Sect. \ref{Sec:Con}.

\section{Method}\label{Sec:Met}
%\subsection{Theoretical description of mixed-modes oscillations}
In its current version, the \egg method relies on the adapted asymptotic description of the mixed-modes pattern given by Eqs.~\eqref{Eq:Shi}-\eqref{Eq:tg}. The core element of the method is the computation of educated initial guesses of the five mixed-mode parameters, enabling a fast adjustment of a reference spectrum via a Levenberg-Marquardt minimisation algorithm.
Before describing the parameter estimation and the fitting procedure, we first recall a few aspects relevant to the subgiant and red-giant spectra. The generation of the models used for illustration is detailed in Sect. \ref{Sec:Ind}.

\subsection{Typical oscillation spectra}
\begin{figure}
\includegraphics[width=\linewidth]{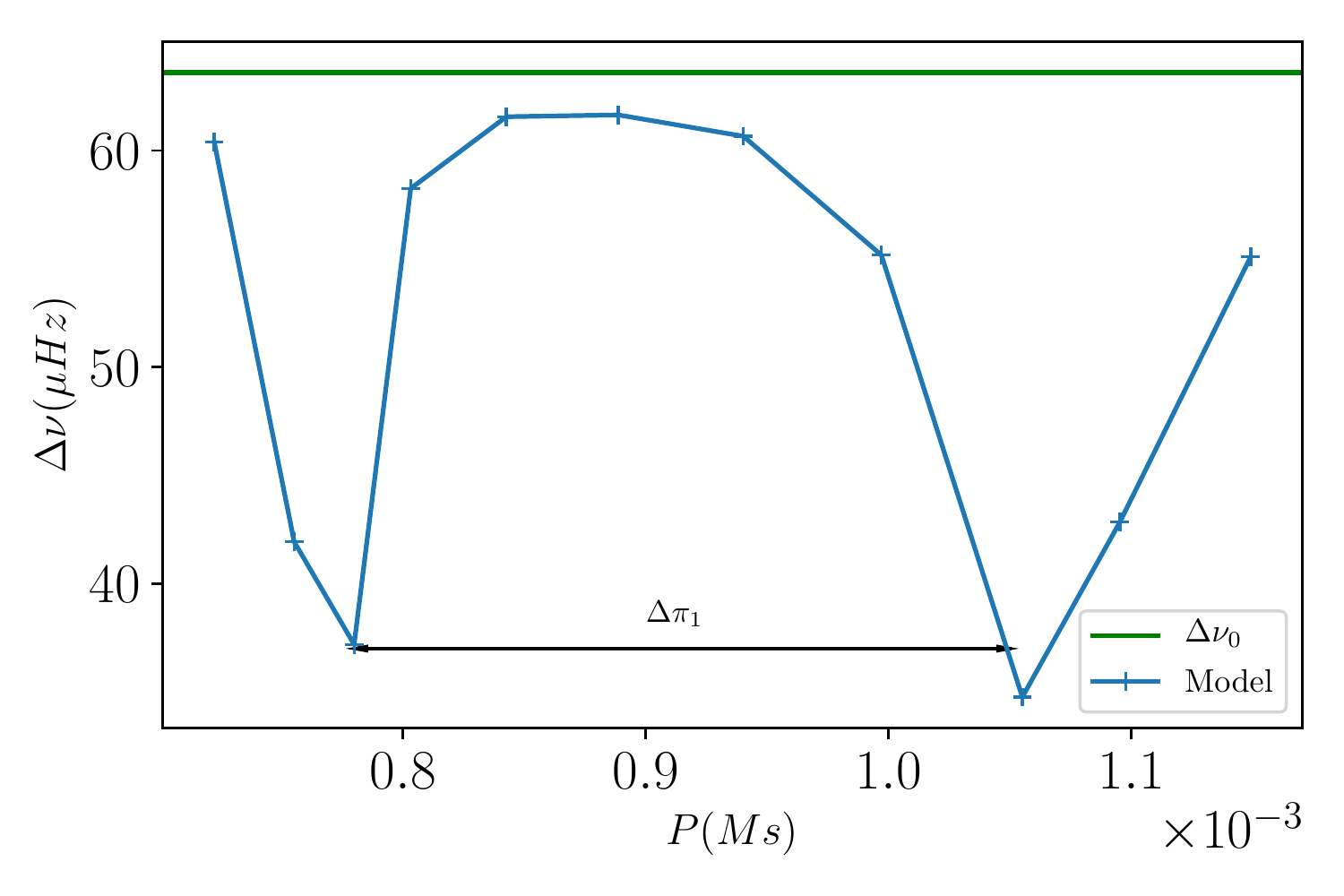}
\caption{Oscillation frequency differences between consecutive modes as a function of the period in the $1M_\odot$ subgiant model presented in \tablename{}~\ref{Tab:ThMod}. The green horizontal line represents the large separation value calculated for radial modes. The double-sided arrow shows the approximate asymptotic period spacing.}\label{Fig:Dnuth}
\end{figure}

\begin{figure}
\includegraphics[width=\linewidth]{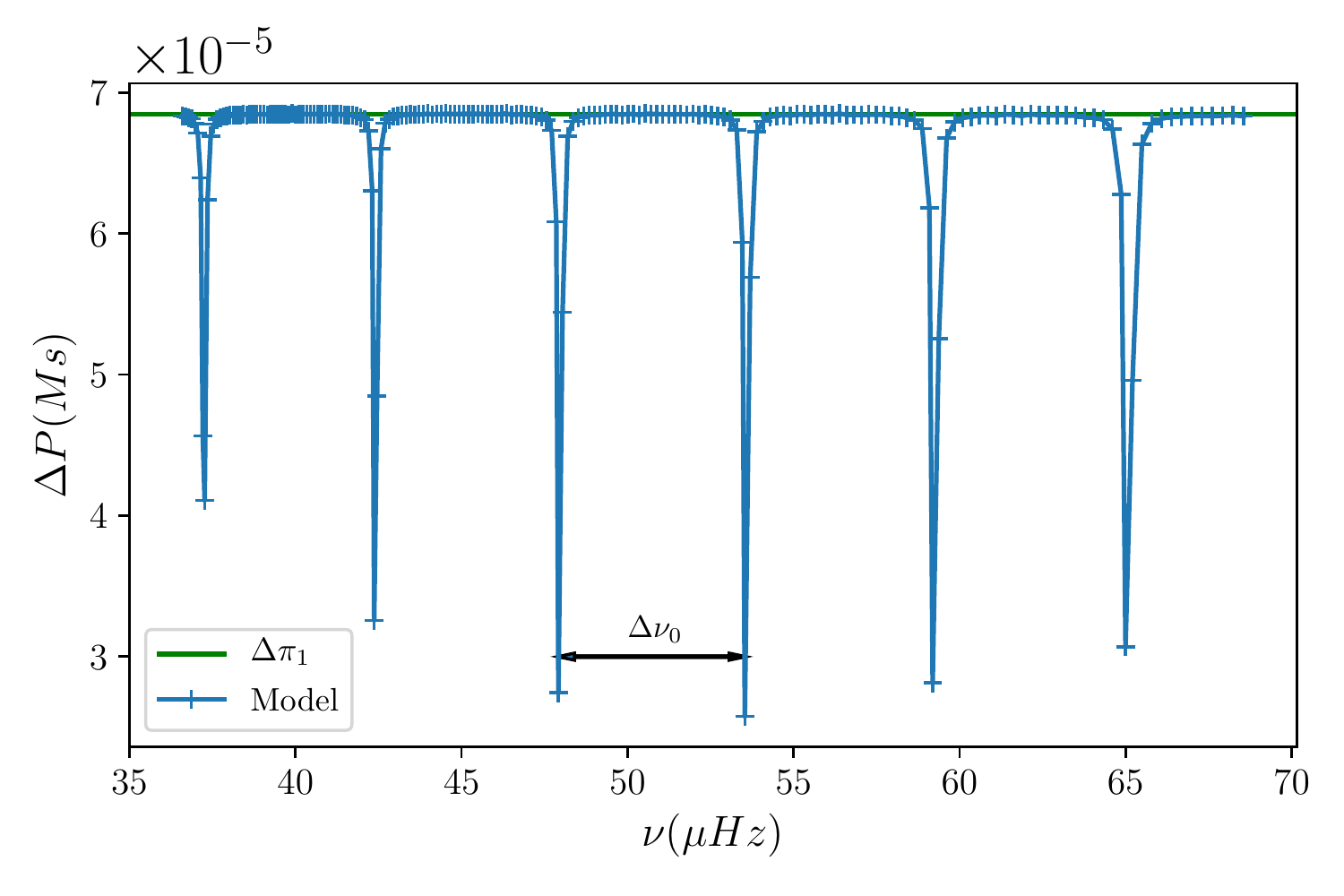}
\caption{Oscillation period differences between consecutive modes as a function of the frequency in the $1M_\odot$ red giant model presented in \tablename{}~\ref{Tab:ThMod}. The green horizontal line represents the asymptotic period spacing. The double-sided arrow shows the large separation value calculated for radial modes.}\label{Fig:DPth}
\end{figure}

%%%%%%%%%%%%%%%%%%%%%%%%%%%%%%%%%%%%%%%%%%%%%%%%%%%%%%%%%%%%%%%%%%%%%%
As a star evolves along the subgiant branch and then rises on the red giant branch, the properties of its spectrum evolve as well. 
As an illustration, we display the theoretical frequency and period differences between consecutive modes for two typical $1M_{\odot}$ solar subgiant and red giant stars in Figs. \ref{Fig:Dnuth} and \ref{Fig:DPth}, respectively. Their parameters are summarised in Table \ref{Tab:ThMod}. The models and their theoretical frequencies were computed with the CLES evolution code and the LOSC pulsation code \citep{2008Ap&SS.316...83S,2008Ap&SS.316..149S}. First, on the subgiant branch, the oscillation spectrum in Fig. \ref{Fig:Dnuth} departs from a pure pressure behaviour, such as solar-like stars display on the main sequence. Nonetheless, the spectrum still shows a majority of pressure-dominated (p-dominated) modes and very few gravity-dominated (g-dominated) modes. As a consequence, successive frequencies are almost evenly spaced. However, the presence of g-dominated modes locally decreases the frequency difference. This results in mode bumping. Conversely, the oscillation spectrum for the red giant star in Fig. \ref{Fig:DPth} displays a greater number of g-dominated modes per p-dominated modes. The modes periods (instead of frequencies) are now predominantly evenly separated. Again, the presence of p-dominated modes locally reduces the period difference, which also corresponds to mode bumping. 

To make the distinction between pressure and gravity dominated spectra the g-dominated modes density has been conveniently defined in \citet{2015A&A...584A..50M} as
\begin{equation}
\mathcal{N}\left(\nu_{\textrm{max}} \right) = \frac{\Delta\nu}{\Delta\pi_1 \nu_{\textrm{max}}^2},
\label{Eq:N}
\end{equation}
with $\nu_{\textrm{max}}$ the frequency of maximum power in the power spectrum. This number represents the ratio of g-dominated modes per p-dominated modes. A g-dominated spectrum will display an $\mathcal{N}$ value greater than unity, while a p-dominated spectrum will have a value lower than unity. For instance, the models plotted in Figs. 1 and 2 have $\mathcal{N}(\nu_{\rm max}) \approx 0.16$ and $30$, respectively.

Moreover, in Fig. \ref{Fig:Dnuth}, we see that the maximum value of the frequency difference in p-dominated spectra is close to the large separation of radial modes, $\Delta\nu_0$ (green horizontal line). We also see that successive bumps are separated by approximately one asymptotic period spacing (black arrow). We recall that the asymptotic period spacing, $\Delta\pi_{1,\textrm{as}}$, is related to the integration of the Brunt-Väisälä frequency, $N$, from the center to the base of the convection zone, $r_{\textrm{BCZ}}$, by the expression
\begin{equation}
\Delta\pi_{1,\textrm{as}} = 2\pi^2\left(\int^{r_{\textrm{BCZ}}}_{0}\frac{N}{r}dr \right)^{-1},
\label{Eq:Dpi}
\end{equation}
with $r$ the distance from the center of the star.
Conversely, in Fig. \ref{Fig:DPth}, we see that the maximum value of the period difference in g-dominated spectra is very close to the asymptotic period spacing (green horizontal line), while the bumps are separated by about the large frequency separation of the radial modes (black arrow). Through an inspection of both figures, we therefore note that we may retrieve estimations for both the large separation and the period spacing directly  from such plots.

\begin{table}
\caption{Parameters of the $1M_{\odot}$ models used to compute the frequencies presented in Figs. \ref{Fig:Dnuth} and \ref{Fig:DPth}.}\label{Tab:ThMod}
\centering
\begin{tabular}{ccc}
\hline 
\hline
\noalign{\smallskip}
 & Subgiant & Red giant \\
 \noalign{\smallskip}
 \hline
 \noalign{\smallskip}
$\mathcal{N}$ & $0.16$ & $29.85$ \\
\noalign{\smallskip}
$\log L/L_{\odot}$ & $0.35$ & $1.39$ \\
\noalign{\smallskip}
$\log T_{\textrm{eff}}$ & $3.74$ & $3.65$ \\
\noalign{\smallskip}
\hline
\end{tabular}
\end{table}

\subsection{Fitting the spectrum}
In the present section, we describe the fitting procedure. In its present version, the goal of the \egg method is to find the values of the five frequency independent mixed-modes parameters ($\Delta\nu$, $\Delta\pi_1$, $\epsilon_p$, $\epsilon_g$, and $q$) in Eqs.~\eqref{Eq:Shi}-\eqref{Eq:tg} that provide the best agreement between the reference and theoretical asymptotic frequencies. Subgiant and red giant stars are known to be slow rotators \citep[e.g.][]{2014A&A...564A..27D,2018A&A...616A..24G}, so that rotation perturbs at first-order only the frequencies of the prograde and retrograde modes. We focus on the m=0 modes in the present paper and thus do not include the contributions of rotation. The adjustment is carried in the following steps:
\textbf{1.} We estimate $\Delta\nu$ and $\epsilon_{p}$ with \who.
\textbf{2.} We estimate the g-dominated mode density.
\textbf{3.} We provide the initial estimates for $\Delta\pi_1$, $\epsilon_g$, and $q$.
\textbf{4.} We adjust frequency (p-dominated spectrum) or period (g-dominated) differences.
\textbf{5.} We adjust individual frequencies.

\begin{figure}
\includegraphics[width=\linewidth]{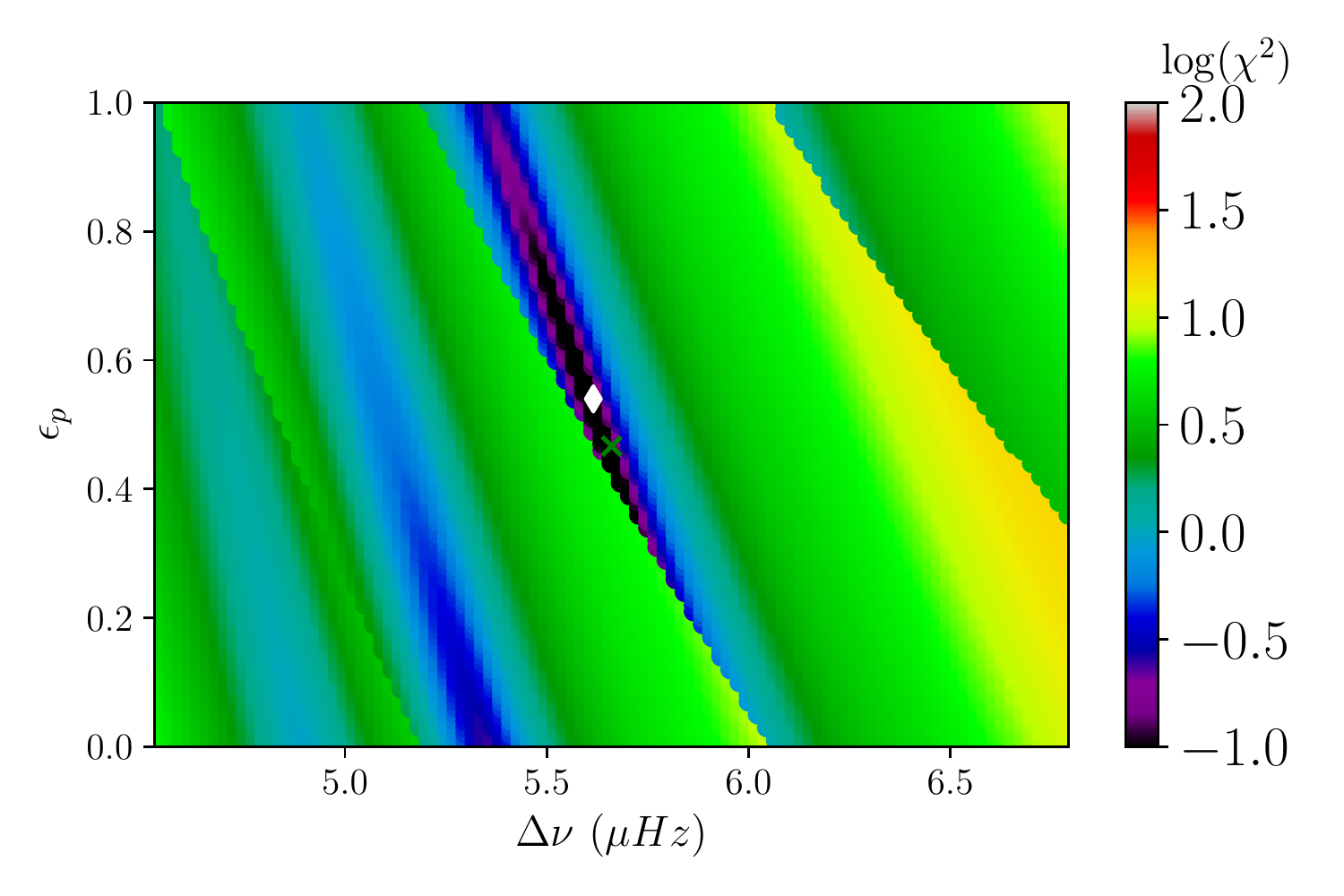}
\caption{Cost function landscape in the neigbourhood of the known solution as a function of the parameters $\Delta\nu$ and $\epsilon_p$. The minimum of the $\chi^2$ landscape is represented by the white diamond and the fitted value by the green cross.}\label{Fig:Dnuep}
\end{figure}

\begin{figure}
\includegraphics[width=\linewidth]{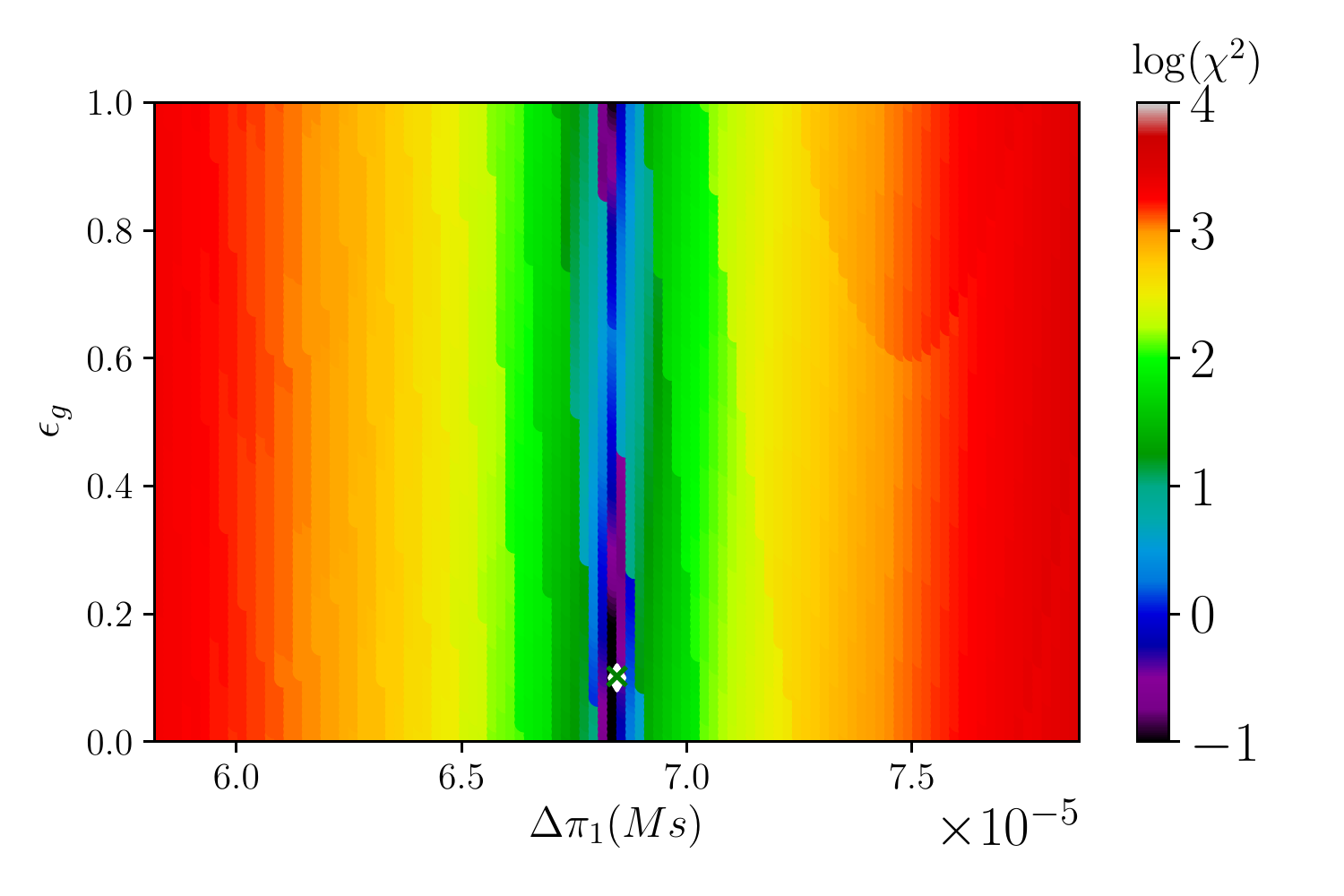}
\caption{Cost function landscape in the neigbourhood of the known solution as a function of the parameters $\Delta\pi_1$ and $\epsilon_g$. The minimum of the $\chi^2$ landscape is represented by the white diamond and the fitted value by the green cross.}\label{Fig:Dpieg}
\end{figure}

As the spectrum adjustment is to be carried via a Levenberg-Marquardt algorithm, which is local, it is crucial to provide proper initial estimates of the parameters. This is even more important as strong correlations exist between the individual parameters of the fit. Indeed, from Eqs. \eqref{Eq:tp} and \eqref{Eq:tg},  we observe tight correlations between $\Delta\nu$ and $\epsilon_p$ and between $\Delta\pi_1$ and $\epsilon_g$. This is also illustrated in Figs. \ref{Fig:Dnuep} and \ref{Fig:Dpieg}. Both figures show the evolution of the $\chi^2$ cost function (measuring the squared difference between the reference and asymptotic frequencies) as a function of two of the five fitted parameters (the three remaining parameters are frozen at their final fitted values). We observe in Fig. \ref{Fig:Dnuep} (respectively Fig. \ref{Fig:Dpieg}) that $\Delta\nu$ and $\epsilon_p$ (resp. $\Delta\pi_1$ and $\epsilon_g$) show an important correlation. In the most extreme case, because of the large value of the pressure (resp. gravity) radial order, a small deviation in the value of $\Delta\nu$ (resp. $\Delta\pi_1$) leads to large differences in $\epsilon_p$ (resp. $\epsilon_g$). Furthermore, we observe steep $\chi^2$ discontinuities. These are the consequence of an improper mode identification caused by the incorrect $\Delta\nu$ and $\epsilon_p$ values. Because of these important correlations, which may impair the convergence of the method, we took special care in devising the initial parameters estimation. In order to provide a first glimpse of the efficiency of the developed method, we represent in these figures the values of the parameters fitted with \egg as green crosses and the minima of the $\chi^2$ in the $2D$ landscape as white diamonds. We see that they greatly match in both cases. We note nevertheless that there is a slight difference, especially in Fig. \ref{Fig:Dnuep}, because both figures constitute a restricted picture of the five-parameter space and the minimum in this restricted space does not necessarily constitute the global five-parameter minimum.

\subsubsection{Estimating $\Delta\nu$ and $\epsilon_p$ with \who}
Since the first detections of solar-like oscillations in red giants \citep[e.g.][]{2002A&A...394L...5F}, it has been known that their spectra always display several radial modes. These may be used in order to estimate a priori the value of the mixed-modes large separation, $\Delta\nu$, and pressure offset, $\epsilon_p$. To do so, we rely on the estimate computed with the \who method \citep{2019A&A...622A..98F} applied to radial modes. This ensures a robust, precise, and fast estimation. This estimation corresponds to a least-squares linear fit of the radial frequencies. We have already highlighted the fact that in order to improve the stability of the method, we maintain the value of $\Delta\nu$ fixed at the first guess.

\subsubsection{Estimating the g-dominated modes density}\label{Sec:NEst}
As the parameter estimation and subsequent steps depend on the g-dominated mode density (Eq. \eqref{Eq:N}), we first need to provide an estimate of this quantity. Nonetheless, as we aim at applying the technique to observed spectra, we cannot assume that we will have access to a measure of $\Delta\pi_1$. We thus provide in this section a technique to recognise g-dominated spectra, p-dominated spectra and intermediate cases. To do so, we take advantage of the following second difference in frequencies:
\begin{equation}
\delta\nu_{2,i} = \frac{\nu_{i+1}-\nu_{i-1}}{\Delta\nu}, \label{Eq:SecDif}
\end{equation}
where the $i$ index is an integer ordering the frequencies in ascending values, and thus the periods in decreasing values. We recall that the $\Delta\nu$ value was previously retrieved via the \who method on radial modes.

In the case of a mixed-mode oscillation spectrum, the second difference is expected to take values between $0$ and $2$. In a pressure-dominated spectrum, as the number of p-dominated modes exceeds that of g-dominated modes, individual modes are almost evenly spaced in frequency of one large separation. As a consequence, the second difference takes a value between $1$ and $2$. Conversely, in a gravity dominated spectrum, the modes are now almost evenly spaced in period of one period spacing. To make the link with the second frequency difference, we can write using $\nu_i=1/P_i$ such that
\begin{align}
\delta \nu_{2,i} & =\frac{\Delta\pi_1\nu_{i+1}\nu_{i-1}}{\Delta\nu} \delta P_{2,i}  \; , \label{Eq:DP2}
\end{align}
where $\delta P_{2,i}=(P_{i-1}-P_{i+1})/\Delta \pi_1$ is defined as the second period difference.
In Eq.~\eqref{Eq:DP2}, the first factor is smaller than $\mathcal{N }(\nu_j)^{-1}$, which is much smaller than unity in g-dominated spectra, and $\delta P_{2,i}=(P_{i-1}-P_{i+1})/\Delta \pi_1\lesssim 2$.
Consequently, the second frequency difference now ranges between $0$ and $1$. Finally, if the g-dominated modes density was exactly equal to $1$, this would mean that the spectrum would alternate between p-dominated and g-dominated modes and the second difference would be exactly equal to $1$ as well. Therefore, using the second difference, we may easily distinguish the different types of spectra. We will consider a spectrum with $\delta\nu_{2,i} > 1$ everywhere as completely p-dominated; a spectrum with $\delta\nu_{2,i} < 1$ everywhere as completely g-dominated; and a spectrum for which the second difference crosses the value of $1$ as an intermediate case. We provide in Appendix~\ref{Sec:SecDif} a visual and mathematical justification of the validity of these previous heuristic arguments.

\subsubsection{$\Delta\pi_1$, $\epsilon_g$ and $q$ initial estimation}\label{Sec:ParEst}

%Nevertheless, in both cases, remaining in close range of the solution ensures the convergence. Furthermore, one must bear in mind that the presented figures only represent a $2D$ of the $5$ dimensional parameters space and that the additional parameters will help constrain the fit.

We now present the estimation of the remaining three parameters. They are estimated according to the nature of the spectrum, that is: completely g-dominated, completely p-dominated or intermediate.

\paragraph{g-dominated estimation ($\mathcal{N} \gg 1$, $\delta\nu_2 < 1$):}
In the g-dominated case, the spectrum presents a majority of gravity dominated modes. Also, the gravity dominated modes closest to pure g-modes are located midway between two dips of the period difference curve. As illustrated in Fig. \ref{Fig:DPth}, the maximum of the local period spacing between consecutive modes, denoted $\Delta P_{\rm max}$, provides a first proper estimate for $\Delta\pi_1$.

Next, we use the $\zeta$ function defined by \cite{2015A&A...584A..50M} to provide an initial value for the coupling factor $q$. This function is defined as
\begin{equation}
\zeta = \left\lbrace 1 + \frac{q}{\mathcal{N}} \frac{1}{q^2 \cos^2 \theta_p + \sin^2 \theta_p} \right\rbrace^{-1},
\label{Eq:zeta}
\end{equation}
such that $\frac{dP}{dn} = \zeta\Delta\pi_1$, with $n=n_p-n_g$ the mixed-mode radial order. In the case of g-dominated spectra, $\mathcal{N} \gg 1$, the $\theta_p$ phase is almost constant between successive modes. Assuming in addition that the $\mathcal{N}(\nu)$ function provided in Eq.~\eqref{Eq:N} does not vary between successive modes, $\zeta$ is thus almost constant, and we may integrate the expression for two successive radial orders so that we obtain $\Delta P_i \simeq \zeta(\nu_i) \Delta\pi_1$. We then use this relation to estimate $q$. First, we define the ratio $Z=\zeta_{\textrm{min}}/\zeta_{\textrm{max}}$ with $\zeta_{\rm min}$ and $\zeta_{\rm max}$ corresponding, respectively, to the minimum (i.e., close to a pure p-mode with $\theta_p = k \pi, k \in \mathbb{N}$) and maximum (i.e., close to a pure g-mode with $\theta_p = \pi/2 + k \pi, k \in \mathbb{N}$) values of the $\zeta$ function. 
From the analytical expressions of $\zeta_{\rm min}$ and $\zeta_{\rm max}$, we can thus get an expression of the coupling factor as a function of $\mathcal{N}$ and $Z$:
\begin{equation}
q = \left[\left(Z-1\right) \mathcal{N} +\sqrt{\left( 1- Z\right)^2\mathcal{N}^2 + 4 Z }\right]/2. \label{Eq:qest}
\end{equation}
Second, as $\Delta P_i \simeq \zeta(\nu_i) \Delta\pi_1$, we can estimate $Z$ from the ratio of the minimum and maximum values of the individual observed period spacings. Note that, as we have a first estimate for $\Delta\pi_1$, we also have an estimate of $\mathcal{N}$. Therefore, according to Eq.~\eqref{Eq:qest}, we can obtain an estimate for the coupling factor.

In addition, having an estimate of $\mathcal{N}$ and $q$, we can now compute the $\zeta$ function for any frequency. This allows us in a final step to correct by iteration the first estimated $\Delta\pi_1$ value using the relation $\Delta\pi_1 \simeq \Delta P_{\rm max} /\zeta(\nu_{\rm max}) $, where $\nu_{\rm max}$ is the frequency at the maximum value of the individual period spacings.

Finally, we note that, for g-dominated spectra, the gravity offset will not be adjusted in the subsequent step as period differences will be adjusted (Sect. \ref{Sec:FitDif}). Therefore, we do not need to provide an estimate for its value in the present step.

\paragraph{p-dominated estimation ($\mathcal{N} \ll 1$, $\delta\nu_2 > 1$):}
For pressure-dominated spectra, the gravity-dominated modes correspond to the dips in the frequency difference curve. In that case, both $\Delta\pi_1$ and $\epsilon_g$ are estimated through a linear fit of the identified gravity-dominated modes. The slope of the fit corresponds to $\Delta\pi_1$ and the intercept to $\epsilon_g$. %We note that the use of the criterion presented in Sect. \ref{Sec:NEst} ensures that we use only gravity-dominated modes.

As the approximation $\Delta P_i \simeq \zeta(\nu_i) \Delta\pi_1$ is only valid for g-dominated spectra where $\mathcal{N} \gg 1$, we need an alternative to estimate the coupling factor in p-dominated spectra. For a p-dominated spectrum, $\mathcal{N} \ll 1$, we define the $\zeta^\prime$ function (see App. \ref{Sec:Zeta}) to express the variation of frequency with the mixed-mode radial order, $n$:
\begin{equation}
\zeta^\prime = \left\lbrace 1 + \frac{q\mathcal{N}}{\cos^2 \theta_g + q^2 \sin^2 \theta_g} \right\rbrace^{-1},\label{Eq:zetap}
\end{equation}
such that $\frac{d \nu}{dn} = \zeta^\prime \Delta\nu$. Because the $\theta_g$ function is almost constant between two dips in p-dominated spectra where $\mathcal{N} \ll 1$, $\zeta^{\prime}$ is almost constant as well within the assumption that $\mathcal{N}$ is quasi constant. We may thus integrate over $n$ the previous expression between two successive modes. This yields $\Delta\nu_i \simeq \zeta^{\prime}\left(\nu_i\right)\Delta\nu$, where the dependency on individual frequencies is shown explicitly. Using Eq. \eqref{Eq:Shi}, we can easily show that $\zeta^\prime = 1-\zeta$.

Similarly to the g-dominated case, we define $Z^\prime = \zeta^\prime_{\textrm{max}}/\zeta^\prime_{\textrm{min}}$ where $\zeta^\prime_{\textrm{max}}$ is the maximum value of $\zeta^\prime$ obtained for $\theta_g = k \pi, k \in \mathbb{N}$ (i.e., close to a pure p-mode) and $\zeta^\prime_{\textrm{min}}$ is the minimum value for $\theta_g = \pi/2 + k \pi, k \in \mathbb{N}$ (i.e., close to a pure g-mode). We thus can get an analytical expression of $q$ based on this ratio, that is,
\begin{equation}
q = \left[\left(1-Z^\prime\right) +\sqrt{\left( Z^\prime - 1\right)^2 + 4 Z^\prime \mathcal{N}^2  }\right]/ \left( 2\mathcal{N} Z^\prime \right). \label{Eq:qest_p}
\end{equation}
Using the fact that $\Delta\nu_i \simeq \zeta^{\prime}\left(\nu_i\right)\Delta\nu$, the maximum and minimum values of the $\zeta^{\prime}$ function can then be estimated with the maximum and minimum values of the individual frequency differences. This provides an estimate of $Z^\prime$ that, combined with the estimate of $\mathcal{N}$ from the estimate of $\Delta \pi_1$, provides an estimate of $q$ according to Eq.~\eqref{Eq:qest_p}.

%Again, we note that, for p-dominated spectra, the pressure offset will not be adjusted in the subsequent step as frequency differences will be adjusted (Sect. \ref{Sec:FitDif}). Therefore, we do not need to provide an estimate for its value in the present step.

\paragraph{Intermediate case ($\mathcal{N} \sim 1$, $\delta\nu_2 \sim 1$):}
When we have comparable p-dominated and g-dominated modes densities, that is, $\mathcal{N} \sim 1$, we cannot rely on the characteristic shape of the frequency or period differences to estimate individual parameters. Nevertheless, we may use the transition in the spectrum where $\delta\nu_2 \simeq 1$ to carry this estimation. From Sect. \ref{Sec:NEst}, we know that $\delta\nu_{2,i} \simeq 1$ and $\delta P_{2,i} \simeq 1$ at the transition, where we have the same amount of p-dominated and g-dominated modes. As a consequence, the first factor in the right-hand side of Eq. \eqref{Eq:SecDif} is close to unity and we may retrieve an estimate for $\Delta\pi_1$:
\begin{equation}
\Delta\pi_1 \simeq \frac{\Delta\nu}{\nu_{t+1} \nu_{t-1}}.
\end{equation}
with $t$ being the mode index closest to the transition, that is, where $\delta \nu_{2,t}$ is the closest to unity.

The coupling factor, $q$, is then estimated on the part of the spectrum being the most dominated by one character. If we note $\nu_{\rm inf}$ and $\nu_{\rm sup}$ the lower and upper bounds of the considered frequency range, this corresponds to the p-dominated part around $\nu_{\rm sup}$ if $1/\mathcal{N}(\nu_{\rm sup}) >\mathcal{N}(\nu_{\rm inf})$ or the g-dominated part around $\nu_{\rm inf}$ if $1/\mathcal{N}(\nu_{\rm sup}) <\mathcal{N}(\nu_{\rm inf})$. We then follow the usual previous procedure associated with the dominant character to estimate $q$.

%This corresponds to parts with the greatest value of either $1/\mathcal{N}$ or $\mathcal{N}$. We then follow the procedure associated with the dominant character.

\subsubsection{Fitting differences}\label{Sec:FitDif}
After providing proper estimates for the mixed-modes parameters, we adjust the values of these parameters that allow us to reproduce individual period spacings or frequency differences between consecutive modes (according to the nature of the spectrum). By doing so, we cancel out the correlation with $\epsilon_g$ (resp. $\epsilon_p$), which remains fixed and will be adjusted in subsequent steps. This differs from most techniques present in the literature as they directly adjust the individual frequencies \citep[e.g.][]{2012A&A...540A.143M,2018A&A...610A..80H}. Techniques that adjust period differences also exist \citep{2015ApJ...805..127C,2019MNRAS.490..909C}, similarly to what we propose, however, those are only valid for red giants, which have a g-dominated spectrum ($\mathcal{N} \gg 1$). The present study therefore represents an extension of such works.

\paragraph{g-dominated spectrum:}
When the spectrum is dominated by the contribution of g-dominated modes, we fit individual period spacings. From Eq. \eqref{Eq:Shi}, it is possible to find an expression for individual period spacings:
\begin{equation}
\Delta P_i = P_{i}-P_{i+1} = \left( \Delta n_g + \Delta\psi_i / \pi \right) \Delta\pi_1,
\label{Eq:DPi}
\end{equation}
with $\Delta n_g$, the difference of gravity radial order between two successive modes, $\Delta\psi_i = \psi_{i} - \psi_{i+1}$ and $\psi_i = \arctan\left( \tan \theta_{p,i}/q \right)$. The $i$ index in the $\theta_{p,i}$ term represents the value of $\theta_p$ evaluated at the period of index $i$. The difference $\Delta n_g$ takes either a value of $1$ when two successive modes are g-dominated or $0$ when encountering a p-dominated mode, resulting in a change of the pressure radial order. In practice, we keep $\Delta n_g=1$ to compute the theoretical period difference $\Delta P_i$ in a first step and then subtract $\Delta\pi_1$ to $\Delta P_i$ where its estimate is greater than unity, which is not permitted.
%The latter case is accounted for in the method by subtracting $\Delta\pi_1$ to the theoretical period difference
A further justification of the value of $\Delta n_g$ is given in Appendix \ref{Sec:Deln}.
%This formulation assumes all the modes to be consecutive. We will discuss this limitation in Sect. \ref{Sec:DisMod}.
The three remaining parameters ($\Delta\pi_1$, $\epsilon_p$ and $q$) may be adjusted to reproduce the reference individual period spacings. 

Second, having adjusted individual period spacings, we may find a value of $\epsilon_g$ such that we minimise the difference between reference and fitted periods expressed with the following function:
\begin{equation}
\chi^2 = \sum\limits^N_{i=1} \frac{(P_{i,\textrm{ref}}-P_{i,\textrm{fit}})^2}{\sigma^2_i},
\end{equation}
with $N$ the number of modes to be adjusted and $\sigma_i$ the uncertainties on the period of each mode.

As there only remains one free parameter to be fitted, $\epsilon_g$, minimising the distance between reference and theoretical periods amounts to compute $\frac{\partial \chi^2}{\partial \epsilon_g} = 0$. This yields an analytical expression for $\epsilon_g$ :
\begin{eqnarray}
\epsilon_g &= & \left[ \sum\limits^N_{i=1} \frac{P_{i,\textrm{ref}}}{\sigma^2_i} + \sum\limits^N_{i=1} \sum\limits^{i-1}_{j=1} \frac{\Delta P_{j,\textrm{fit}}}{\sigma^2_i} \right] \frac{1}{\Delta\pi_1 \sum\limits^N_{i=1} 1/\sigma^2_i } \nonumber \\
& - &\left( n_{g,1} -1/2 +\psi_1/\pi \right),
\label{Eq:eg}
\end{eqnarray}
where $\Delta P_{j,\textrm{fit}}$ represent individual period spacings from the previous step and $n_{g,1}$ is the gravity radial order of the first mode in the observed set. Because $\epsilon_g$ is defined modulo $1$ and $n_{g,1}$ is an integer, its actual value does not impact $\epsilon_g$. % As $\Delta\pi_1$ has been estimated through the previous steps, it is possible to identify the individual modes and compute the value of $n_g$ via the resonance condition $\theta_g = n_g\pi$. Given the expression \eqref{Eq:tg}, the integer part of $P/\Delta\pi_1-1/2$ provides a value for $n_g$.

\paragraph{p-dominated spectrum:}
In the case of a pressure dominated spectrum, we proceed in a very similar fashion. First, to avoid the correlation between $\Delta\nu$ and $\epsilon_p$, the individual frequency spacings are adjusted. Their expressed as follows:
\begin{equation}
\Delta \nu_i = \nu_{i+1}-\nu_{i} = \left( \Delta n_p + \Delta\phi_i / \pi \right) \Delta\nu,
\label{Eq:Dnui}
\end{equation}
where $\Delta n_p$ is the difference of pressure radial order between two successive modes and $\phi_i = \arctan\left( q \tan \theta_{g,i} \right)$. Similarly to the g-dominated case, $\Delta n_p$ takes either a value of $1$, for two successive p-dominated mixed-modes, or $0$ when alternating between p-dominated and g-dominated character.

Finally we get the following expression for $\epsilon_p$, minimising the difference between reference and asymptotic frequencies:
\begin{eqnarray}
\epsilon_p &= & \left[ \sum\limits^N_{i=1} \frac{\nu_{i,\textrm{ref}}}{\sigma^2_i} - \sum\limits^N_{i=1} \sum\limits^{i-1}_{j=1} \frac{\Delta \nu_{j,\textrm{fit}}}{\sigma^2_i} \right] \frac{1}{\Delta\nu \sum\limits^N_{i=1} 1/\sigma^2_i } \nonumber \\
& - &\left( n_{p,1} +\phi_1/\pi \right),
\label{Eq:ep}
\end{eqnarray}
with $n_{p,1}$ the radial order of the first mode in the set. As $\epsilon_p$ is defined modulo $1$ and $n_{p,1}$ is an integer, its actual value is not important. We note that, in this context, the $\sigma_i$ now represent uncertainties on the frequencies of each mode. % Its value is retrieved from the resonance condition of pressure modes $\theta_p = n_p \pi$, and is given by the integer part of $\nu/\Delta\nu$. We note that, in this context, the $\sigma_i$ now represent uncertainties on the frequencies of each mode.

\subsubsection{Fitting frequencies}
Independently of the nature of the spectrum, a last Levenberg-Marquardt minimisation step is carried to simultaneously adjust the four parameters $\Delta\pi_1$, $\epsilon_p$, $\epsilon_g$, and $q$ in such a way that the individual theoretical frequencies, that are solutions of Eq. \eqref{Eq:Shi}, reproduce at best the reference frequencies. This last complete adjustment further improves the agreement with the data and also ensures the reduction the uncertainties on the parameters of the adjustment. Before assessing the probing potential of the individual parameters of the adjustment, we tested the ability of the \egg method to retrieve parameters from frequencies that were generated with the asymptotic formulation and known parameters. The results were excellent and did not introduce unwanted biases.

\section{Seismic indicators}\label{Sec:Ind}
In this section we apply the above-described method to several sequences of giant models and display the evolution of the individual parameters to assess their probing potential as relevant proxies of the stellar structure and evolution. The models were computed with the CLES evolution code \citep{2008Ap&SS.316...83S} as described in \cite{2019A&A...622A..98F}. The reference model has a mass of $1M_{\odot}$, with an initial hydrogen abundance of $X_0=0.72$ and metal abundance of $Z_0=0.015$. Oscillation modes are computed using the LOSC oscillation code \citep{2008Ap&SS.316..149S}. Therefore, the reference modes are not the solution of the asymptotic formulation.
Regarding the frequency range considered for each model, \cite{2012A&A...537A..30M} estimated that the extent around $\nu_{\textrm{max}}$ of the modes that are efficiently excited, therefore observable, in red giant stars follow the simple relation $0.66 \nu_{\textrm{max}}^{0.88}$. Typical observations from \cite{2012A&A...543A..54A} for a subgiant star with $\nu_{\textrm{max}}\sim 1000 \mu Hz$ show that a little more than ten radial modes may be clearly identified. Therefore, to match such observed ranges and ensure computing a sufficient amount of modes, we chose a slightly broader range of about $\nu_{\textrm{max}} \pm 0.4 \nu_{\textrm{max}}^{0.88}$. This corresponds, for red giants (resp. subgiants) to approximately three (resp. $5$) p-dominated modes on both sides of $\nu_{\textrm{max}}$, as expected from the observations.

\subsection{Individual spectra}
\begin{figure}
\includegraphics[width=\linewidth]{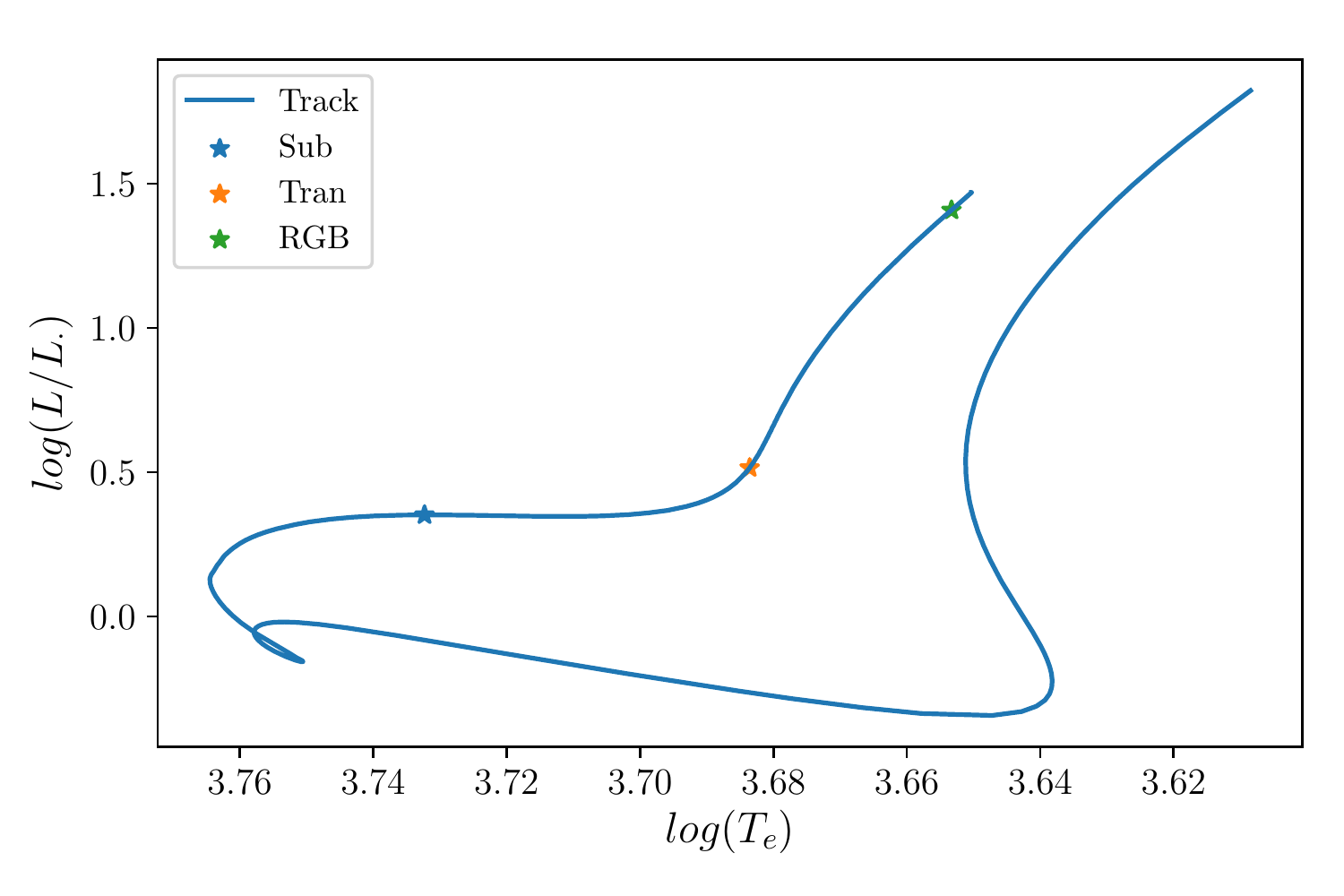}
\caption{Position of the models presented in Figs. \ref{Fig:DnuBeg} to \ref{Fig:DPEnd}}\label{Fig:HR}
\end{figure}

\begin{figure}
\includegraphics[width=\linewidth]{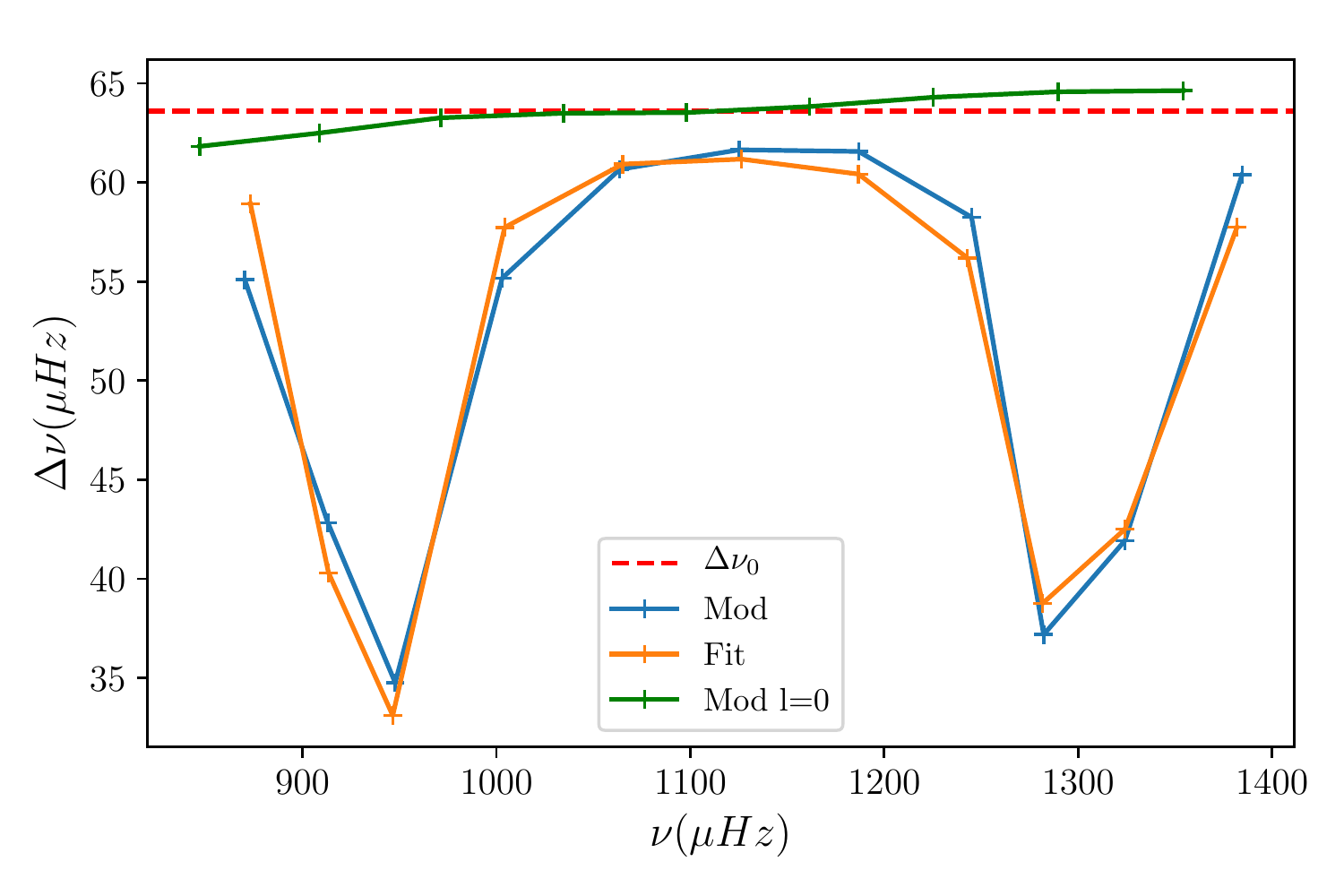}
\caption{Fitted frequency differences as a function of frequency for an early $1M_{\odot}$ subgiant model, denoted  `Sub' in Fig. \ref{Fig:HR}.}\label{Fig:DnuBeg}
\end{figure}

\begin{figure}
\includegraphics[width=\linewidth]{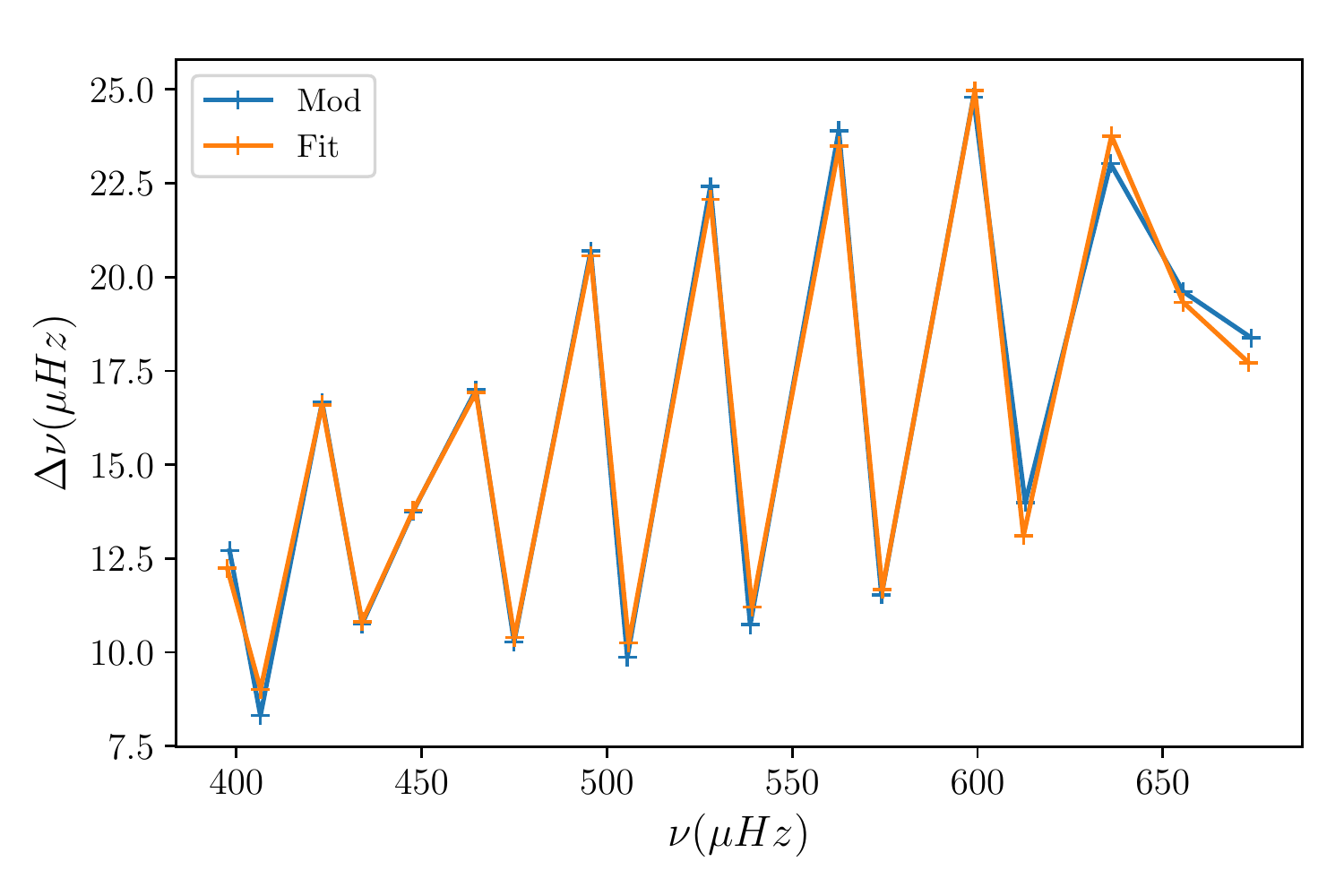}
\caption{Fitted frequency differences as a function of frequency for a $1M_{\odot}$ model with similar numbers of p-dominated and g-dominated modes. It is denoted `Tran' in Fig. \ref{Fig:HR}.}\label{Fig:DnuTran}
\end{figure}

\begin{figure}
\includegraphics[width=\linewidth]{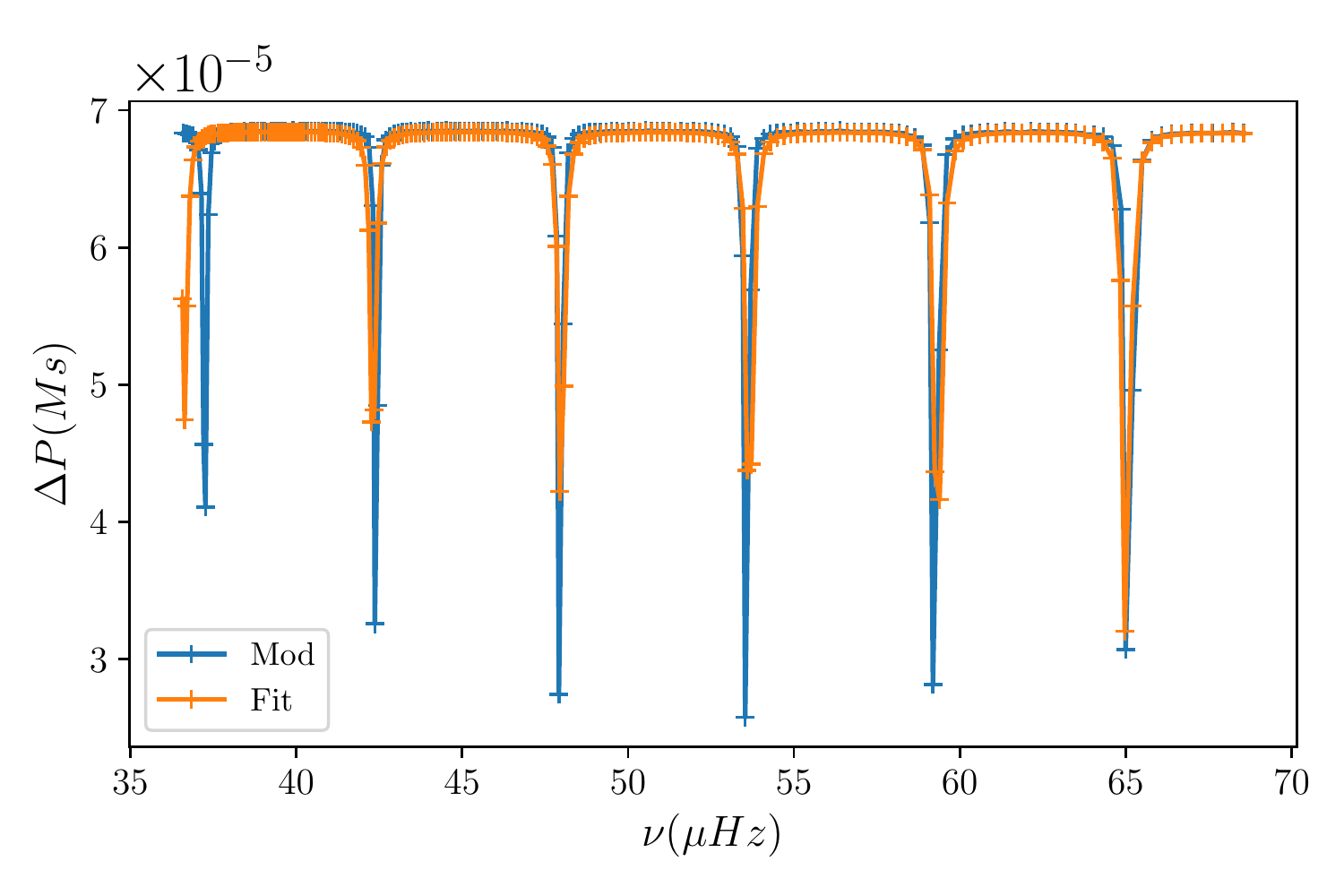}
\caption{Fitted individual period spacings as a function of frequency for a late $1M_{\odot}$ red giant model. It is denoted `RGB' in Fig. \ref{Fig:HR}.}\label{Fig:DPEnd}
\end{figure}

In the present section we display adjusted oscillation spectra of models typical of the Sun ($1M_{\odot}$, $X_0=0.72$ and $Z_0=0.015$) at different stages of evolution: at the beginning of the subgiant phase (subsequently referred to as `Sub'), at the transition between subgiant and red giant phases (`Tran') and at the tip of the red giant branch, before the luminosity bump (`RGB'). These spectra are represented as frequency or period differences (resp. for pressure or gravity dominated spectra) as a function of the frequency. Those stages are represented in a HR diagram in Fig. \ref{Fig:HR} and correspond to $\mathcal{N}$ values of respectively $0.16$, $0.98$, and $29.85$. Figures \ref{Fig:DnuBeg} to \ref{Fig:DPEnd} compare the reference spectra obtained with LOSC (in blue) with the fitted spectra (in orange). To produce these results, the adjustment was undertaken in an automated fashion following the methodology described in Sect. \ref{Sec:Met}. %We immediately observe that the agreement is striking in the three cases. The shape and position of the bumps are well accounted for. Let us now inspect each individual spectra.

\paragraph{Early subgiant:}
Figure \ref{Fig:DnuBeg} corresponds to an early subgiant model (marked `Sub' in Fig. \ref{Fig:HR}), the first one on the sequence displaying two local minima in the individual period spacing, corresponding to mixed modes. This is the much lower threshold of applicability with regard to the \egg technique. Nonetheless, we observe that it is very efficient at providing a qualitative adjustment of the data. Both the shape and position of individual bumps are properly accounted for. However, we note a slight offset in the bump height around a frequency of $1000 \mu Hz$. This offset is similar in amplitude to the error made by assuming the large separation of radial modes to be constant even though it presents a slight dependency with the frequency. This is illustrated by the dashed red line, corresponding to the constant estimate of the large separation of radial modes obtained with \who \citep{2019A&A...622A..98F}, compared to the local value in green. We observe that the offset between the green and red curves is similar to that between the blue and orange ones.

\paragraph{Late subgiant:}
We represent a model presenting a comparable amount of pressure dominated modes and gravity dominated ones in Fig. \ref{Fig:DnuTran} (marked `Tran' in Fig. \ref{Fig:HR}). It corresponds to $\mathcal{N} \simeq 1$. Although the shape of the spectrum is complex, we find a proper fit to the data. This is possible thanks to the proper estimation of the parameters beforehand.

\paragraph{Evolved red giant:}
For the more evolved star displayed in Fig. \ref{Fig:DPEnd}, we again observe a very good agreement with the data. However, we note a slight shift in the position of dips towards low frequencies. Furthermore, we also observe that the adjusted dips tend to be shallower than the data suggests. Possible reasons for such discrepancies will be discussed in Sects. \ref{Sec:DisHig} and \ref{Sec:DisSec}. 

\subsection{Variation with mass along the evolution}
We present in the current section the variation of the parameters of the adjustment with stellar evolution and mass. The models were computed from the beginning of the subgiant phase up to the RGB-bump. The results are displayed in Figs. \ref{Fig:DpiEv} through \ref{Fig:evEv}. To ease the comparison with other works, stellar evolution goes from right to left.

\subsubsection{Period spacing, $\Delta\pi_1$}
\begin{figure}
\includegraphics[width=\linewidth,trim= 0 35 0 0,clip=true]{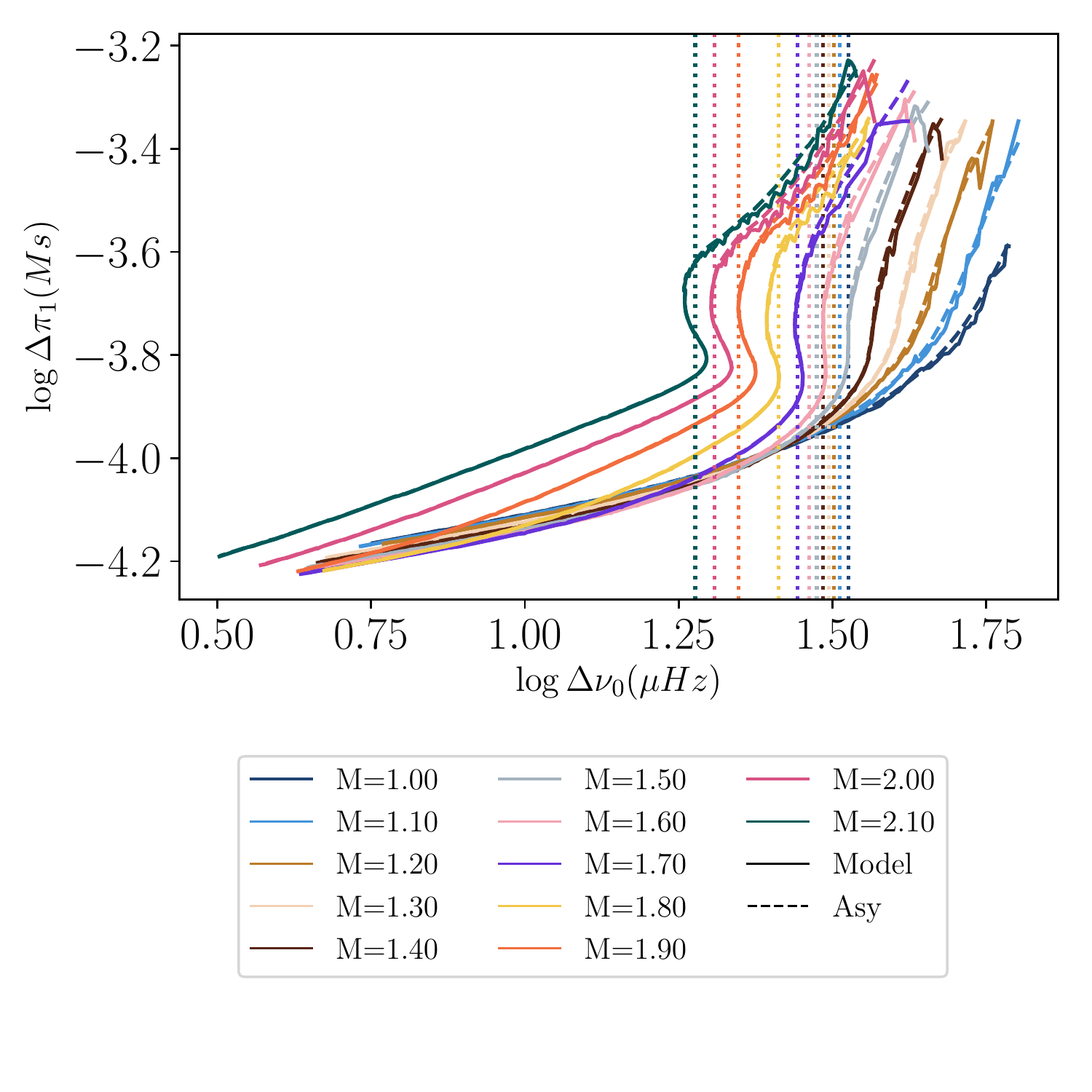}
\caption{Variation of $\Delta\pi_1$ with $\Delta\nu_0$ for different masses, depicted by the colours. The dashed lines correspond to the asymptotic value. The dotted vertical lines correspond to the transition between subgiant and red-giant phases.}\label{Fig:DpiEv}
\end{figure}

Figure \ref{Fig:DpiEv} represents the evolution of the period spacing as a function of the large separation of radial modes which decreases with evolution. The large separation is indeed a proxy of the mean density \citep{1986ApJ...306L..37U,2019A&A...622A..98F} which decreases with evolution during the subgiant and red giant phases. We represent tracks for stellar masses in the range $\left[1.0M_{\odot},2.1M_{\odot}\right]$ ($0.1M_{\odot}$ step) in different colours. We represent the transition between subgiant and red giant phases (at $\mathcal{N} = 1$) by dotted lines. We thus observe that the period spacing decreases with the evolution, at different rates according to the evolutionary phase, the decrease on the subgiant phase being the steepest.

\begin{figure}
\includegraphics[width=\linewidth]{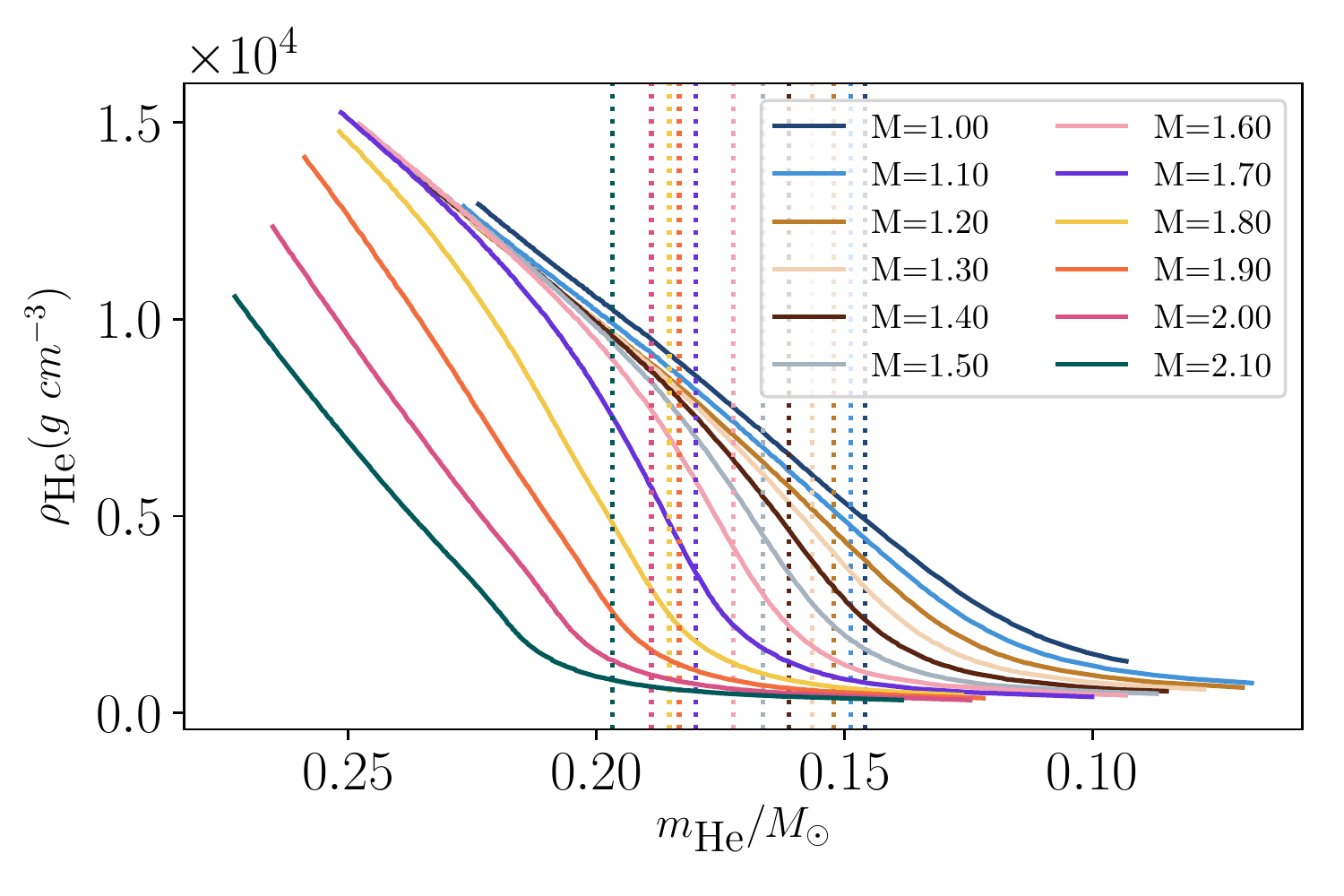}
\caption{Evolution of the helium core density as a function of its mass. The colours and different line styles have the same indications as in Fig. \ref{Fig:DpiEv}.}\label{Fig:rhoHe}
\end{figure}

We first note that our computations in Fig.~\ref{Fig:DpiEv} qualitatively agree with the observations of \citet[][see Fig. 1]{2014A&A...572L...5M}. This confirms that subgiant and red-giant stars occupy distinct regions in a seismic HR diagram. We also note an excellent agreement between the fitted period spacing and its asymptotic value, $\Delta\pi_{1,\textrm{as}}$, represented by dashed lines \citep[see also][]{2016MNRAS.457L..59L}. Assessing the normalised difference between the fitted and asymptotic values of the large separation, given by $\delta\Delta\pi_1=\frac{\left\vert\Delta\pi_{1,\textrm{fit}}-\Delta\pi_{1,\textrm{as}} \right\vert}{\Delta\pi_{1,\textrm{as}}}$, we observe that it never exceeds $0.2\%$ on the red-giant phase. On the subgiant phase, this difference is greater and decreases as the star evolves. It is below $10-15\%$ at the beginning of the subgiant phase and quickly drops below $5-10\%$. This demonstrates that as $n_g$ increases, the validity of the asymptotic analysis improves. Finally, only a few models exceed the $15\%$ disagreement and they correspond to models with only two g-dominated modes, which stands as the very limit of applicability of our technique. This suggests that the adjusted value is a valid proxy of the asymptotic one. This agreement demonstrates that, although the asymptotic approximation is questionable for g-dominated modes in the subgiant phase (the number of nodes of the g-dominated mode eigenfunction in the buoyancy cavity is $n_g \sim 3$ in an early subgiant, and its wavelength is thus large), it globally yields valid results. Using the asymptotic expression in Eq.~\eqref{Eq:Dpi}, we can crudely estimate that $\Delta \pi_{1}$ is about inversely proportional to the maximum of the Brunt-Väisälä frequency in the radiative region, which was shown by \cite{2020A&A...634A..68P} to be approximately proportional to the square root of the helium core density. The evolution of the helium core density as a function of its mass is plotted in Fig. \ref{Fig:rhoHe} for different stellar masses. During these stages, the helium core mass increases as $\Delta \nu$ decreases. We can thus see that the helium core density progressively increases during evolution, leading to the global decrease in the period spacing, as expected.
% The asymptotic period spacing is defined as:
%\begin{equation}
%\Delta\pi_{1,\textrm{as}} = 2\pi^2\left(\int^{r_2}_{r_1}\frac{N}{r}dr \right)^{-1},
%\label{Eq:Dpi}
%\end{equation}
%with $N$ the Brunt-Väisälä frequency, $r_2$ and $r_1$ the inner and outer turning points in the g-cavity.

\begin{figure}
\includegraphics[width=\linewidth]{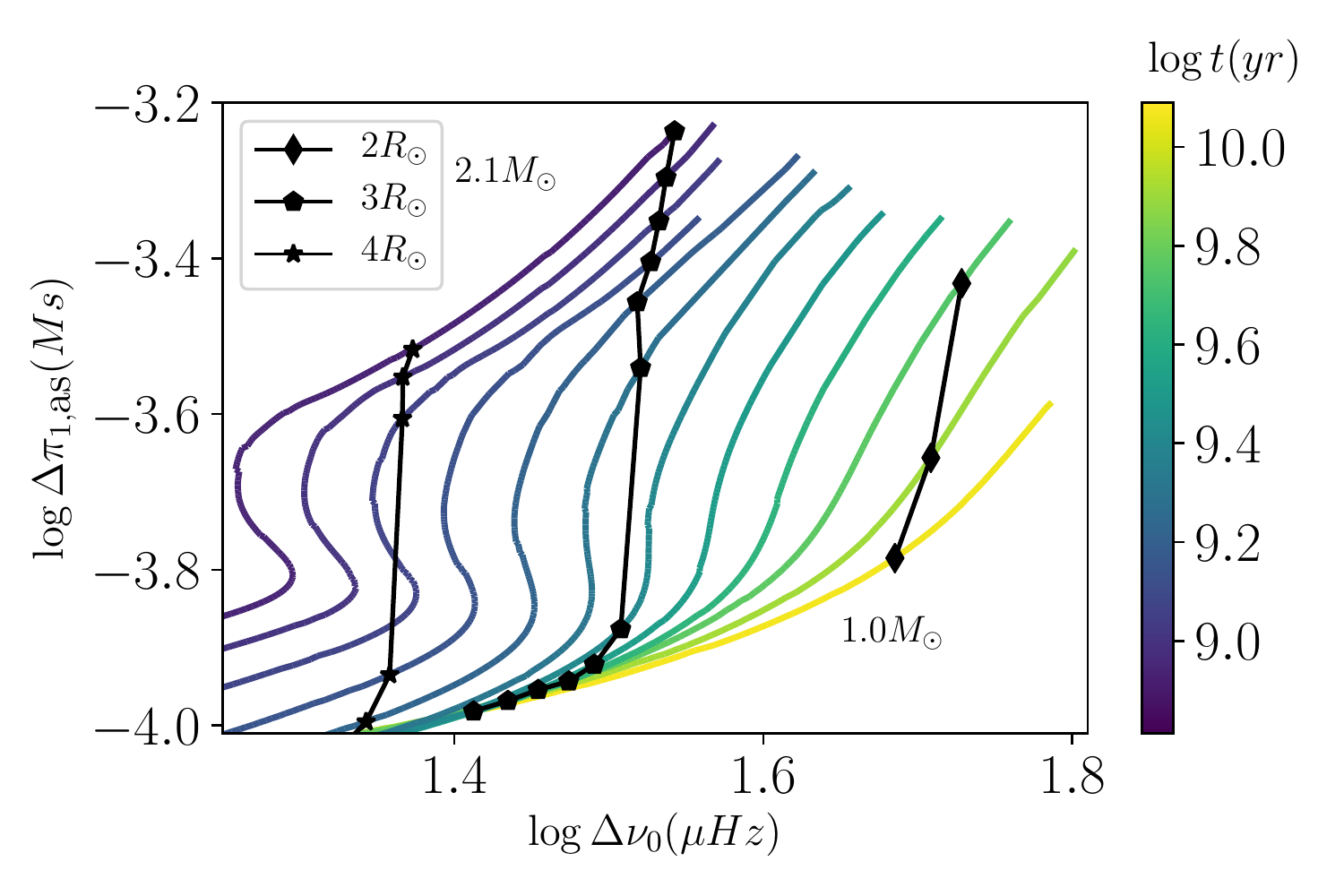}
\caption{Variation of $\Delta\pi_{1,\textrm{as}}$ with $\Delta\nu_0$ on the subgiant phase for different masses. The colour gradient represents the age. The black symbols correspond the models at fixed radius. The diamonds correspond to models of $2R_{\odot}$, pentagons to $3R_{\odot}$, and stars to $4R_{\odot}$.}\label{Fig:DpiEvSub}
\end{figure}

We further note that the subgiant tracks in Fig.~\ref{Fig:DpiEv}, corresponding to different masses, are separated to a significant extent. This trend with the stellar mass can again be explained by the dependence of the helium core density on the stellar mass during the subgiant branch as illustrated in Fig. \ref{Fig:rhoHe}. We see in Fig.~\ref{Fig:DpiEv} that the $\Delta\pi_1$ separation between successive tracks is much larger than the typical observed relative uncertainties from \cite{2020A&A...642A.226A}, which are smaller than $1\%$ in most cases. This demonstrates that the measure of both $\Delta\nu_0$ and $\Delta\pi_1$ should allow us to infer the mass of an observed star with a precision much better than $0.1M_\odot$. Consequently, because the age of a subgiant star is dominated by the duration of the main sequence phase, which is a function of the mass, we may in turn constrain the stellar age. This holds great promises for the accurate characterisation of stellar populations. To further demonstrate that the age of a subgiant may indeed be constrained by the measure of $\Delta\nu_0$ and $\Delta\pi_1$, we display in Fig. \ref{Fig:DpiEvSub}, the evolution of the asymptotic period spacing, $\Delta\pi_{\textrm{as}}$, with $\Delta\nu_0$ along the subgiant phase. The colour gradient corresponds to the stellar age. We observe that individual tracks indeed represent distinct ages. We also show iso-radius values with the black symbols. Models with $2R_{\odot}$ are symbolised by a diamond, models with $3R_{\odot}$ by a pentagon and those with $4R_{\odot}$ by a star. We observe that measuring both $\Delta\pi_1$ and $\Delta\nu_0$ allows us to position a star on this diagram and to constrain its mass, radius and age at a given metallicity. Nevertheless, assuming the duration of the main sequence to be mainly a function of the stellar mass only holds when there is no overshooting during this phase, as is the case for solar-like stars. However, stars with a mass greater than $\sim 1.2M_{\odot}$ have a convective core, and the overshooting may therefore impact the inferred age. For example, \cite{2021A&A...647A.187N} demonstrated in the specific case of the \object{KIC10273246} subgiant that models with a finite amount of overshooting are in better agreement with observed data that models without overshooting. Including the effect of overshooting will thus be mandatory in more quantitative studies that will follow the preliminary exploratory work presented here.

\begin{figure}
\includegraphics[width=\linewidth]{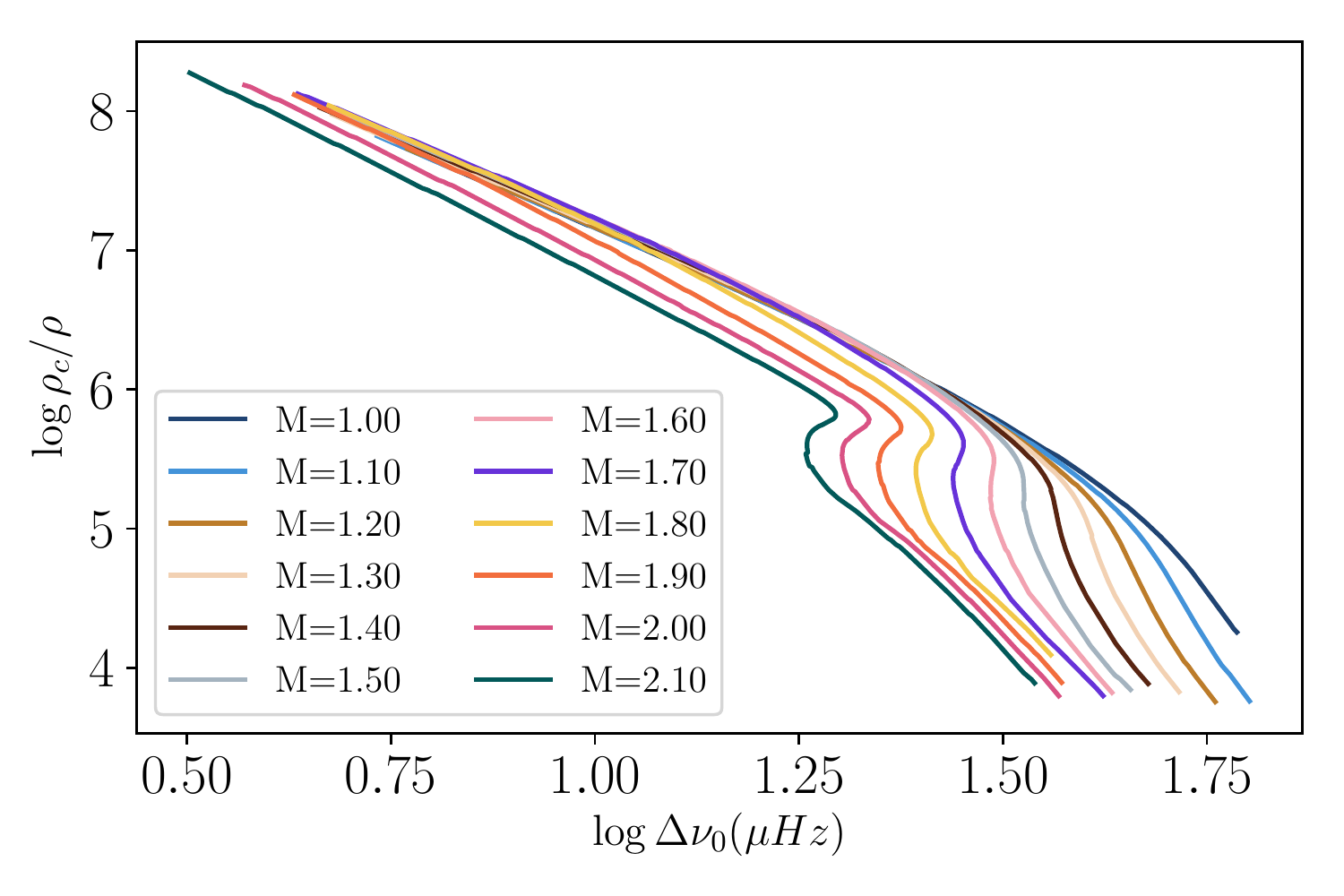}
\caption{Evolution of the ratio of the central density to the mean stellar density as a function the large separation of radial modes. The colours represent different total stellar masses, as in Fig. \ref{Fig:DpiEv}.}\label{Fig:rhoc}
\end{figure}

In red giants with masses $\lesssim 1.8M_{\odot}$, we see in Fig.~\ref{Fig:DpiEv} that the evolution of $\Delta\pi_1$ as a function of $\Delta \nu$ converges to a degenerate track. This degeneracy is actually the result of the electron degeneracy in the helium core at these low masses. In these evolved stars, the density contrast between the core and the envelope is such that the mass of the envelope is negligible compared to that of the core. Therefore, we may show by homology that the properties of the shell are determined by the mass and radius of the helium core \citep{1970A&A.....6..426R,2012sse..book.....K}. Furthermore, because of the central electron degeneracy, the mass and radius of the core are related and the density of the core is a function of the core mass only. As a consequence, the evolution of the helium core density, in these stars with a degenerate core, should be independent of the total stellar mass and vary only with the mass of the helium core. In particular, this is what we observe in Fig. \ref{Fig:rhoHe}. The low-mass tracks indeed converge to an identical evolution once the transition to the red-giant phase, represented by the dotted vertical lines, has been crossed. The consequence of this relation between the core mass and radius is that the properties of the shell are solely determined by the mass of the helium core. The temperature and luminosity of the shell, which, in turn determine the total luminosity, are then only a function of the mass of the core. As the effective temperature is almost constant on the red-giant branch, the stellar radius thus also predominantly depends on the mass of the core. This is also true for the mean density, $\overline{\rho}$, as it is predominantly a function of the stellar radius. Consequently, the same goes for the large frequency separation $\Delta \nu$ that is a proxy of the mean density. This results in a helium core density and a density contrast $\rho_c/\overline{\rho}$, with $\rho_c$ the central density, which only depend on the mass of the helium core.  These quantities are therefore degenerate as well as a function of the the stellar mass for low mass stars with a degenerate core. This is indeed what we observe in Figs. \ref{Fig:rhoHe} and \ref{Fig:rhoc}. The consequence of this degeneracy in the core helium density as a function of $\Delta \nu$ is the degeneracy in period spacing observed in Fig. \ref{Fig:DpiEv}. 
Finally, as the degeneracy is lifted in red-giant stars with masses $\gtrsim 1.8M_{\odot}$, it is theoretically possible to constrain the mass, radius and age of these stars by measuring $\Delta\nu_0$ and $\Delta\pi_1$, similarly to the case of the subgiants. However, in practice, it might not be possible to observe such stars as they evolve fast.

%\begin{figure}
%\centering
%\includegraphics[width=\linewidth]{Figures/Dpi1MeanDiffAll.pdf}
%\caption{Mean normalised difference between asymptotic and fitted values of $\Delta\pi_1$ along the different tracks represented in Fig. \ref{Fig:DpiEv} as a function of the mass and using the same colour code. We define the normalised difference as $\delta\Delta\pi_1=\frac{\left\vert\Delta\pi_{1,\textrm{fit}}-\Delta\pi_{1,\textrm{as}} \right\vert}{\Delta\pi_{1,\textrm{as}}}$. The mean over a complete track is represented by a cross, over the subgiant phase as a diamond, and over the red-giant phase as a triangle.}\label{Fig:DpiDiff}
%\end{figure}

\subsubsection{Pressure offset, $\epsilon_p$}
\begin{figure}
\includegraphics[width=\linewidth]{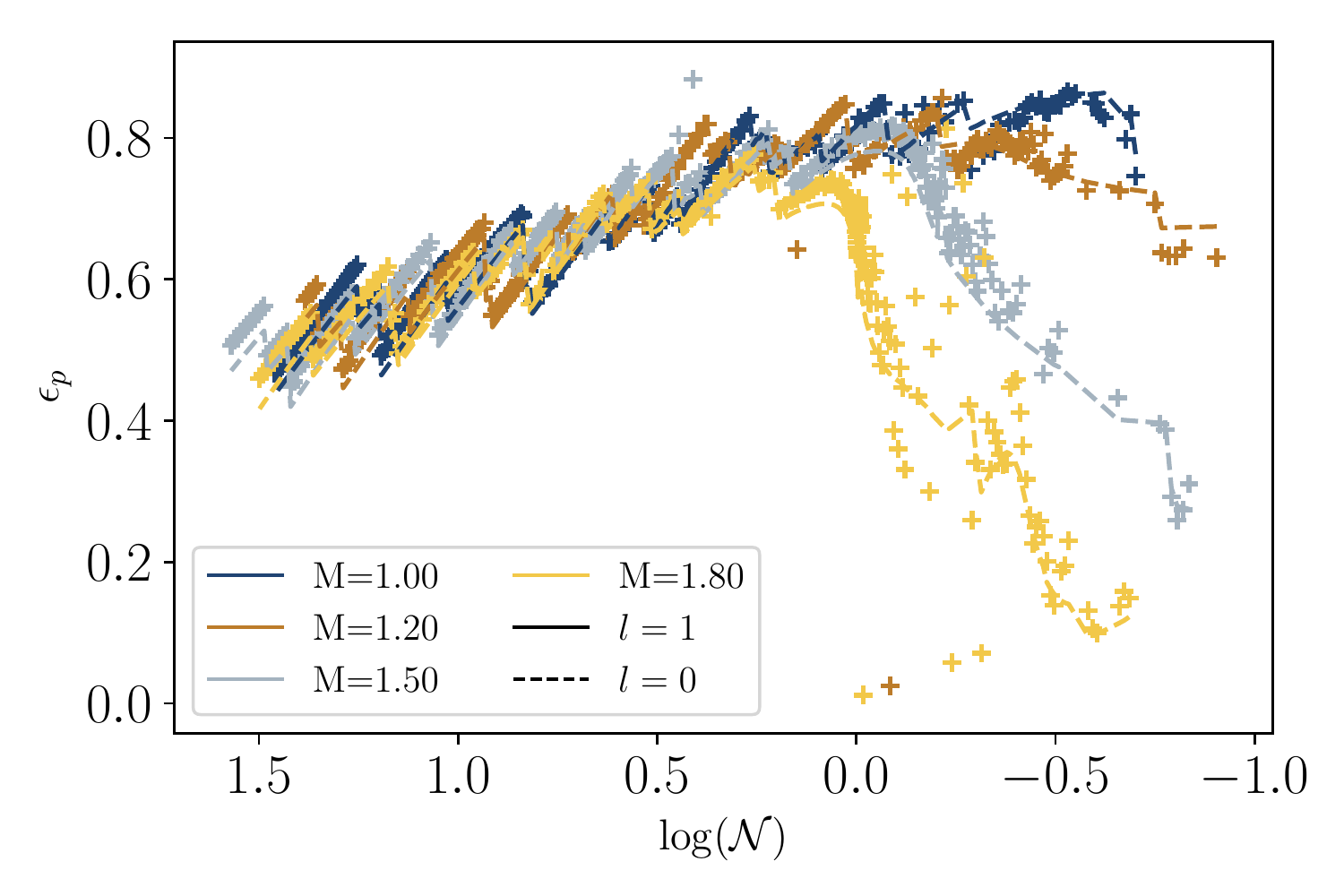}
\caption{Variation of $\epsilon_p$ as a function of $\mathcal{N}$. The dashed line correspond to the value estimated with \who on the radial modes.}\label{Fig:epEv}
\end{figure}

Figure \ref{Fig:epEv} shows the evolution of the pressure offset as a function of $\mathcal{N}$, which increases with evolution. From now on, we restrict the sample of masses to a sub-sample ($1.0$, $1.2$, $1.5$ and $1.8M_{\odot}$) for better clarity. The same trend is followed by models with masses above $1.8M_{\odot}$. The first striking feature is that there exist two regimes, depending on the evolution stage. During the subgiant phase, we observe that $\epsilon_p$ mostly displays an increasing trend, of which the slope as a function of $\mathcal{N}$ increases with mass. This increase is followed by a steady decrease along the red giant phase with a slope that is independent of the mass. This is in qualitative agreement with the measured evolution from \cite[][Fig.7]{2013A&A...550A.126M}. We note that their measurements are shifted up by $0.5$. This is to be expected as they consider radial modes while we consider $l=1$ dipolar modes, which introduces a shift of $l/2$. 

\begin{figure}
\includegraphics[width=\linewidth]{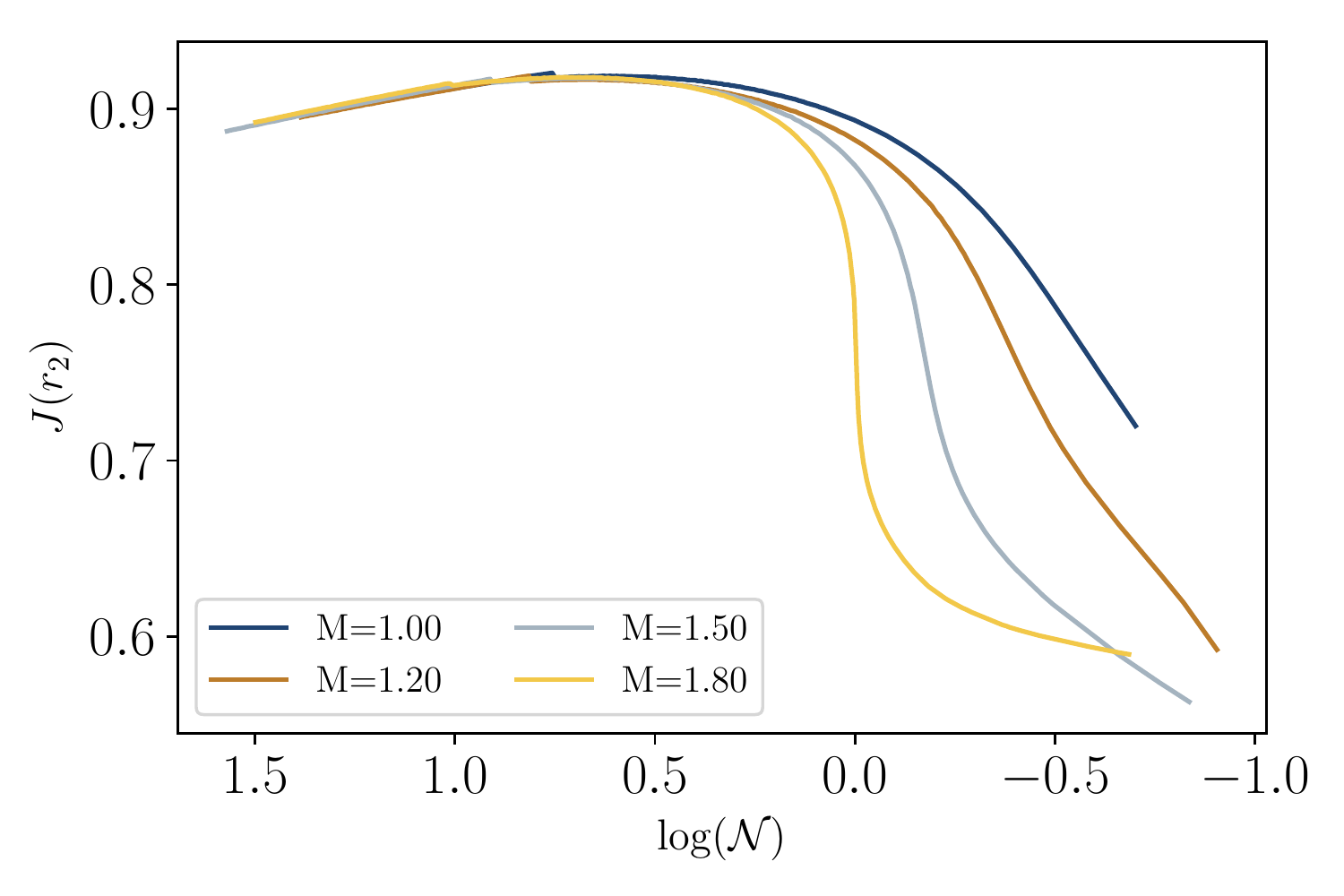}
\caption{Variation of the density contrast compared to the inner encapsulated sphere at the outer edge of the evanescent region, $r_2$, as in Fig. \ref{Fig:DpiEv}.}\label{Fig:JEv}
\end{figure}

We further investigate the two apparent regimes in the evolution of the pressure offset. As it represents the phase lag induced at the boundaries of the pressure cavities, we expect its behaviour to be influenced by their properties. As a consequence, we display in Fig. \ref{Fig:JEv} the density contrast compared to the inner sphere at the lower boundary of the pressure cavity, corresponding to the outer edge of the evanescent region. The local density contrast is defined in \citep{2016PASJ...68...91T} by:
\begin{equation}
J(r)=1-\rho\left(r\right)/\bar{\rho}\left(r\right),
\end{equation}
which compares the local density $\rho$ and local mean density $\bar{\rho}(r)=\frac{m(r)}{4/3\pi r^3}$, with $m(r)$ the mass encapsulated by the sphere of radius $r$.
As an example, a value of $J=0.7$ means the inner sphere is in average three times denser than the local layer whereas a value of $J=0.9$ means the inner sphere is, on average, ten times denser. As the density contrast compared to the inner sphere tends to zero, $J$ tends towards unity.
We observe that the density contrast at the outer edge of the evanescent region, $r_2$, is moderate and strongly varies with the mass in the subgiant phase. 
Then, all the tracks converge towards a similar and high density contrast during the red giant phase (i.e., $J\sim 0.9$).
This matches the observations for the pressure offset, indicating that the pressure offset holds an information about the density contrast and the structure in the evanescent region. Indeed, \cite{2020A&A...634A..68P} showed that the structure of the intermediate evanescent region behaves as power laws of the radius when the density contrast between the core and the evanescent region is large, independently of the stellar mass. This also goes for the Brunt-Väisälä and Lamb frequencies, In contrast, the structure deviates from such a configuration for lower core-envelope density contrast as observed in subgiant stars (see also discussion in Sect.~\ref{Sec:q}). This suggests that the evolution of the core-envelope density contrast between the subgiant and the red giant branches is the main responsible for the different regimes observed in the pressure offset.

In Fig. \ref{Fig:epEv}, during the red giant phase, we observe discontinuities, that result in a seesaw behaviour. This is a direct consequence of the set of modes considered and does not question the quality of the adjustment. Indeed, for such an extended evolution, we may not consider a fixed set of modes, that is, of fixed radial orders. As a consequence, the set shifts towards lower pressure modes orders and discontinuities in the evolution are representative of this shift. Such an effect is discussed in more details in Sect. \ref{Sec:DisMod}.

In this figure, we also represent (as dashed lines) the $\epsilon_{p,0}$ value retrieved for the radial modes via \who. The displayed values account for the $l/2$ shift in value compared with dipolar modes. We observe that the trends of radial and dipolar modes are in excellent agreement, with a slight offset for the most evolved stars. This illustrates that it is a proper estimate for the pressure offset of the dipolar mixed modes. The seesawing of the radial value of $\epsilon_p$ further demonstrates that this is not caused by any improper convergence of the technique. 

In addition, we note that the behaviour is rather erratic during the subgiant phase. This may be a direct consequence of the need to include higher order contributions to the pressure phase, $\theta_p$, because of the extended set of modes. This aspect is further discussed in Sect. \ref{Sec:DisSec}.

\subsubsection{Gravity offset, $\epsilon_g$}
\begin{figure}
\includegraphics[width=\linewidth]{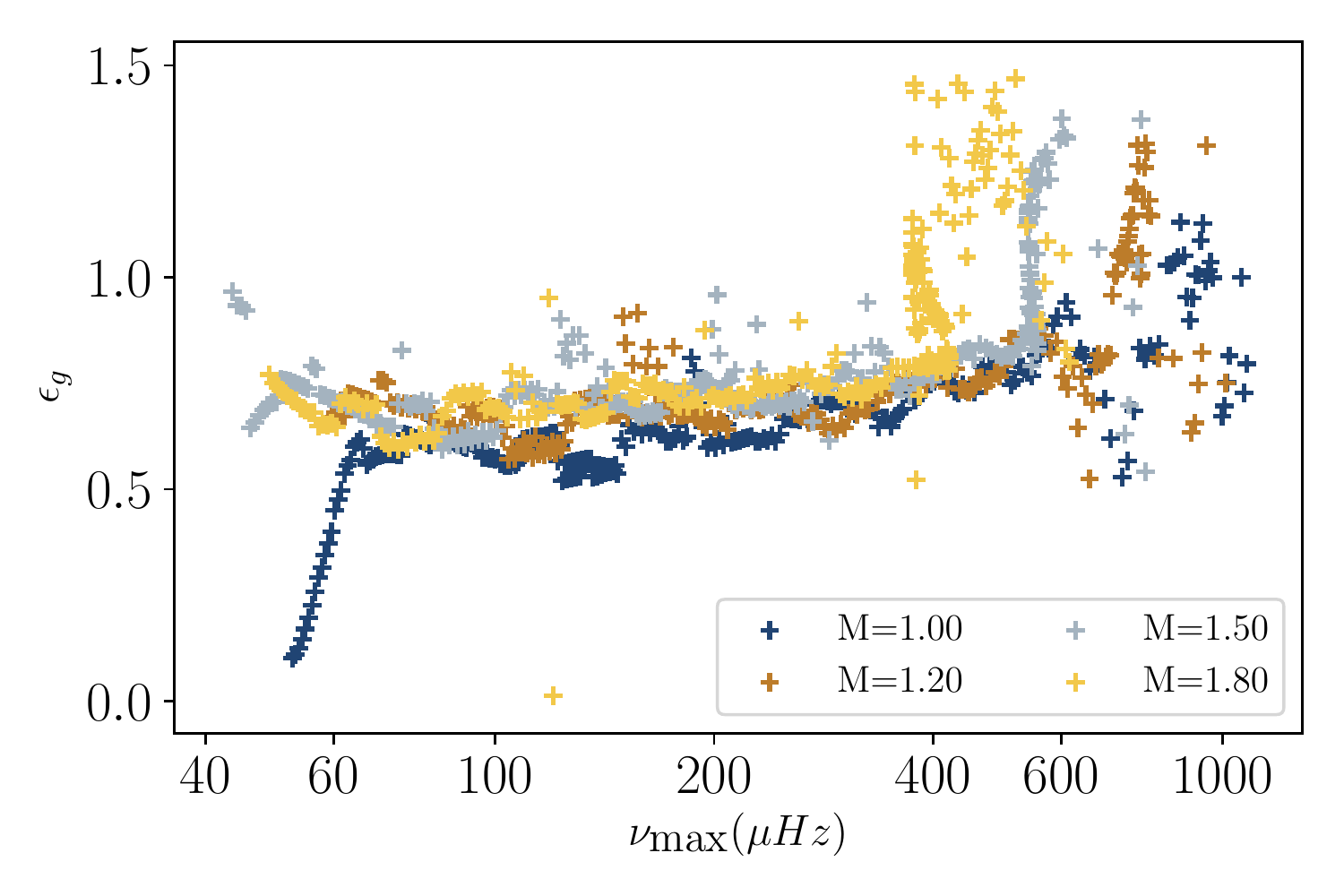}
\caption{Variation of $\epsilon_g$ as a function of $\nu_{\textrm{max}}$ for different masses.}\label{Fig:egEv}
\end{figure}

Figure \ref{Fig:egEv} represents the evolution of $\epsilon_g$ with $\nu_{\textrm{max}}$ \citep[to ease the comparison with][Fig. 4]{2019A&A...626A.125P}. This is, to our knowledge, the first representation of the gravity offset on a grid of models from the subgiant phase to the red giant phase. As for the case of the pressure offset, we observe two regimes, each depending on the evolutionary phase. We expect that this also stems from the qualitative difference in the evolution of the density contrast in the evanescent region.

On the red giant branch, when comparing our results with Fig. 4 of \cite{2019A&A...626A.125P} (which confronts their asymptotic computations with observations from \citealp{2018A&A...618A.109M}) the agreement is convincing. We must bear in mind that we include an additional $1/2$ term in the $\theta_g$ phase compared to their study. As a consequence, the values of $\epsilon_g$ we measure will be shifted up of that same factor compared to theirs. Indeed, in the red giant phase, we observe a plateau at a value of approximately $0.75$ of the gravity offset. Accounting for the shift in values of $0.5$, this is in excellent agreement with their observation of a plateau at an approximate value of $0.25$. This plateau is then followed by a sudden drop of the gravity offset happening in the range of $\nu_{\textrm{max}} \in \left[50 \mu Hz,110 \mu Hz\right]$.

The constant value of the gravity offset during the first part of the red giant branch comes from the fact that, as mentioned earlier, the profiles of the Brunt-Väisälä and Lamb frequencies may be assumed to be parallel and represented by a power-law of radius in the evanescent region because of the high density contrast between the core and the surface. The slope of the Brunt-Väisälä frequency is then constant and determines the gravity offset value. As the star evolves, $\nu_{\textrm{max}}$ decreases along with the set of excited modes. Therefore, the evanescent region moves outwards, up to the point where it penetrates the convective zone. The Brunt-Väisälä frequency then suddenly drops. Both frequencies can no longer be considered parallel to one another. The gravity offset then drops, as observed in Fig. \ref{Fig:egEv} and predicted by \cite{2019A&A...626A.125P}.

In the subgiant phase, we first note that the evolution of $\epsilon_g$ depends on the stellar mass. Similarly to the pressure offset, we expect this dependence to stem from the low and mass-dependent density contrasts displayed by these stars in the evanescent region (see Fig. \ref{Fig:JEv}), in opposition to the high and almost mass-independent density contrasts in red giant stars. We also note that the behaviour is less regular than in the red giant phase. However, individual spectra are properly adjusted, as illustrated in Fig. \ref{Fig:DnuBeg} for the most extreme case. We thus expect this effect to either results from structural features or the necessity to extend the asymptotic formulation to higher orders. Another feature in the subgiant phase is the apparent oscillation for low-mass stars, which should be caused by variations in the evanescent region.% as asymptotically demonstrated in \cite{2019A&A...626A.125P}.

Finally, similarly to the case of the pressure offset, we note a seesaw behaviour. This is again a consequence of the varying set of modes. This will be addressed in the discussion (Sect. \ref{Sec:DisMod}).

\subsubsection{Coupling factor, $q$} \label{Sec:q}
\begin{figure}
\includegraphics[width=\linewidth]{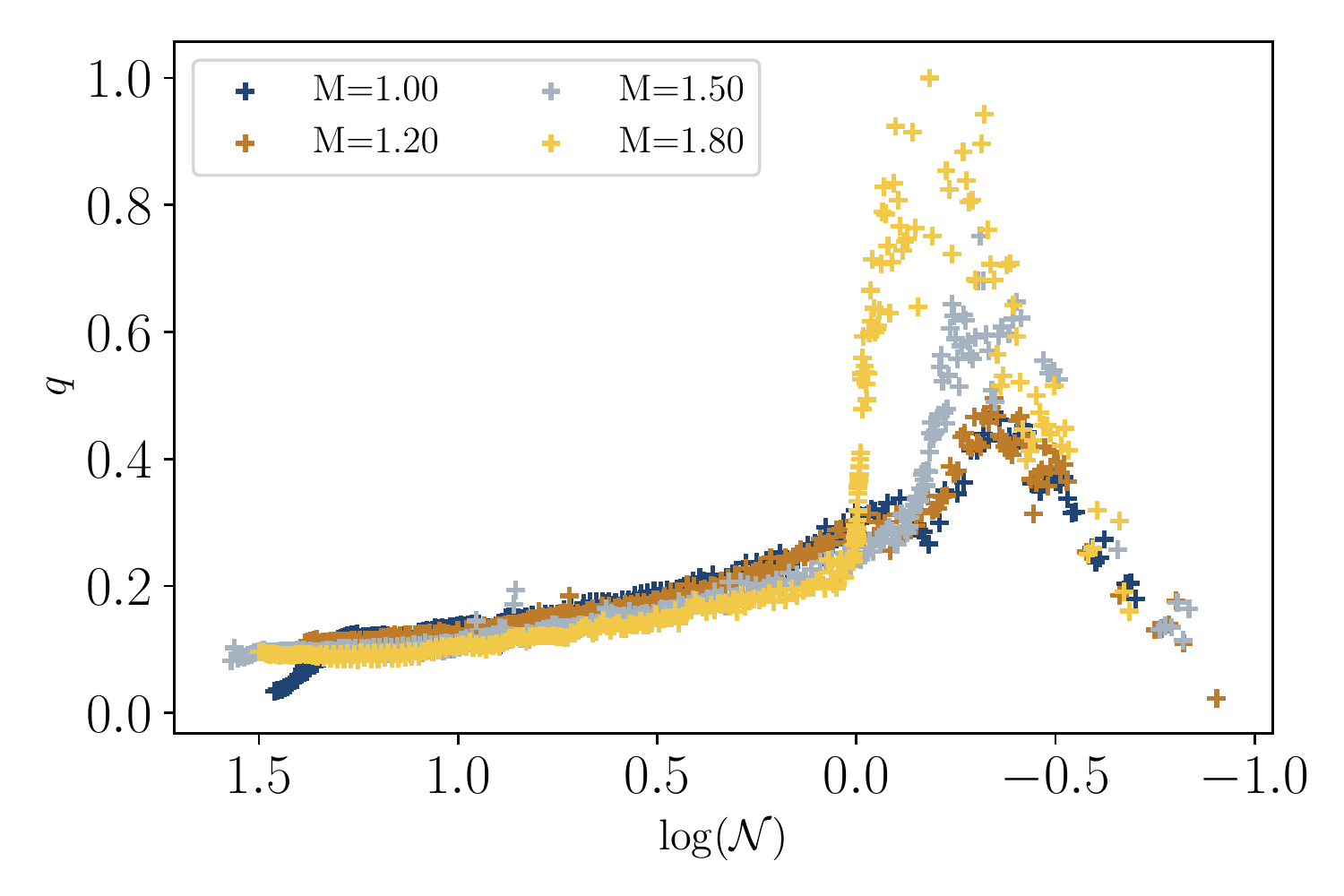}
\caption{Variation of $q$ as a function of $\mathcal{N}$ for different masses.}\label{Fig:qEv}
\end{figure}

The evolution of the coupling factor is displayed in Fig. \ref{Fig:qEv}. We see that the value of $q$ first increases to a high value in the subgiant phase. This corresponds to the case of a strong coupling \citep{2016PASJ...68...91T}. To be complete, we also note that, for the lowest masses ($1.0M_{\odot}$ and $1.2M_{\odot}$), there is a local minimum of the coupling factor before the sharp increase at the end of the subgiant phase. Then it suddenly drops before the red giant phase. Finally, during the red giant phase, the coupling factor steadily decreases from a value of about $0.25$ to approximately $0.10$, corresponding this time to a weak coupling. Eventually, the value of $q$ further drops by the end of the RGB phase. As our sequences stop at the RGB bump, this drop is not visible for all of them. 

\begin{figure}
\includegraphics[width=\linewidth]{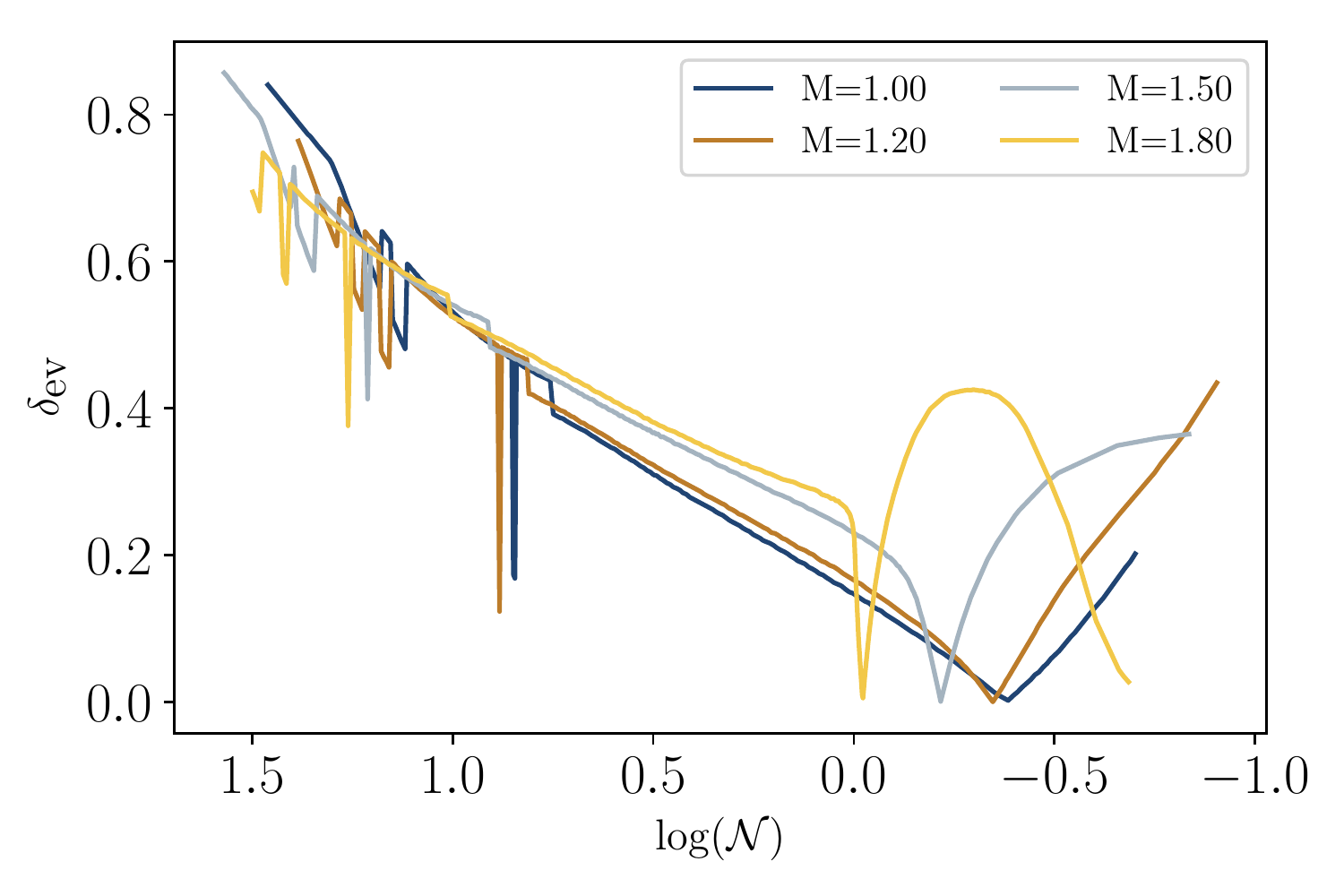}
\caption{Variation of width of the evanescent zone as a function of $\mathcal{N}$ for different masses.}\label{Fig:evEv}
\end{figure}

This predicted evolution of the coupling factor is very similar to observations made by \citet[namely Fig. 6]{2017A&A...600A...1M}. As demonstrated by \cite{2020A&A...634A..68P} under the assumption that the Brunt-Väisälä frequency and the Lamb frequency are log-parallel, the coupling strength should be a proxy of the width of the evanescent zone; thus the larger the evanescent zone, the lower the coupling. To check whether the width of the evanescent region is correlated with the coupling factor, we display in Fig. \ref{Fig:evEv} the evolution of this relative width at $\nu_{\textrm{max}}$, denoted $\delta_{\textrm{ev}}$, as a function of the g-dominated modes density. It is defined as:
\begin{equation}
\delta_{\textrm{ev}} = \frac{r_{2}-r_{1}}{\left(r_{1}+r_{2}\right)/2}, \label{Eq:dr}
\end{equation} 
with $r_{1}$ and $r_{2}$ the positions of the inner and outer edges of the evanescent region (respectively).
For all masses, we indeed observe a global rapid decrease in the size of the evanescent zone during the subgiant phase (with $\mathcal{N} < 1$) followed by a steady increase of this size during the red giant phase (with $\mathcal{N} > 1$). This coincides with the evolution of the coupling factor. We note that the discontinuities in the evolution on the red giant branch come from the discontinuity in composition at the base of the convective envelope, which, in turn, creates a peak of the Brunt-Väisälä frequency \citep[see for example][]{2015ApJ...805..127C}.

Regarding the dependence of $q$ with the stellar mass, the two regimes are again observed, as expected from the behaviour of the density contrast in the evanescent region. During the subgiant phase, the coupling factor strongly depends on the mass while, on the red giant branch, the coupling factor is much less sensitive to the stellar mass. In both cases, the same global trend is nevertheless observed: the higher the mass, the lower the $q$ value. Firstly, on the red giant branch, the density contrast in the evanescent region compared to the inner sphere is large enough for the profiles of the Brunt-Väisälä and Lamb frequencies to be assumed to be parallel and the structure of the evanescent region is quite comparable for all the masses at a given value of $\mathcal{N}$. The width of the evanescent region nevertheless depends slightly on the stellar mass, as seen in Fig.~\ref{Fig:evEv}, explaining the slight dependence on $q$ observed in Fig. \ref{Fig:qEv} on the red giant branch. Only the position of the ultimate drop of the coupling factor by the end of the sequences appears to be significantly affected by the stellar mass. However, as we restricted ourselves to models before the luminosity bump, this drop is not visible for every track. Secondly, on the subgiant branch, the density contrast is moderate and depends on the stellar mass (see Fig. \ref{Fig:JEv}). Because of this lower density contrast than on the red giant branch, the Brunt-Väisälä profile does not follow a simple power-law relation with the radius and may not be assumed to be parallel to the profile of the Lamb frequency. This impacts the evolution of the width of the evanescent region for the different masses, as shown in Fig.~\ref{Fig:evEv}, and thus explains the significant mass dependence of the coupling factor on the subgiant branch. We even note that the $1.8M_{\odot}$ model exhibits an oscillation with regard to the size of its evanescent region.  Indeed, both critical frequencies may cross in this model. As the star evolves, $\nu_{\textrm{max}}$ decreases. It therefore reaches this crossing of the frequencies, corresponding to a very narrow evanescent region. Then, as the star continues to evolve, the evanescent region increases in size again. Furthermore, as the frequency profiles also evolve with time (mainly due to the evolution of the density contrast), the point at which they cross may evolve as well and other minima of the width of the evanescent zone may occur, as we observe in Fig. \ref{Fig:evEv}. This phenomenon will be further discussed in Pin\c{c}on et al. (2021, in prep.).

\subsection{Variation with chemical composition along the evolution}\label{Sec:Comp}
\begin{figure}
\includegraphics[width=\linewidth]{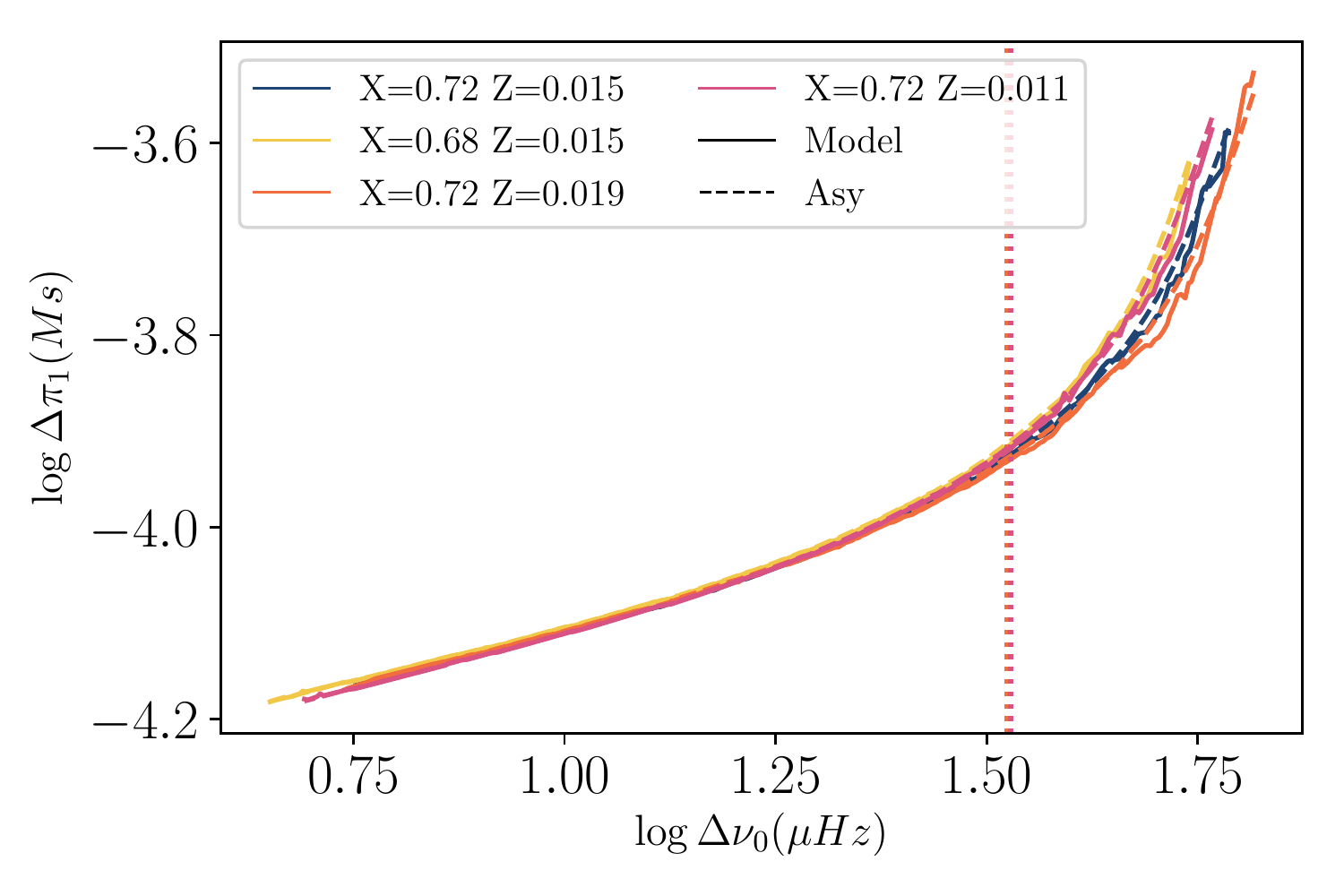}
\caption{Variation of $\Delta\pi_1$ as a function of $\Delta\nu_0$ for $1M_{\odot}$ models with several compositions, represented by the colours. The dashed lines correspond to the asymptotic values and the vertical dotted lines to the transition at $\mathcal{N}=1$.}\label{Fig:DpiCompEv}
\end{figure}

\begin{figure}
\includegraphics[width=\linewidth]{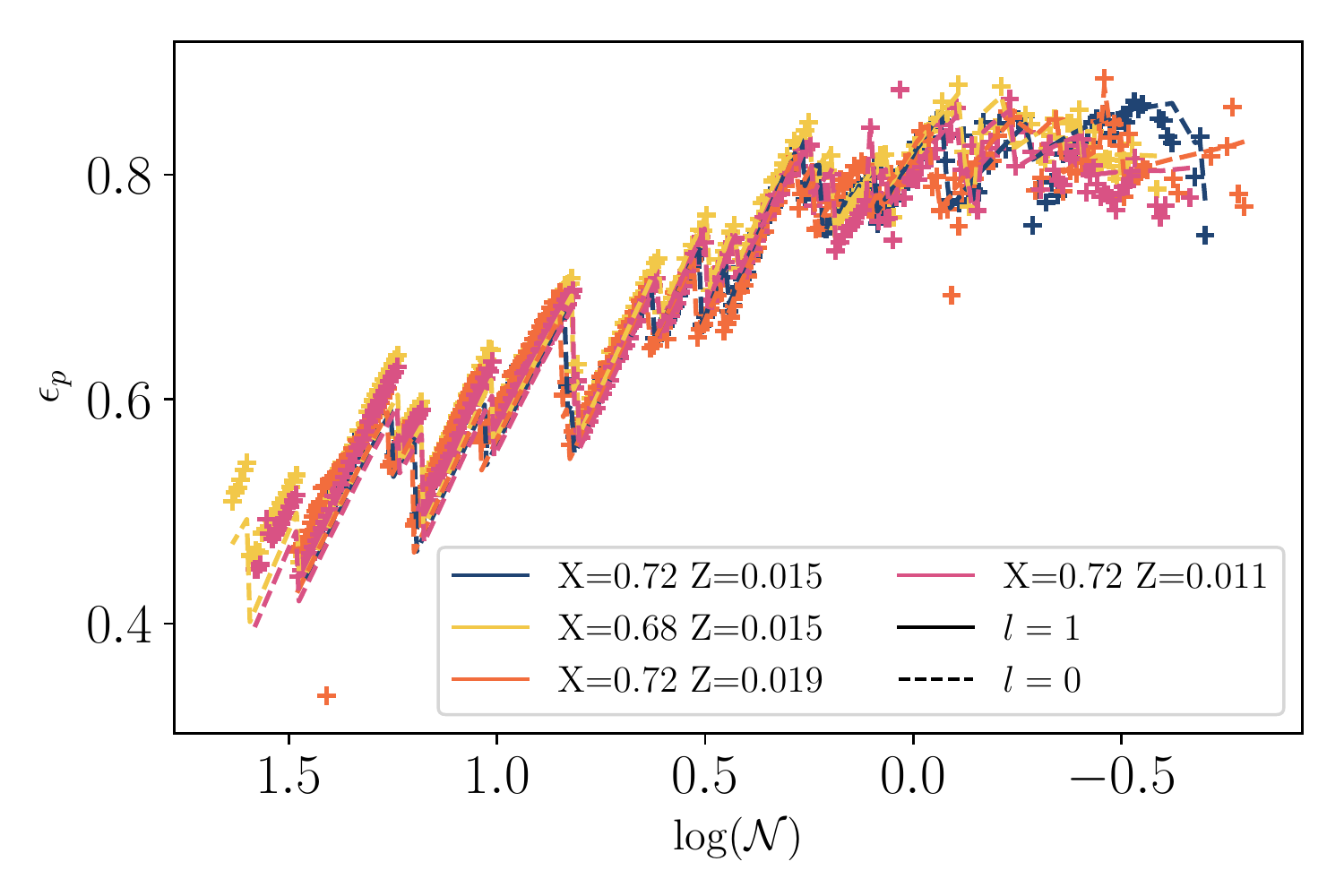}
\caption{Variation of $\epsilon_p$ as a function of $\mathcal{N}$ for $1M_{\odot}$ models with several compositions, represented by the colours. The dashed lines correspond to the values computed with \who on the radial modes.}\label{Fig:epCompEv}
\end{figure}

\begin{figure}
\includegraphics[width=\linewidth]{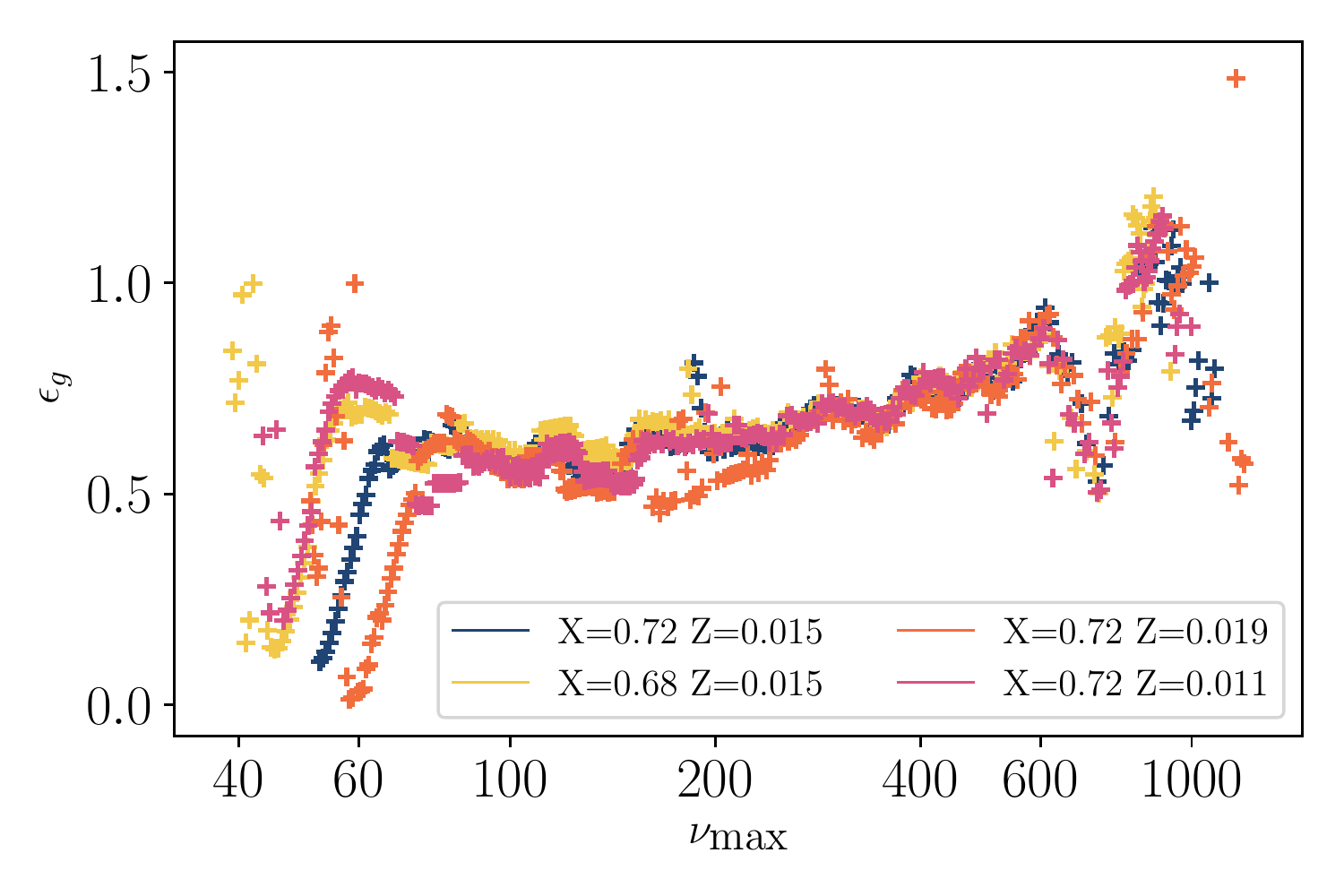}
\caption{Variation of $\epsilon_g$ as a function of $\nu_{\textrm{max}}$ for $1M_{\odot}$ models with several compositions, represented by the colours.}\label{Fig:egCompEv}
\end{figure}

\begin{figure}
\includegraphics[width=\linewidth]{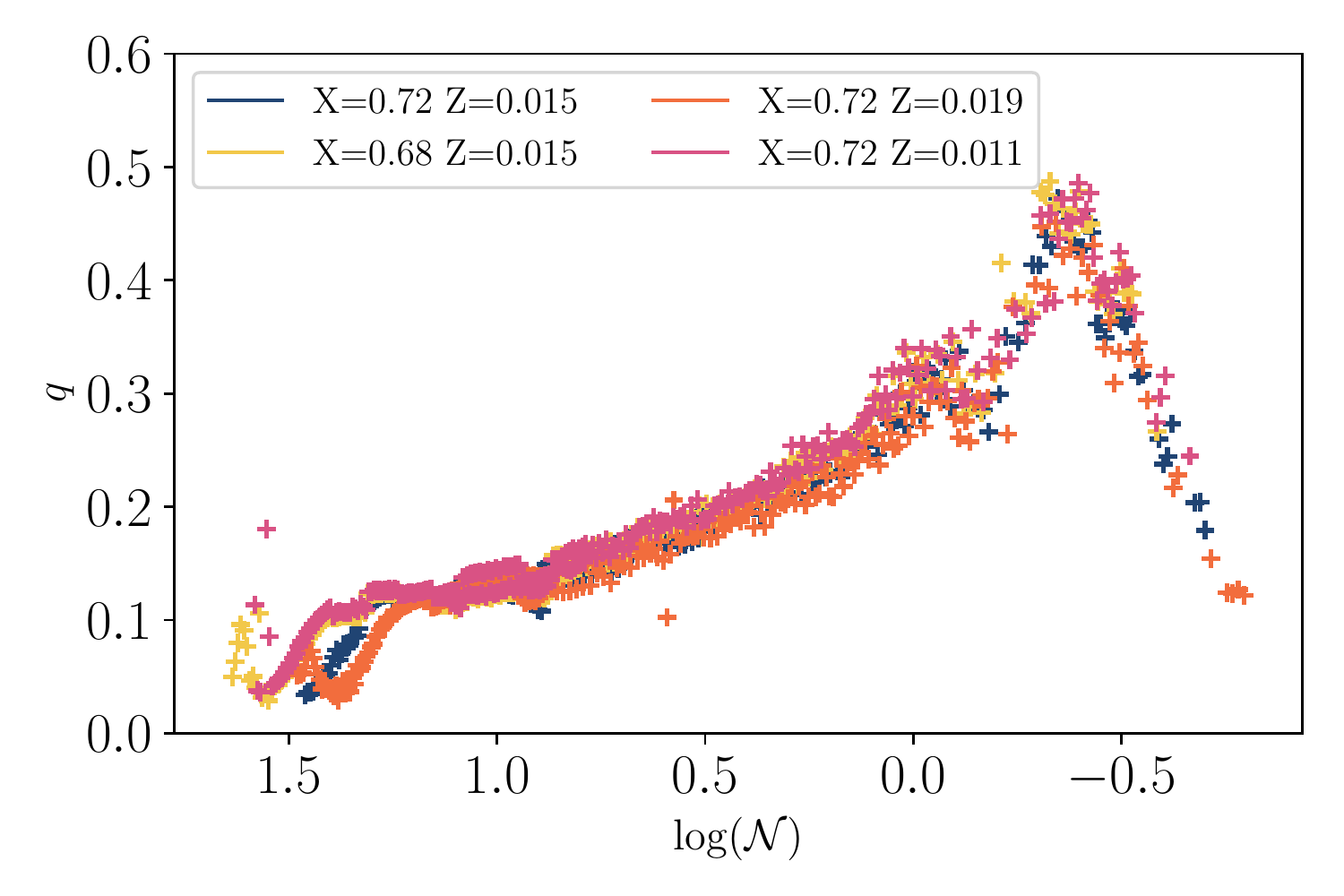}
\caption{Variation of $q$ as a function of $\mathcal{N}$ for $1M_{\odot}$ models with several compositions, represented by the colours.}\label{Fig:qCompEv}
\end{figure}

To study the impact of the chemical composition on the fitted parameters, we computed several tracks for a $1M_{\odot}$ star with different initial hydrogen and metals abundances. We consider pairs of initial hydrogen and metal abundances in $X_0 \in \left[0.68,0.72\right]$ and $Z_0 \in \left[0.011,0.019\right]$. The results are shown in Figs. \ref{Fig:DpiCompEv} to \ref{Fig:qCompEv} for $\Delta \pi_1$, $\epsilon_p$, $\epsilon_g$ and $q$, respectively. In Fig. \ref{Fig:DpiCompEv}, we observe that the several tracks for $\Delta \pi_1$ are almost indistinguishable from one another during the red giant phase. Only a small difference is visible on the subgiant phase. Nevertheless, thanks to a close inspection of our Fig. \ref{Fig:DpiCompEv}, alongside Fig.1 of \cite{2019A&A...622A..98F}, we expect that an improper determination of the metallicity will impact the inferred mass in a similar way as it does in the main sequence case. Indeed, at fixed $\Delta\pi_1$ and $\Delta\nu_0$ values, a variation of $0.008$ in $Z_0$ could change the estimated mass of about $0.1M_{\odot}$. This suggests that, in addition to the measurements of $\Delta\nu$ and $\Delta\pi_1$, a spectroscopic measurement of the composition will be necessary to a good determination of the mass, radius and age of subgiant stars. Indeed, the initial composition may impact the inferred stellar mass, thus the inferred age.  Regarding $\epsilon_p$, we do not note any significant impact of the chemical composition on the evolution of this indicator, as illustrated in Fig. \ref{Fig:epCompEv}. Finally, concerning $q$ and $\epsilon_g$, we note in Figs. \ref{Fig:egCompEv} and \ref{Fig:qCompEv} that only the position (in either $\nu_{\textrm{max}}$ or $\log \mathcal{N}$) of the drop in the values of $\epsilon_g$ and $q$ just before the luminosity bump is significantly affected by the composition. As shown by \cite{2020svos.conf..313P}, this likely results from a modification of the position of the base of the convective envelope. 

The impact of the metallicity on the measured value of the period spacing and coupling factor has already been studied by \cite{2020MNRAS.495..621J}. In this work, they looked at the evolution of these indicators on a grid of red-giant models, but their fits were made around a fixed value of the pressure radial order $n_p$ only. While they also observe that there is no significant impact of the metallicity on the evolution of $\Delta\pi_1$, they note a slight impact of the metallicity on the rate of decrease of $q$. A close look at their Fig. 9 also seems to indicate that this dependency with metallicity mostly appears for the youngest stars. The individual trends seem to settle to a common one as the stellar evolution goes on during the red giant branch. Nevertheless, we do not observe such distinction with the composition. A possible reason for this difference might stem from the fact that they consider the coupling factor to depend on the radial order, $n_p$, and represent its evolution following specific modes; whereas we consider the coupling factor to be constant over the spectrum with a typical set of frequencies representative of the observations all along the subgiant and red giant branches. This is further discussed in Sect. \ref{Sec:DisSec}.

%\cite{2020MNRAS.495..621J} also studied the impact of the metallicity on the measured value of the period spacing and coupling factor at fixed $n_p$ on a grid of red-giant models. While they also observe that there is no significant impact of the metallicity on the evolution of $\Delta\pi_1$, they note a slight impact of the metallicity on the rate of decrease of $q$. A close look at their Fig. 9 also seems to indicate that this dependency with metallicity mostly appears for the youngest stars. The individual trends seem to settle to a common one as the stellar evolution goes on. Nevertheless, we do not observe such distinction with the composition. A possible reason for this difference might also stem from the fact that they consider the coupling factor to depend on the the radial order $n_p$ and represent its evolution following specific modes while we consider the coupling factor to be constant over the spectrum with a typical set of frequencies representative of observations. This is further discussed in Sect. \ref{Sec:DisSec}. \textcolor{red}{\sout{Finally, we do not note any significant impact of the chemical composition on the evolution of $\epsilon_p$, as Fig. \ref{Fig:epCompEv} illustrates.}}

Finally, what is striking in Figs. \ref{Fig:DpiCompEv} to \ref{Fig:qCompEv} is that some subtle features are present for every composition considered. For example, looking at the evolution of the gravity offset in Fig. \ref{Fig:egCompEv}, it stands out that the oscillation present on the subgiant phase is present for all the compositions. Furthermore, in Fig. \ref{Fig:qCompEv}, we also observe that the local minimum in the coupling factor, right before the transition at $\mathcal{N} = 1$, is present for every track. This might result from the fact that the changes in composition considered might not significantly affect the evolution on the subgiant phase and, therefore, the evolution of the indicators during this phase. Another striking feature is the homology between the track with $X_0=0.68$ and $Z_0=0.015$ and the one with $X_0=0.72$ and $Z_0=0.011$. The tracks are almost identical.

Overall, we may assert that the indicators are degenerate with the chemical composition on the red giant phase, except for the latest stages of evolution: for instance the drops in $\epsilon_g$ and $q$ before the luminosity bump.

%\subsubsection{$q$}

\section{Discussion}\label{Sec:Dis}
In the present section, we further discuss the results presented in Sect. \ref{Sec:Ind} as well as possible improvements of the \egg method.

\subsection{Impact of the considered set of modes}\label{Sec:DisMod}
\begin{figure}
\includegraphics[width=\linewidth]{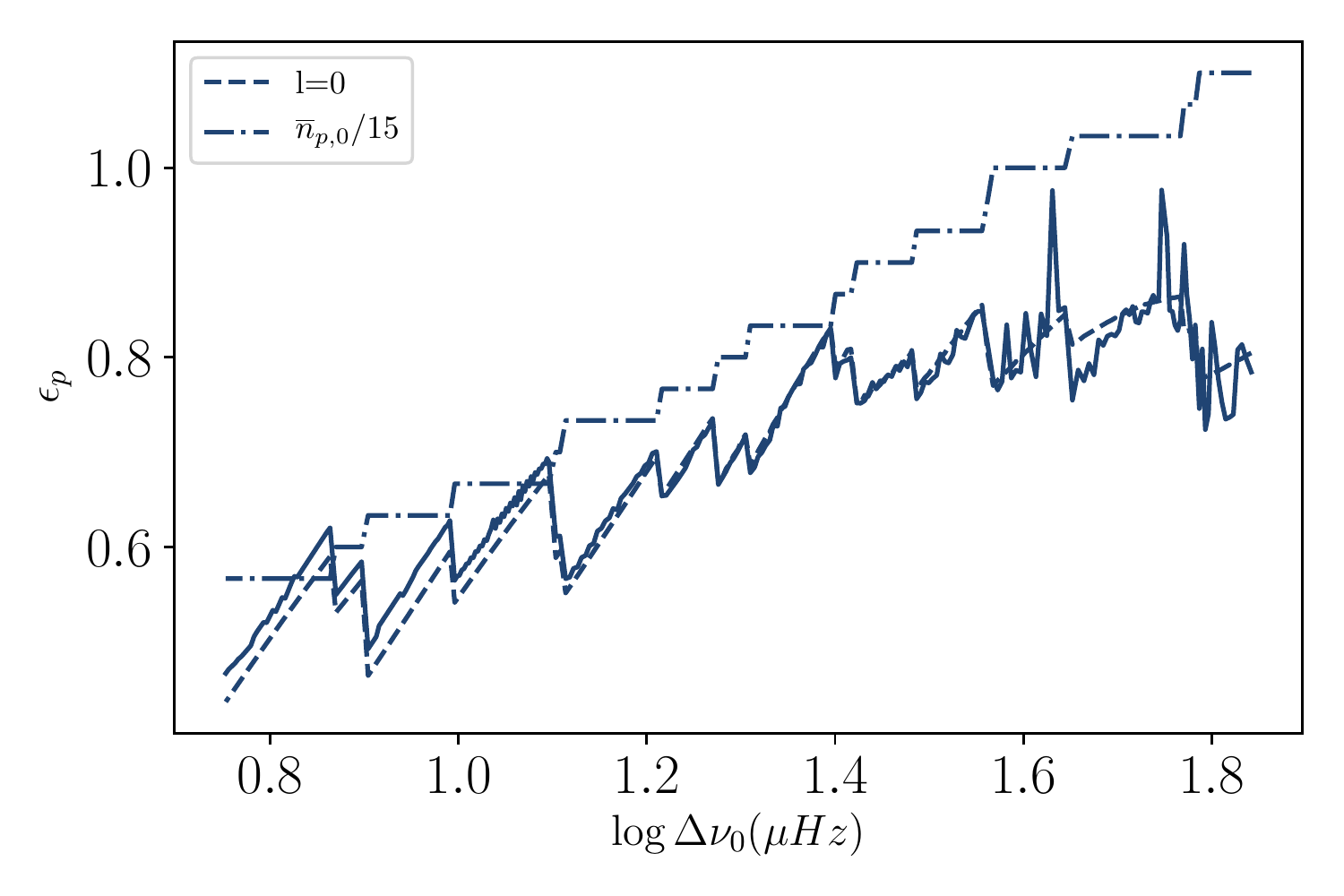}
\caption{Variation of $\epsilon_p$ with evolution for the $1M_{\odot}$ track. The dashed line represents the variation of the estimation with \who on radial modes and the dot-dashed line the mean pressure radial order of modes considered divided by $15$.}\label{Fig:epMod}
\end{figure}

In the present paper, we considered modes in the range of the width $0.8\nu_{\textrm{max}}^{0.88}$ around $\nu_{\textrm{max}}$, determined to include at least ten radial modes for the youngest subgiant models \citep{2012A&A...540A.143M,2020A&A...642A.226A}. We immediately see that this range evolves with $\nu_{\textrm{max}}$, both in terms of its central frequency $\nu_{\textrm{max}}$ and in the number of modes. As the number of modes is discrete its evolution experiences discontinuities. This creates the saw-like pattern we observe in the pressure and gravity offsets (Figs. \ref{Fig:epEv}, \ref{Fig:egEv}, \ref{Fig:epCompEv} and \ref{Fig:egCompEv}). To illustrate this effect, we plot in Fig. \ref{Fig:epMod} the evolution of $\epsilon_p$ for the $1M_{\odot}$ track as well as the mean radial order of pressure modes, $\bar{n}_p$. This value is divided by $15$, an arbitrary value, such that $\epsilon_p$ and $\bar{n}_p$ have comparable values. We observe that both behave as a seesaw and that the discontinuities in the values are synchronous along evolution. In the case of the observations, the set of modes also changes with evolution, which should also create the discontinuities we observe theoretically. Nevertheless, when attempting to carry stellar modelling of a given star considering $\epsilon_p$ as a constraint, this will not constitute a problem as the set of modes will be fixed by the observations.

\subsection{Generalisation to spectra with holes}\label{Sec:GenHol}

When adjusting the spectrum, we assume in this paper that the modes that are adjusted are successive, that is, the difference in radial order between the considered modes, $\Delta n_{g}$ and $\Delta n_{p}$, are equal to either $0$ or $1$. However, when applying the method on observational spectra, it may be the case that some modes are not detected. Consequently, the period and frequency difference formulations in Eqs. \eqref{Eq:DPi} and \eqref{Eq:Dnui} will have to be adapted considering proper values for the $\Delta n_{g}$ and $\Delta n_{p}$ parameters in these equations. This will thus require that a proper identification of the modes has been carried out. 

Moreover, regarding the initial estimation of the parameters to be adjusted, the position and number of holes might be problematic in some specific cases. For example, as we estimated $\Delta\pi_1$ via the maximum of the local period differences in g-dominated spectra, missing several modes in the central region between dips would lead to an underestimation of its value. In addition, the coupling factor is estimated from the ratio between the maximum and minimum of the local difference in period (for g-dominated spectra) or in frequency (p-dominated). Therefore, missing modes close to these minima or maxima might severely impact the initial estimate of $q$. However, \citet{2018A&A...618A.109M} have showed that, with \textit{Kepler} data, g-dominated mixed modes should be below the limit of observability only for evolved giant stars with $\Delta\nu \leq 6 \mu \textrm{Hz}$, for which only p-dominated mixed modes would be detected. This actually corresponds to the most evolved stars, which are close to the luminosity bump, considered in this study. The number of observed g-dominated modes should increase as we go down the red giant branch, meaning that younger stars should constitute less of a problem. Therefore, it will be necessary in future studies to test the ability of the method to provide correct results in such evolved cases.

\subsection{Higher order contributions to the asymptotic formulation}\label{Sec:DisHig}
In the present paper, we considered the pressure phase, $\theta_p$, to depend linearly on the frequency. However, because the set of modes is broad in the case of subgiant stars (about $10\Delta\nu$ wide),  the large separation may not be considered to be constant over this interval. As an illustration, its relative variation in the subgiant star considered in Fig. \ref{Fig:DnuBeg} is of about $\frac{\delta \Delta\nu}{\Delta\nu}\sim 5\%$.
Furthermore, the mean value of the pressure radial order is of $n_p \sim 20$ in such stars. As a consequence, the product of both quantities, corresponding to the error made by considering only a linear pressure phase, is on the order of unity. It is therefore not negligible compared to typical observed uncertainties on $\epsilon_p$. In such a case, the assumed formulation for the pressure phase may not be valid any more. Therefore, it may be necessary to include second-order contributions to this phase. This effect may be so important that it may result in the addition or removal of a p-dominated mode to the set of considered frequencies. 
Following this discussion, the case of the gravity phase of evolved red-giants naturally comes to mind, as such stars span a large range of gravity radial orders. However, \cite{2019A&A...626A.125P} analytically showed that the second order contribution to the gravity phase remains small compared to the current observed uncertainties on the gravity offset for stars typically observed before the luminosity bump.

For the evolved red giant stars, it may again be necessary to include higher order contributions to the pressure phase \citep{2013A&A...550A.126M}. Indeed, we noted in Fig. \ref{Fig:DPEnd} that there is a slight shift in the position of the $\Delta P$ dips as well as small differences in their exact magnitude. This can now be caused by the fact that the hypothesis that the number of nodes in the pressure cavity is large and thus that the local wavelength is small is not verified for the pressure dominated modes. Indeed, the radial order of p-dominated modes is very low, namely, $n_p \sim 5$.

The inclusion of such higher order contributions to the pressure phase might be necessary to improve the robustness of the method and of the measured seismic indicators. As a consequence, it will be implemented and tested in subsequent papers of this series.

\subsection{Frequency dependence of the coupling factor in evolved red giants}\label{Sec:DisSec}
\begin{figure}
\includegraphics[width=\linewidth]{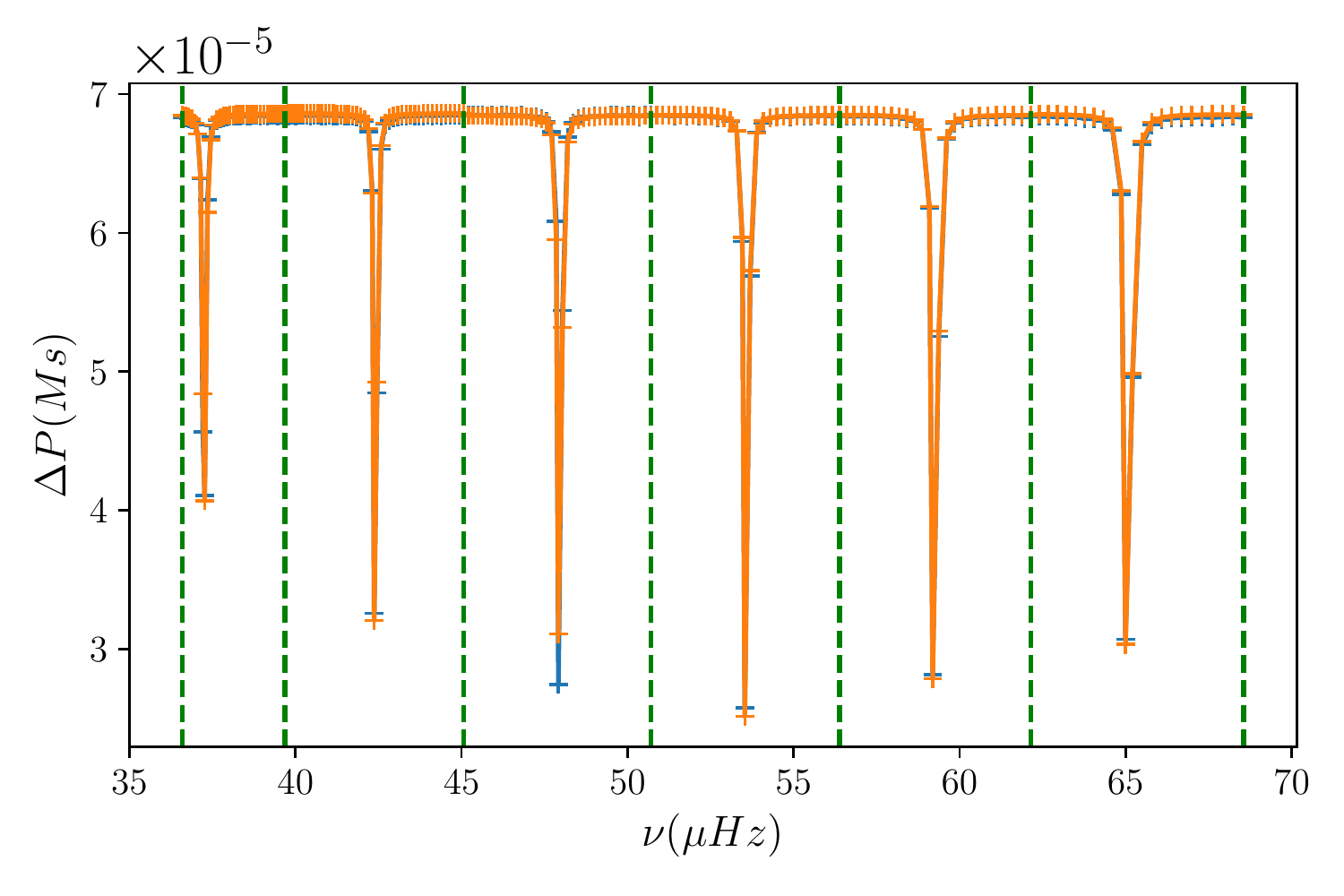}
\caption{Fitted individual period spacings as a function of frequency for the same red-giant model as in Fig. \ref{Fig:DPEnd}, but the spectrum has been binned for each individual bump. The green vertical dashed lines delimit each bin.}\label{Fig:DPEndBin}
\end{figure}

\begin{figure}
\includegraphics[width=\linewidth]{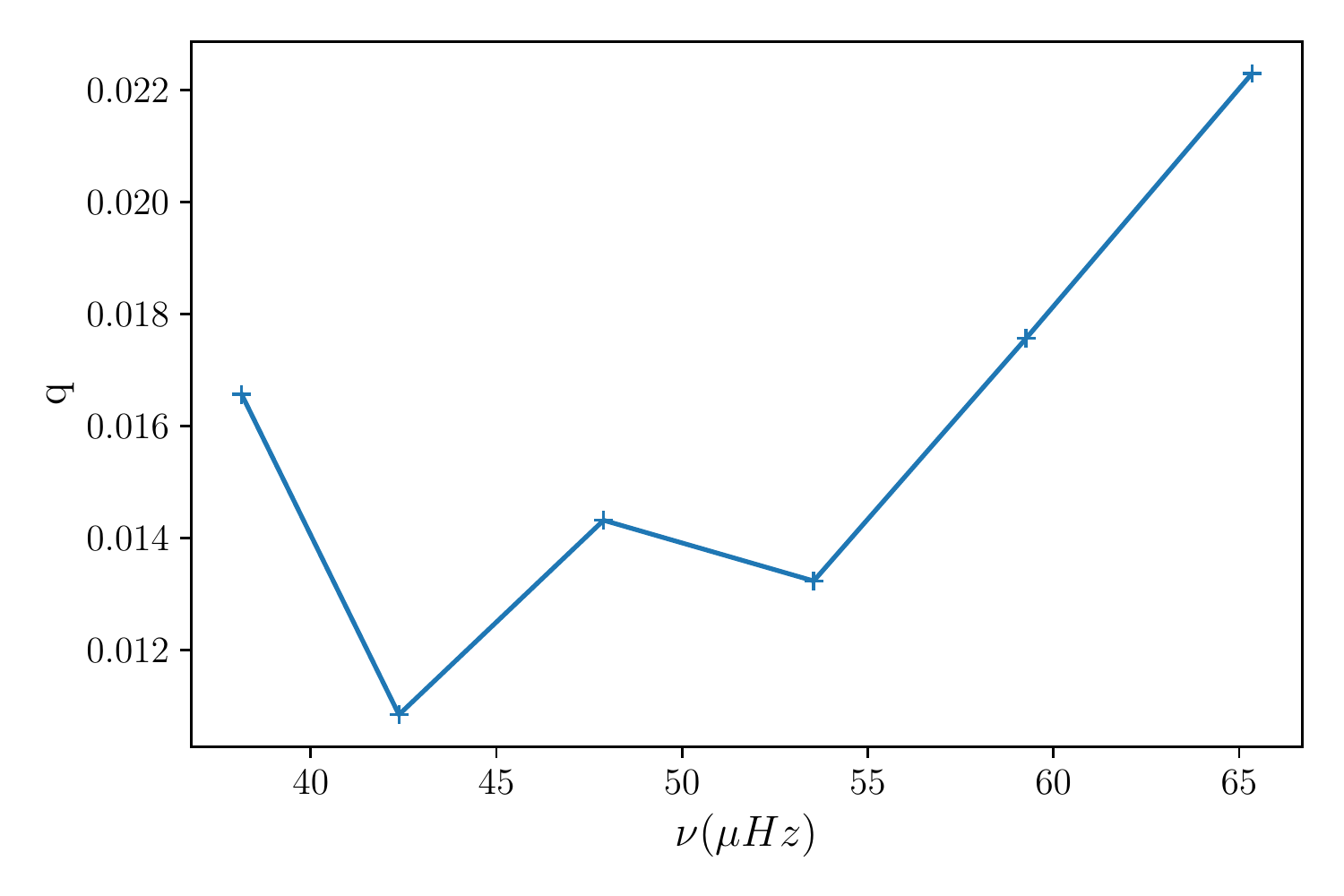}
\caption{Evolution of $q$ throughout the binned spectrum presented in Fig. \ref{Fig:DPEndBin}}\label{Fig:DPEndq}
\end{figure}

\citet{2019MNRAS.490..909C} showed that, for evolved models, the coupling factor may depend on the frequency. This is due to the fact that the evanescent zone has penetrated into the convective zone. As a consequence, the Brunt-Väisälä frequency drops to zero and is no longer log-parallel to the Lamb frequency. The relative width of the evanescent region defined in Eq.~\eqref{Eq:dr} may not be considered constant with respect to the frequency any longer. Therefore, the coupling factor may in turn depend on the frequency \citep{2020A&A...634A..68P}. To mimic this effect, we binned the oscillation spectrum of the evolved giant presented in Fig. \ref{Fig:DPEnd} into sub-spectra containing only one dip each. The binned spectrum is shown in Fig. \ref{Fig:DPEndBin}. We then fitted individual $q$ values in each sub-spectra. The evolution of the coupling factor as a function of the central frequency of each bin is displayed in Fig. \ref{Fig:DPEndq}. We indeed observe that it may vary with the frequency in an almost linear fashion. Only the coupling factor in the lowest frequency bin strays far from the linear trend. This may result from the asymmetric number of modes around the dip. Finally, we note that the variation of the coupling factor on the spectrum is significant when compared to the constant fitted value. Indeed, while the fitted value is of about $0.12$ \citep[comparable to values in the literature, see][]{2017A&A...600A...1M}, it changes from $\sim 0.11$ to $\sim 0.22$ along the spectrum. This illustrates the necessity to account for its dependency with the frequency in order to properly interpret its value.

\subsection{Glitches}
A further refinement of the technique would be the inclusion of glitches in the formulation used. These glitches are the result of a sharp variation (compared to the wavelength of the incoming mode) in the stellar structure. Their signature is an oscillating feature in the oscillation spectrum. \cite{2015ApJ...805..127C} showed that buoyancy glitches, caused by a sharp variation in the Brunt-Väisälä frequency, are mainly found for red giant stars at the luminosity bump, at the early phases of helium core burning and at the beginning of helium shell burning. In this paper, we only consider models before the luminosity bump. Therefore, we should not expect the detection of such glitches in these models. Nonetheless, their inclusion will be a necessary step to the application of the \egg method to more evolved stellar models and data. Furthermore, such glitches carry essential information for constraining the stellar cores of giants as well as the transport processes of chemical elements.

Aside from buoyancy glitches, there are the acoustic glitches, found in the pressure part of the spectrum. In the case of red giants, we may observe the signature of the helium glitch, created by the second ionisation zone of helium. Therefore, it holds information about the surface helium content, providing additional constraints to stellar models. The study of such glitches in giant stars has been carried in the past \citep[e.g.][]{2010A&A...520L...6M,2020MNRAS.497.1008D}. Combining the present method with \who \citep{2019A&A...622A..98F} we will be able to retrieve this signature in the p-dominated modes in a robust way. The inclusion of both the buoyancy and acoustic glitches in the dipolar modes will be discussed in subsequent papers of this series.

\section{Conclusion}\label{Sec:Con}
With the aim of defining relevant seismic indicators and relying on a prior modes extraction \citep[e.g.][]{2015A&A...584A..50M,2018A&A...616A..24G,2020A&A...642A.226A}, we present a method of automated, consistent,
robust, and fast adjustment of observed and theoretical mixed-mode oscillation spectra. Theoretical oscillations spectra of low-mass subgiant and red-giant stars are well adjusted, as illustrated in Figs. \ref{Fig:DnuBeg} - \ref{Fig:DPEnd}. 

We explored the probing potential of the mixed-mode parameters ($\Delta\nu$, $\Delta\pi_1$, $\epsilon_p$, $\epsilon_g$, and $q$) as indicators of the stellar structure of subgiant and red giant stars, along a grid of models for masses between $1.0M_{\odot}$ and $1.8M_{\odot}$ (extended to $2.1M_{\odot}$ in the case of $\Delta\pi_1$) and initial chemical compositions in $X_0 \in \left[ 0.68,0.72 \right]$ and $Z_0 \in \left[ 0.011, 0.019 \right]$. Overall, the evolution of the indicators displays clear trends and the chemical composition has only a slight impact. In contrast, we note that the evolution of the parameters with the mass follows two regimes, depending on the evolutionary stage of the star. 

During the subgiant phase, because of a moderate core-envelope density contrast, the mixed-mode parameters evolve differently with $\Delta\nu$ according to the stellar mass. Notably, the evolution of $\Delta\pi_1$ in subgiants is such that it may be used, combined with $\Delta\nu$ and a proper measurement of the metallicity, to infer the stellar mass, radius and age (Fig. \ref{Fig:DpiEvSub}). We also demonstrate that the asymptotic period spacing tightly agrees with the fitted one. This came as a surprise as the contribution of the gravity modes departs from the asymptotic regime for these stars.

As the stars evolve to the red giant phase, the core-envelope density contrast becomes large. As a consequence, the structure of the evanescent region is almost independent of the stellar mass and the evolutions of the pressure offset, gravity offset, and coupling factor as a function of $\Delta \nu$ are not really affected by the stellar mass in this phase. We showed that this is also true for $\Delta\pi_1$ in stars with masses $\lesssim 1.8M_{\odot}$ because of the core electron degeneracy, which makes the helium core density quasi independent of the stellar mass at a given value of $\Delta \nu$. Above this threshold, the electron degeneracy is lifted and the evolution of $\Delta\pi_1$ again depends on the mass. Observing stars in that region would therefore allow us to constraint their masses, radii, and ages, similarly to the case of subgiants. However, such stars evolve swiftly and might not be observed.

Here, we provide the first depiction, to our knowledge, of the gravity offset evolution along a grid of models during both the subgiant and red giant phases. The evolution during the red-giant phase agrees with the observations of \citet{2018A&A...618A.109M} and the asymptotic computations from \citet{2019A&A...626A.125P}. As the gravity offset corresponds to the phase lag of the g-dominated modes induced at the inner edge of the evanescent region, we expect it should hold information about this region. However, some issues remain to be tackled as the behaviour of this indicator remains erratic in the subgiant phase.

The evolution of the coupling factor along our grid of models also qualitatively agrees with the observations of \citet{2017A&A...600A...1M}. We also show, based on the study of \citet{2020A&A...634A..68P}, that its evolution is concordant with that of the width of the evanescent region (see Fig. \ref{Fig:evEv}).

Owing to the use of the asymptotic formulation and appropriate estimation of the mixed-mode parameters, the \egg technique offers a robust and fast\footnote{Computation times are much smaller than those necessary to compute theoretical adiabatic frequencies.} adjustment of the mixed-mode spectra displayed by subgiant and red giant stars. Furthermore, we also plan on extending the method to include refinements of the asymptotic formulation such as higher order contributions and glitches. Finally, we expect that the technique would represent a great asset to the automated treatment of large samples of data as will be generated by spacecrafts such as PLATO \citep{2014ExA....38..249R}, which will observe a great number of subgiant stars (core program) and red-giant stars (secondary science program). Indeed, after a proper modes extraction, the measured seismic indicators can be used as constraints on stellar models to automatically compute stellar parameters with model search algorithms such as AIMS \citep{2019MNRAS.484..771R}. % A few benchmark stars must be selected, both on the subgiant and red giant phases. 

%\section{Some references}
%See. \citet{2019MNRAS.490..909C}\\
%AIMS: \citet{2019MNRAS.484..771R}\\
%prediction: \citet{1974A&A....36..107S}\\
%dupret: \citet{2009A&A...506...57D}\\
%detection: \citet{2011Natur.471..608B}\\
%GAIA: \citet{2016A&A...595A...1G}\\
%Y-M degeneracy: \citet{2014A&A...569A..21L}\\
%KIC-444: \citet{2019A&A...630A.126B}\\
%Kepler LEGACY: \citet{2017ApJ...835..172L}\\
%Rosu: \citet{2020arXiv200908658R}\\
%Buldgen: \citet{2019A&A...630A.126B}\\
%\who: \citet{2019A&A...622A..98F}\\
%Vienne: \citet{2020svos.conf..281F}\\
%Vienne Buldgen: \citet{2020svos.conf..249B}\\
%GN93: \citet{1993oee..conf...15G}\\
%AGSS09: \citet{2009ARA&A..47..481A}\\
%K2: \citet{2014PASP..126..398H}\\
%CESTAM :\citet{2013A&A...549A..74M}\\
%CESAM: \citet{2008Ap&SS.316...61M}\\
%cles: \citet{2008Ap&SS.316...83S}\\
%$\gamma$ dor rotation: \citet{2019A&A...626A.121O}\\
%Deformation from rotation \citet{2003PhDT.......136D}\\
%Giant core mixing constraint \citet{2013ApJ...766..118M}\\
%Red giants fast core rotation \citet{2012Natur.481...55B}\\
%Deheuvels subgiants \citet{2011A&A...535A..91D}\\
%Deheuvels giants \citet{2017A&A...605A..75D}\\
%611 $\gamma$ dor rotation rates and period spacings \citet{2020MNRAS.491.3586L}

\section{Acknowledgments}
M.F. is supported by the FRIA (Fond pour la Recherche en Industrie et Agriculture) - FNRS PhD grant. C.P. is supported by the F.R.S - FNRS  as a Charg\'e de Recherche.

\bibliographystyle{aa}
%\nocite{*}
\bibliography{bibli}

\begin{appendix}\label{Sec:App}
%\section{Asymptotic quantities}\label{Sec:AsyQua}
%In this section we recall the definitions of several asymptotic quantities.
%
%The asymptotic large separation is defined as:
%\begin{equation}
%\Delta\nu = \left(2\int^{R_*}_0 \frac{dr}{c}\right)^{-1},
%\label{Eq:Dnu}
%\end{equation}
%where $c$ is the local sound speed, $r$ the local radius and $R_*$ the total stellar radius.

%The asymptotic period spacing of dipole gravity modes is give by:
%\begin{equation}
%\Delta\pi_1 = 2\pi^2\left(\int^{r_2}_{r_1}\frac{N}{r}dr \right),
%\label{Eq:Dpi}
%\end{equation}
%with $N$ the Brunt-Väisälä frequency, $r_2$ and $r_1$ the inner and outer turning points in the g-cavity;\\
%the pressure and gravity offsets: $\epsilon_p$ and $\epsilon_g$ and the coupling factor $q$, representing the strength of the coupling between the two cavities. 

\section{Bounds of the asymptotic frequency differences}\label{Sec:SecDif}

The method presented in Sect.~\ref{Sec:Met} takes advantage of the theoretical bounds of the first and second frequency differences of the asymptotic frequency pattern. On the one hand, the first frequency (resp. period) difference between two consecutive modes normalised by $\Delta \nu$ (resp. $\Delta \pi_1$) is always smaller than unity. On the other hand, the second frequency difference in (Eq. \eqref{Eq:SecDif}) displays values greater than $1$ in a p-dominated spectrum while it presents values lower than $1$ in g-dominated spectra. In the current section, we mathematically demonstrate these statements.

\subsection{Case $\mathcal{N} < 1$ over the spectrum}\label{Sec:pDom}

In a first step, we study the properties of the asymptotic frequency pattern focusing on the case where the local g-dominated mode density $\mathcal{N}(\nu) < 1$.
Using the expression of the pressure phase (Eq. \eqref{Eq:tp}), we may first rewrite the asymptotic resonance condition (Eq. \eqref{Eq:Shi}) as a function of the independent variable $x=\nu/\Delta\nu$. In this form, the asymptotic frequency pattern is obtained by solving the implicit relation
\begin{equation}
x = \mathcal{F}(n_p,x) = n_p + \epsilon_p +\frac{1}{\pi} \arctan\left[q \tan\left( \theta_g \left(x\right) \right)\right]\; ,\label{Eq:PhaFun}
\end{equation}
where $n_p$ is the pressure radial order and the gravity phase $\theta_g$ (Eq. \eqref{Eq:tg}) is also expressed as a function of the variable $x$, that is,
\begin{equation}
\theta_g (x) = \pi \left[x \mathcal{N}(x)-\epsilon_g +1/2 \right],\label{Eq:theta_g app}
\end{equation}
with $\mathcal{N}(x)=\left(x^2 \Delta\nu\Delta\pi\right)^{-1}$ the local g-dominated modes density defined in Eq. \eqref{Eq:N} but rewritten in terms of the $x$ variable.

\begin{figure}[h]
\includegraphics[width=\linewidth]{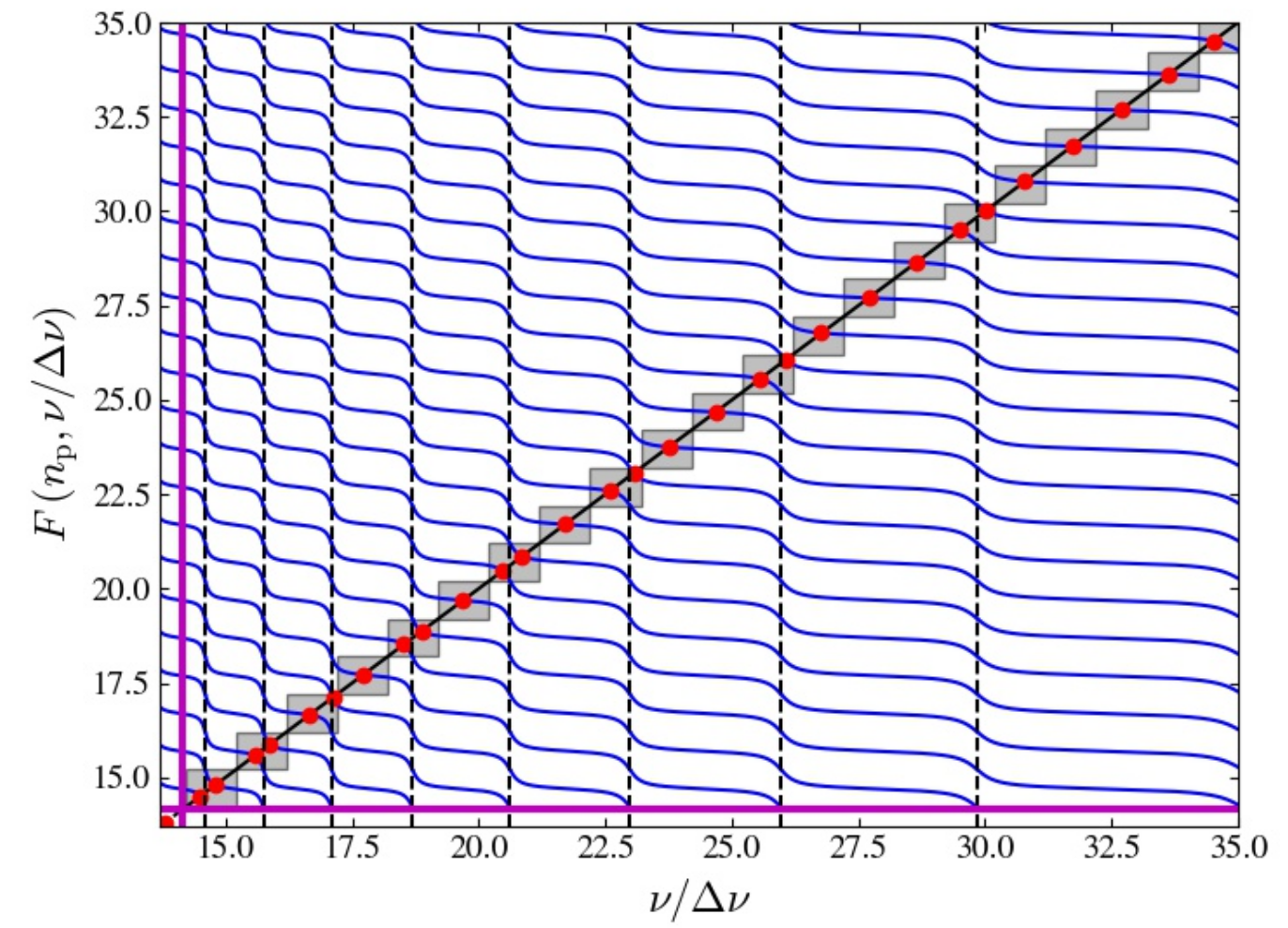}
\caption{Evolution of the phase function with the frequency over the large speration. The phase function for different pressure radial orders is represented in blue. The straight line represents the identity function $f(x)=x$. Its intersections with the phase function are the solutions, in red. The grey square domains represent regions of constant $n_p$ values. The vertical dashed lines are the positions of pure g-modes. The mixed-mode parameters used are $q = 0.2$ and $\Delta\nu\Delta\pi_1 = 200$}\label{Fig:PhaFun}
\end{figure}

As an illustration, the $\mathcal{F}$ function in the case $\mathcal{N}(x)<1$ is plotted as a function of $x$ in Fig. \ref{Fig:PhaFun} for different values of $n_p$. To plot this figure, we choose $\Delta \pi_1 \Delta \nu \approx 200$, which is a typical value for an observed subgiant star. The solutions of the implicit equation in Eq.~\eqref{Eq:PhaFun} are provided by the intersection between the $\mathcal{F}$ function and the identity function $f(x)=x$ represented by the solid black line. These solutions are shown as red filled circles. In this figure, a given value of $n_p$ is associated with a horizontal strip located in the range $[n_p+\epsilon_p-1/2, n_p+\epsilon_p+1/2[$ in the vertical axis. In such a horizontal strip, we see that the $\mathcal{F}$ function exhibits discontinuities as a function of $x$. These discontinuities occur at values, $x_{n_g}$, which correspond to the frequencies of pure g-modes verifying the condition $\theta_{g} = (n_{g} + 1/2)\pi$ with $n_{g} \in \mathbb{N}$ the gravity radial order. The values of $x_{n_g}$ are thus provided by
\begin{equation}
x_{n_g} = \frac{1}{\Delta\pi_1\Delta\nu}\left(n_g + \epsilon_g \right)^{-1}. \label{Eq:gMod}
\end{equation}
The positions, $x_{n_g}$, for different values of $n_g$ are represented by vertical dashed lines in Fig~\ref{Fig:PhaFun}. As the gravity phase $\theta_g$ has a local period in $x$ of $\mathcal{N}(x) < 1$, it is obvious that two consecutive pure gravity modes are such that: $x_{n_g}-x_{n_g+1}>1$, as confirmed in Fig.~\ref{Fig:PhaFun}.
Over a range $I_{n_g}=]x_{n_g+1},x_{n_g}]$ (referred to as `g-subset') and for a given value of $n_p$, we also note that $\mathcal{F}(n_p,x)$ is continuous and monotonically decreasing as a function of $x$, which can be easily checked by deriving this function with respect to $x$.

With framework set out thus far, it is now possible to study the bounds of the first and second differences of the solution pattern in a simple way. For the sake of convenience, we start the investigation with the first difference. Firstly, we focussed on a g-subset $I_{n_g}$. Over such an interval, we distinguished three cases:
\begin{enumerate}
\item On each subset $I_{n_p}=\left[n_p + \epsilon_p -1/2, n_p + \epsilon_p +1/2\right[$ (referred to as `p-subset') such that $I_{n_p}\subset I_{n_g}$, the $f$ function monotonically and continuously increases from $n_p + \epsilon_p -1/2$ to $n_p + \epsilon_p +1/2$. In contrast, the $\mathcal{F}(n_p,x)$ function monotonically and continuously decreases and is such that: $n_p + \epsilon_p -1/2<\mathcal{F}(n_p,x) < n_p + \epsilon_p +1/2$. Therefore, both functions intersect only once and there is only one solution in the p-subset $I_{n_p}$.\\
\item Over the subset $I_{n_p^-}=]x_{n_g+1},n_p^- + \epsilon_p +1/2]$, where $n_p^-$ is the lowest integer such as $x_{n_g+1} \le n_p^- + \epsilon_p +1/2$, according to the continuity and the monotonic behaviour of the $\mathcal{F}$ and $f$ functions, we still have only one solution since $\mathcal{F}(n_p^-,x_{n_g+1})=n_p^- + \epsilon_p +1/2\ge x_{n_g+1}$ and $\mathcal{F}(n_p^-,n_{p}^- + \epsilon_p +1/2)\le n_p^- + \epsilon_p +1/2$.\\
\item Over the subset $I_{n_p^+}=[n_p^+ + \epsilon_p -1/2,x_{n_g}]$ where $n_p^+$ is the largest integer such as $x_{n_g}\ge n_p^+ + \epsilon_p -1/2$, according to the continuity and the monotonic behaviour of the $\mathcal{F}$ and $f$ functions, we again have only one solution since $\mathcal{F}(n_p^+,x_{n_g})=n_p^+ + \epsilon_p -1/2\le x_{n_g}$ and \smash{$\mathcal{F}(n_p^+,n_{p}^+ + \epsilon_p -1/2)\ge n_p^+ + \epsilon_p -1/2$.}
\end{enumerate}
As a result, over a g-subset $I_{n_g}$, each solution is associated with a unique value of $n_p$. Now, because $n_p^+ -n_p^-\ge 1$ since $x_{n_g}-x_{n_g+1}>1$ when $\mathcal{N}<1$, there are at least two solutions over $I_{n_g}$, and we call $x_k$ the solution associated with the pressure radial order $n_{p,k}=n_p^-+k$ with $0\le k\le n_p^+ -n_p^-$. We thus have for each successive solutions
\begin{align}
x_{k+1}-x_k &= \mathcal{F}(n_{p,k+1},x_{k+1})-\mathcal{F}(n_{p,k},x_k)\nonumber\\
&=1+\mathcal{F}(n_{p,k},x_{k+1})-\mathcal{F}(n_{p,k},x_k),
\end{align}
where the last equality comes from the definition of the phase function (Eq. \eqref{Eq:PhaFun}) and the fact that the difference of pressure radial order between two successive solutions is $\Delta n_p=n_{p,k+1}-n_{p,k}=1$. Because $\theta_g(x)$ monotonically decreases over $I_{n_g}$, the $\mathcal{F}$ function continuously decreases as well and we have:
\begin{equation}
\theta_g(x_k) > \theta_g(x_{k+1}) \Rightarrow \mathcal{F}(n_{p,k},x_{k+1}) - \mathcal{F}(n_{p,k},x_k) < 1 \; ,
\end{equation}
such that $x_{k+1}-x_{k} < 1$. Finally, the last case to tackle is when two successive solutions belong to two successive distinct g-subsets $I_{n_g}$ and $I_{n_g-1}$. We have just shown that these two solutions belong to the same p-subset $I_{n_p}$ such as $x_{n_g}\in I_{n_p}$. The difference of pressure radial order between these two successive solutions is thus $\Delta n_p=0$ and it is trivial to conclude that the difference between the two solutions remains lower than unity. All these findings are well illustrated in the square grey domains in Fig.~\ref{Fig:PhaFun}, which represent the `p-domain' $I_{n_p}\times I_{n_p}$ that contains the solutions of the implicit relation. To summarize this first part, we have thus shown that the difference between two successive solutions of the implicit equation is smaller than unity. Converting this result as a function of the frequency $\nu_j$ (listed in ascending order with respect to the subscript $j)$, we therefore obtain in pressure-dominated spectra such as $\mathcal{N}(\nu)<1$, such that
\begin{equation}
\frac{\nu_{j+1}-\nu_{j}}{\Delta\nu} < 1 \; ,
\end{equation}
which in terms of period $P_{j}=1/\nu_{j}$ is equivalent to
\begin{align}
\frac{P_{j}-P_{j+1}}{\Delta\pi_1} & = \frac{\nu_{j+1}-\nu_{j}}{\Delta\nu}\frac{\Delta\nu}{\Delta\pi_1\nu_{j+1}\nu_{j}} \\ \nonumber
 & < \frac{\nu_{j+1}-\nu_{j}}{\Delta\nu}\mathcal{N}\left(\nu_j\right) < 1 \; ,
\end{align}
because $\mathcal{N}(x)<1$ in the present case. Moreover, we have shown that the difference of pressure radial orders between two successive modes is either equal to unity when the modes belong to the same g-subset associated with a unique gravity radial order $n_g$ (i.e., $\Delta n_p =1$ and $\Delta n_g=0$), or equal to zero when the modes belong to two successive distinct g-subsets associated with successive gravity radial orders $n_g$ and $n_{g}-1$, respectively (i.e., $\Delta n_p=0$ and $\Delta n_g=-1$).

Secondly, we now search the bounds of the second difference between two solutions $x_{j+1}$ and $x_{j-1}$. According to the previous paragraph, we always have $x_{j+1}-x_{j} < 1$ when $\mathcal{N}<1$, so that the upper bound of the second difference is directly $x_{j+1}-x_{j-1} < 2$. Regarding the lower bound, we first note that when the three considered solutions, $x_{j-1}$,  $x_j$, and $x_{j+1}$, are part of the same g-subset $I_{n_g}$, they are associated with successive values of the pressure radial order; thus, it is obvious that $1 < x_{j+1} - x_{j-1}$ since $\Delta n_{p}=n_{p,j+1}-n_{p,j-1}=2$.
When the three solutions are spread over two g-subsets, such that $x_j \in I_{n_g}$ and $x_{j+1} \in I_{n_g-1}$, we cannot directly draw a conclusion. To demonstrate that the result also holds in that case, we define the functions around the pure g-mode $x_{n_g}$:
\begin{align}
\tilde{\mathcal{F}}\left(n_p,x;x_{n_g} \right) &=  ~n_p + \epsilon_p  \label{tilde F} \\ \nonumber 
& + \frac{1}{\pi}\arctan\left[q \tan\left\lbrace \tilde{\theta_g}\left(x;x_{n_g} \right) \right\rbrace\right] \; ,
\end{align}
and
\begin{equation}
\tilde{\theta_g} (x;x_{n_g} ) = \pi \left(x_{n_g}-x \right) +\theta_g(x_{n_g}) \; , \label{tilde theta}
\end{equation}
where we recall that $\theta_g(x_{n_g})= \pi (1/2 + n_{g} )$. In an analogous way to the previous steps, we define the three consecutive solutions $\tilde{x}_{j-1}$, $\tilde{x}_{j}$, and $\tilde{x}_{j+1}$ around $x_{n_g}$ of the new implicit equation:
\begin{equation}
\tilde{x}=\tilde{\mathcal{F}}\left(n_p,\tilde{x};x_{n_g} \right). \label{Eq:tgTil}
\end{equation}
As shown previously, the solutions $\tilde{x}_{j+1}$ and $\tilde{x}_{j-1}$ around $x_{n_g}$ are respectively associated with the radial orders $n_{p,j+1}$ and $n_{p,j-1}=n_{p,j+1}-1$, as consecutive solutions on both sides of $x_{n_g}$ verify $\Delta n_p = 0$. In this case, the $\tilde{F}$ function has a period of $1$ and it is obvious that $\tilde{x}_{j+1}-\tilde{x}_{j-1}=\Delta n_{p} = 1$. To go further, we then express the $\theta_g$ phase in Eq. \eqref{Eq:theta_g app} as
\begin{align}
\frac{\theta_g (x)}{\pi} & = x \mathcal{N}(x) -\epsilon_g +1/2  \label{Eq:theta_g app 2}\\ \nonumber
 & = x \mathcal{N}\left(x\right) - x_{n_g}\mathcal{N}(x_{n_g}) + n_{g} +1/2 \\ \nonumber
 & = \frac{x_{n_g}}{x} \mathcal{N}(x_{n_g})\left(x_{n_g} -x\right) +  n_{g}+1/2,
\end{align}
where the second equality comes from Eq. \eqref{Eq:gMod}. By comparing Eq.~\eqref{Eq:theta_g app 2} with the definition of $\tilde{\theta}_g$ in Eq. \eqref{Eq:tgTil}, we can determine that:
\begin{equation}
\tilde{\theta}_{g}(\tilde{x}_{j+1};x_{n_g}) < \theta_g(\tilde{x}_{j+1}),
\end{equation}
since $\tilde{x}_{j+1} > x_{n_g}$ and $\mathcal{N}(x_{n_g})<1$,
which implies that
\begin{equation}
\tilde{x}_{j+1} = \tilde{\mathcal{F}}(n_{p,j+1},\tilde{x}_{j+1};x_{n_g}) < \mathcal{F}(n_{p,j+1},\tilde{x}_{j+1}).
\end{equation}
In other words, this means that the identity function and the $\mathcal{F}$ function do not intercept for $x< \tilde{x}_{j+1}$ inside the considered interval $I_{n_{p,j+1}} \cap I_{n_g-1}$, so that they will necessarily intercept at a higher value (since we have shown before that there is a unique solution in such an interval), that is,
\begin{equation}
x_{j+1} > \tilde{x}_{j+1}.
\end{equation}
Similarly, for $\tilde{x}_{j-1} < x_{n_g}$, we have:
\begin{equation}
\tilde{\theta}_{g}(\tilde{x}_{j-1}; x_{n_g}) > \theta_g(\tilde{x}_{j-1}),
\end{equation}
implying
\begin{equation}
\tilde{x}_{j-1} = \tilde{\mathcal{F}}(n_{p,j-1},\tilde{x}_{j-1};x_{g,i}) > \mathcal{F}(n_{p,j-1},\tilde{x}_{j-1})\; .
\end{equation}
This means that the identity function and the $\mathcal{F}$ function do not intercept for $x> \tilde{x}_{j+1}$ inside the considered interval $I_{n_{p,j-1}} \cap I_{n_g}$, so that they will necessarily intercept at a lower value, that is,
\begin{equation}
x_{j-1} < \tilde{x}_{j-1}.
\end{equation}
As a result, we find that:
\begin{equation}
x_{j+1}-x_{j-1} > \tilde{x}_{j+1}-\tilde{x}_{j-1} = 1.
\end{equation}
In summary, we thus conclude that for $\mathcal{N}<1$, the second frequency difference is bounded, such that:
\begin{equation}
1 < \delta \nu_{2,j}=\frac{\nu_{j+1}-\nu_{j-1}}{\Delta\nu} <2.
\end{equation}

\subsection{Case $\mathcal{N} > 1$ over the spectrum}\label{Sec:gDom}
In a g-dominated spectrum, the analysis can be put in a similar form to the case treated in App.~\ref{Sec:pDom} if we apply the following substitutions
\begin{align}
\nu  \leftarrow P\; , ~~~~&& \Delta\pi_1  \leftarrow \Delta\nu \; ,&&\Delta\nu  \leftarrow \Delta\pi_1\; ,&& \nonumber\\ 
\epsilon_g -1/2  \leftarrow \epsilon_p \; ,~~~~& &\epsilon_p  \leftarrow \epsilon_g -1/2 \; ,&&~~~~\mathcal{N} \leftarrow \mathcal{N}^{\dprime}\equiv\frac{1}{\mathcal{N}}\; ,&& \nonumber\\ 
q  \leftarrow \frac{1}{q} \; ,~~~~& &n_p  \leftarrow n_g \; ,&& n_g  \leftarrow n_p \; ,&& \nonumber \\ 
\theta_g  \leftarrow \theta_p \; .~~~~&& 
\end{align}
Indeed, in the case $\mathcal{N} >1$, we have $\mathcal{N}^{\dprime} < 1$, and it is possible to follow the same reasoning as in the previous section. 

One the one hand, for the first difference, when $\mathcal{N}>1$ over the considered spectrum, we obtain:
\begin{align}
\frac{P_{j}-P_{j+1}}{\Delta\pi_1}&<1\\
\frac{\nu_{j+1}-\nu_{j}}{\Delta\nu}&=\frac{P_{j}-P_{j+1}}{\Delta\pi_1}\frac{\Delta\pi_1\nu_{j}\nu_{j+1}}{\Delta\nu}\nonumber\\
&<\frac{P_{j}-P_{j+1}}{\Delta\pi_1} \mathcal{N}\left(\nu_{j+1}\right)^{-1}<1 \; . \label{Eq: Dnu g}
\end{align}
Moreover, the difference of gravity radial orders between two successive modes (i.e., still listed in ascending order with frequency) is either equal to $-1$ when the modes belong to the same p-subset associated with a unique pressure radial order $n_p$ (i.e., $\Delta n_g =-1$ and $\Delta n_p=0$), or equal to zero when the modes belong to two successive distinct p-subsets associated with successive pressure radial orders $n_p$ and $n_{p}+1$, respectively (i.e., $\Delta n_g=0$ and $\Delta n_p=1$).

On the other hand, for the second period difference, we obtain when $\mathcal{N}>1$ over the considered spectrum
\begin{align}
&1<\frac{P_{j-1}-P_{j+1}}{\Delta\pi_1}<2\; .
\end{align}
For the second frequency difference, when $\mathcal{N}>1$, we can solely determine:
\begin{align}
\delta\nu_{2,j}=\frac{\nu_{j+1}-\nu_{j-1}}{\Delta\nu}=\frac{P_{j-1}-P_{j+1}}{\Delta\pi_1} \mathcal{N}^\prime_j{}^{-1} \; , \label{Eq: D2nu g}
\end{align}
where an alternative definition for the g-dominated mode density naturally appears, namely,
\begin{equation}
\mathcal{N}^\prime_j= \frac{\Delta\nu}{\Delta\pi_1\nu_{j+1}\nu_{j-1}} \; . \label{Eq:N prime}
\end{equation}
Since $\mathcal{N}(\nu_{j+1})<\mathcal{N}^\prime_j$, we have $\mathcal{N}^\prime_j>1$ over the considered spectrum. As soon as
$\mathcal{N}^\prime> 2$,
Eq.~\eqref{Eq: D2nu g} shows us that $\delta \nu_{2,j}<1$.
In the case $1<\mathcal{N}^\prime < 2$,
we can adapt the reasoning made in App. \ref{Sec:pDom}. Indeed, either the solutions $x_{j-1}$ and $x_{j+1}$ are associated with the same pressure radial order (i.e., $\Delta n_p=0$ and $\Delta n_g=-2$) and $x_{j+1}-x_{j-1}<1$, or the difference in $n_p$ is equal to unity (i.e., with \smash{$\Delta n_g=-1$})\footnote{We recall that $\Delta n_p<2$ between $x_{j+1}$ and $x_{j-1}$ when $\mathcal{N}>1$, in a similar way that $\Delta n_g>-2$ when $\mathcal{N}<1$.}. In the second case, this means that the solutions $x_{j-1}$ and $x_{j+1}$ are located on both sides of a pure g-mode $x_{n_g}$. Applying the reasoning as in App. \ref{Sec:pDom} to determine the bounds of the second frequency difference, we obtain $x_{j+1}-x_{j-1}<1$ as $\mathcal{N}>1$ in the present case. We therefore conclude that over a spectrum such as $\mathcal{N}>1$, we have
\begin{equation}
\delta \nu_{2,j}=\frac{\nu_{j+1}-\nu_{j-1}}{\Delta\nu}<1.
\end{equation}

\subsection{Case $\mathcal{N}=1$ somewhere in the spectrum}
The last case to tackle is when the two solutions that are compared are located from each side of the transition point $x^\star$ where $\mathcal{N}(x^\star)=1$. 

For the first difference, the demonstration is simple. We denote $x_q$ (resp. $x_{q+1}$) as the largest (the smallest) solution lower (resp. greater) than $x^\star$. If we note $n_g^\star$ as the lowest gravity radial order such a $x_{n_g^\star} \le x^\star$, we have \smash{$x^\star - x_{n_g^\star}<1$} since otherwise $n_g^\star$ would not be the highest lowest gravity radial order such that \smash{$x_{n_g^\star} \le x^\star$ as $x_{n_g}-x_{n_g+1}<1$} when $\mathcal{N}>1$.
Therefore, we have two cases. If \smash{$x_q<x_{n_g^\star}$}, $x_{q+1}$ is then  necessarily comprised in the same p-subset as $x_q$. If \smash{$x_q>x_{n_g^\star}$}, either $x_{q+1}>x_{n_g^\star-1}$ and $x_{q+1}$ is then again necessarily comprised in the same p-subset as $x_q$; or $x_{q+1}<x_{n_g^\star-1}$ and $x_{q+1}$ then belongs to an adjacent p-subset to that of $x_q$. In all cases, following the same reasoning as in Appendix \ref{Sec:pDom}, we have $x_{j+1}-x_j<1$, which is therefore unconditionally met over the whole spectrum. This is obviously also true for the difference in period.

For the second difference, we consider two solutions $x_{j-1}$ and $x_{j+1}$ such as $x_{j-1}<x^\star$ and $x_{j+1}>x^\star$. We also consider the solutions of the implicit equation:
\begin{align}
\bar{x}=\tilde{\mathcal{F}}\left(n_p,\bar{x};x_{j-1} \right) \; , \label{Eq:xbar}
\end{align}
where $\tilde{\mathcal{F}}$ is defined in Eqs.~\ref{tilde F}. It can be straightforward to see that $x_{j-1}$ is solution of Eq.~\eqref{Eq:xbar}. By considering the consecutive solutions \smash{$\bar{x}_j$ and $\bar{x}_{j+1}$} of Eq.~\eqref{Eq:xbar}, we obviously have \smash{$\bar{x}_{j+1}-x_{j-1}=1$} since the \smash{$\tilde{\mathcal{F}}$} function has a period of 1. To go further, we then compute from Eq.~\ref{tilde theta}
\algn{
&\frac{\tilde{\theta}_g(x_{j+1};x_{j-1})}{\pi}-\frac{\theta_g(x_{j+1})}{\pi}=(\mathcal{N}^\prime_j-1) (x_{j+1}-x_{j-1}) \; , \label{Eq:difftheta}
}
with $\mathcal{N}_j^\prime$ defined as in Eq.~\eqref{Eq:N prime}. Therefore, if $\mathcal{N}^\prime_j>1$, we have \smash{$\tilde{\theta}_g(x_{j+1};x_{j-1})>\theta_g(x_{j+1})$} according to Eq.~\eqref{Eq:difftheta}. This means that the $\tilde{\mathcal{F}}$ function and the identity function intercept at higher values than $x_{j+1}$. In other words, $x_{j+1}<\bar{x}_{j+1}$, and thus $x_{j+1}-x_{j-1}<1$. Conversely, if $\mathcal{N}^\prime_j<1$, we have \smash{$\tilde{\theta}_g(x_{j+1};x_{j-1})<\theta_g(x_{j+1})$.} This means that the $\tilde{\mathcal{F}}$ function and the identity function can intercept at lower values than $x_{j+1}$. In other words, $x_{j+1}>\bar{x}_{j+1}$, and thus $x_{j+1}-x_{j-1}>1$.

Therefore, $\mathcal{N}^\prime_j$ appears to be a relevant proxy of the g-dominated modes density over the whole spectrum. Indeed, we remind that in the previous cases considered in Sects. \ref{Sec:pDom} and \ref{Sec:gDom}, when $\mathcal{N}(\nu_{j+1})>1$, then $\mathcal{N}^\prime_j>1$ and $\delta \nu_{2,j}<1$, and when $\mathcal{N}(\nu_{j-1})<1$, then $\mathcal{N}^\prime_j<1$ and $\delta \nu_{2,j}>1$. Adding in the results of the present section, we thus conclude that over the whole spectrum:
\algn{
&0<\delta \nu_{2,j}<2 \nonumber \\
&{\rm sgn}\left( \delta \nu_{2,j}-1\right)={\rm sgn}\left(1-\mathcal{N}^\prime_j \right) \; ,
}
where sgn() denotes the sign function.

\subsection{Illustration}
\begin{figure}[h]
\includegraphics[width=\linewidth]{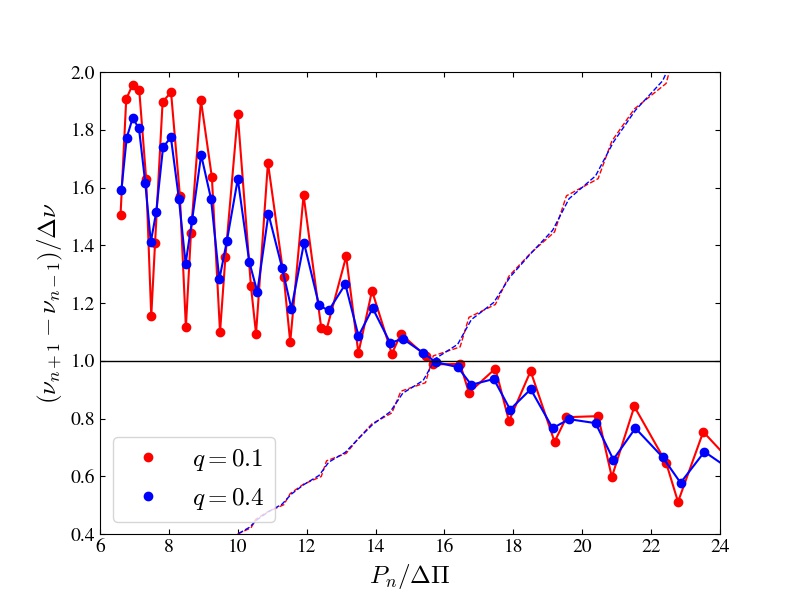}
\caption{Second frequency difference $\delta\nu_2$ as a function of the normalised period of mixed modes (dots). The red and blue colours are for $q=0.1$ and $q=0.4$, respectively. It is plotted for a typical value of $\Delta \nu \Delta \pi_1=200$. The corresponding dashed curves show the evolution of $\mathcal{N}^\prime$.}\label{Fig:Tran}
\end{figure}

We illustrate in Fig. \ref{Fig:Tran} the evolution of the second difference (Eq. \eqref{Eq:SecDif}) with the reduced period, obtained by solving Eq. \eqref{Eq:Shi}. We show this evolution for two choices of the coupling factor $q=0.1$ and $q=0.4$ in red and blue, respectively. We also show, as dashed lines, the evolution of the alternate definition for the g-dominated modes density $\mathcal{N}^\prime$ in Eq.~\eqref{Eq:N prime}.
We observe that $\mathcal{N}^\prime$ and the second frequency difference cross at a value of $1$, as expected from the previous sections. As a consequence, we may locate the transition where $\mathcal{N}_j=1$ using the second frequency difference.

\section{Deriving $\zeta^\prime$}\label{Sec:Zeta}
With analogous reasoning as in \cite{2015A&A...584A..50M}, we may express the variation of frequency with the mixed-mode radial order, $n=n_p-n_g$. Assuming the spectrum to be dominated by the pressure modes, $\mathcal{N} \ll 1$, we consider that the frequency of a mixed mode experiences a perturbation from the evenly space frequencies, $\eta$. We write:
\begin{equation}
\nu = n \Delta\nu + \eta.
\end{equation}
Because of periodicity, when introducing this relation in the phase of pressure modes, $\theta_p$ (Eq. \eqref{Eq:tp}), $\tan \theta_p$ becomes
\begin{equation}
\tan \left[\pi \left(\frac{\eta}{\Delta\nu} - \epsilon_p\right)\right].
\end{equation}
The derivation of Eq. \eqref{Eq:Shi} with respect to $n$, assuming the five mixed-mode parameters to be constant with $n$, then yields
\begin{equation}
\frac{1}{\Delta\nu \cos^2 \theta_p}\frac{d \eta}{d n} = -\frac{q}{\Delta\pi_1 \nu^2 \cos^2 \theta_g}\frac{d \nu}{d n}.
\end{equation}
Finally, using the relation $\cos^2 \theta_p = \frac{\cos^2 \theta_g}{q^2 \sin^2 \theta_g+\cos^2\theta_g}$ (obtained from Eq. \eqref{Eq:Shi}), using $\eta = \nu-n \Delta\nu$, and the definition of the g-dominated modes density evaluated in $\nu$ (Eq. \eqref{Eq:N}), we retrieve the final expression:
\begin{equation}
\frac{d \nu}{d n} = \Delta\nu \left[1 + \frac{q\mathcal{N}}{\cos^2 \theta_g + q^2 \sin^2\theta_g} \right]^{-1}. \label{Eq:zetap demo}
\end{equation}
We note that $\theta_g$ and $\mathcal{N}$ in Eq.~\eqref{Eq:zetap demo} are two functions of frequencies provided by Eqs.~\eqref{Eq:tg} and \eqref{Eq:N}.

\section{Radial order difference between successive modes}\label{Sec:Deln}
Based Eq. \eqref{Eq:DPi}, along with the fact that between two g-dominated modes there may exist a p-dominated mode, it is not obvious that $\Delta n_g$ should be equal to zero or one. By carefully studying the behaviour of first and second frequency differences, Appendix \ref{Sec:SecDif} provides a justification for its value. Nevertheless, to focus only on the $\Delta n_g$ parameter, we follow a slightly different but equivalent approach in the present section.

From Appendix \ref{Sec:SecDif}, we know that the local value of the period spacing is at most equal to the asymptotic value $\Delta\pi_1$. Furthermore, from the ordering of frequencies, $\Delta P_i$ must be positive. We thus have (from Eq. \eqref{Eq:DPi}):
\begin{equation}
0 < \Delta n_g + \Delta\psi_i / \pi \leq 1.
\end{equation}
Also, we have that $\theta_p$ is an increasing function of the frequency (see Eq. \eqref{Eq:tp}), thus $\theta_{p,i} < \theta_{p,i+1}$. As the $\arctan$ function is continuous and monotonous and the $\tan$ function is continuous and monotonous over a given interval $\theta_p\in\left[k\pi -\pi/2, k\pi +\pi/2 \right], k \in \mathbb{N}$, the $\psi_i = \arctan\left(\tan \theta_{p,i}/q \right)$ is continuous and monotonous over the same interval. In addition, in such an interval, $\psi_i$ increases with $\theta_{p,i}$. Thus, $\Delta\psi_i < 0$. Using the definition of the $\arctan$ function, we know that $\psi \in \left]-\pi/2,\pi/2\right[$ and $\Delta\psi_i/\pi$ must be greater than $-1$.
We may thus conclude that:
\begin{equation}
0 < \Delta n_g < 2,
\end{equation}
and, as $n_g$ only takes integer values, $\Delta n_g = 1$.
This demonstration does not hold in the case where two successive modes span over different intervals $\theta_p\in \left[k\pi -\pi/2, k\pi +\pi/2 \right], k \in \mathbb{N}$, the $\tan$ function is discontinuous. In that case, $\Delta \psi_i>0$ and it is necessary to have $\Delta n_g=0$ to ensure $\Delta P_i/\Delta \pi_1 <1$. Physically speaking, this corresponds to the case when we alternate between a g-dominated and a p-dominated mode and this is the pressure radial order that changes, keeping a constant $n_g$ value.

Finally, with an analogous reasoning, we may conclude, for the case of p-dominated spectra that $\Delta n_p$, appearing in Eq. \eqref{Eq:Dnui}, must also equal $1$ for two successive p-dominated modes. Again, when two successive modes span over different intervals $\theta_g\in \left[k\pi -\pi/2, k\pi +\pi/2 \right], k \in \mathbb{N}$, we alternate between a p-dominated mode and a g-dominated mode, and the gravity radial order changes, keeping in a constant $n_p$ value.

\end{appendix}

\end{document}